\documentclass[11pt, a4paper]{article}

\usepackage{jheppub}

\usepackage{mathrsfs}
\usepackage[T1]{fontenc}
\usepackage{mathpazo}
\usepackage{setspace}
\usepackage{amsfonts}
\usepackage{amssymb}
\usepackage{amsmath}
\usepackage{epsfig}
\usepackage{latexsym}
\usepackage{color}
\usepackage{graphicx}
\usepackage{nicefrac}
\usepackage[latin1]{inputenc}
\usepackage{slashed}
\usepackage{multirow}
\usepackage{hyperref}

\def\be{\begin{equation}}
\def\ee{\end{equation}}
\def\ba{\begin{eqnarray}}
\def\ea{\end{eqnarray}}

\def\del{\partial}

\def\a{\alpha}

\def\d{\delta}

\def\e{\epsilon}

\def\C{\Chi}
\def\th{\theta}

\def\m{\mu}

\def\s{\sigma}
\def\S{\Sigma}

\def\cR{{\cal R}}

\def\IR{\relax{\rm I\kern-.18em R}}

\def\inv{^{\raise.0ex\hbox{${\scriptscriptstyle -}$}\kern-.05em 1}}

\def\diag{{\rm diag}}

\def\arctanh{{\rm arctanh}}

\title{Resurgence in $\eta$-deformed Principal Chiral Models}

\author[a]{Saskia Demulder,}
\author[b]{Daniele Dorigoni}
\author[a]{and Daniel C. Thompson}

\affiliation[a]{Theoretische Natuurkunde, Vrije Universiteit Brussel, and
The International Solvay Institutes\\
Pleinlaan 2, 1050, Brussels, Belgium}
 \affiliation[b]{Centre for Particle Theory \& Department of Mathematical Sciences,\\
Durham University, Durham, UK}

 \emailAdd{Saskia.Demulder@vub.ac.be}
  \emailAdd{daniele.dorigoni@gmail.com}
\emailAdd{Daniel.Thompson@vub.ac.be}

\abstract{We study the $SU(2)$ Principal Chiral Model (PCM) in the presence of an integrable $\eta$-deformation.
We put the theory on $\mathbb{R}\times S^1$ with twisted boundary conditions and then reduce the circle to obtain an effective quantum mechanics associated with the Whittaker-Hill equation. Using resurgent analysis we study the large order behaviour of perturbation theory and recover the fracton events responsible for IR renormalons. The fractons are modified from the standard PCM due to the presence of this $\eta$-deformation but they are still the constituents of uniton-like solutions in the deformed quantum field theory. We also find novel $SL(2,\mathbb{C})$ saddles, thus strengthening the conjecture that the semi-classical expansion of the path integral gives rise to a resurgent transseries once written as a sum over Lefschetz thimbles living in a complexification of the field space. 
We conclude by connecting our quantum mechanics to a massive deformation of the $\mathcal{N}=2~$ $4$-d gauge theory with gauge group $SU(2)$ and $N_f=2$. }


\def\beq{\begin{equation}}
\def\eeq{\end{equation}}
\def\bea{\begin{eqnarray}}
\def\eea{\end{eqnarrat}}
\begin{document}

\def\C{{\cal C}_\zeta}
\def\S{{\cal S}_\zeta}
\def\T{{\cal T}}
\def\R{{\cal R}}
\def\ii{{\textrm{i}}}
\def\tr{{\textrm{tr}}}
\maketitle
\flushbottom

\section{Introduction}

\subsection{The Resurgence Paradigm}
The calculation of  the anomalous magnetic moment of the electron to ten significant figures stands as testament to efficacy of perturbation theory in quantum field theory (QFT) \cite{Aoyama:2012wj}. The fact that perturbative expansions, even in the simpler setting of quantum mechanics (QM), are quite generally divergent may then be a cause of some cognitive dissonance to the practicing physicist. Resurgence theory is perhaps the therapy that perturbation theory demands.

 Consider, for example, a ground state energy obtained as a perturbative expansion in some real and positive coupling $g^2$,
\be\label{eq:Epert}
E^{pert} = \sum_{n=0}^\infty c_n (g^2)^n \ .
\ee
Due to the proliferation of Feynman diagrams it is quite generic to find factorial growth $c_n \sim A^n n!$ and the perturbative expansion, whilst initially providing increasingly accurate results, will begin to diverge.  Borel summation can provide a way to attach a meaning to  such asymptotic series. In this  method the factorial growth is essentially traded for an integral via $n! = \int_0^\infty \mathrm{d}t e^{-t} t^n$.  One first constructs the Borel transform\footnote{In practice one might not have access to the full infinite series \eqref{eq:Epert} but only a truncation in which case it is necessarily to use some approximation to the full series, e.g.  a  Pad\'e approximant, as an input to this Borel transformation. }, 
\be
{\cal B}[E^{pert} ](t) = \sum_{n=0}^\infty \frac{c_n}{n!} t^n\ , 
\ee 
which will have a finite radius of convergence and then one can perform the resummation, 
 \be\label{eq:borelresum}
{\cal S}[E^{pert} ](g^2) = \frac{1}{g^2} \int_0^\infty {\cal B}[E^{pert} ](t) e^{-\frac{t}{g^2}} \mathrm{d}t  \ .
\ee 
There is some danger here!  The integral \eqref{eq:borelresum} may not be well defined; the Borel transformed series can have poles along the integration path. It is oft said that this occurs when the coefficients $c_n$ are non-alternating but even an alternating series can be non-Borel summable and indeed we will encounter this.   In this case it is natural to deform the integration contour by extending $t$ to the complex plane and integrating along a ray i.e.,
  \be\label{eq:borelresumv2}
{\cal S}_\theta[E^{pert} ](g^2) = \frac{1}{g^2} \int_0^{e^{\ii\theta} \infty} {\cal B}[E^{pert} ](t) e^{-\frac{t}{g^2}} \mathrm{d}t  \ ,
\ee 
which is called directional Borel resummation.
If $ {\cal B}[E^{pert} ](t) $ is regular for $\mbox{arg} (t) = \theta$ (and decays sufficiently fast at infinity), then ${\cal S}_\theta[E^{pert} ](g^2) $ defines an analytic function in the wedge of the complex coupling plane $-\pi/2+\theta\leq \mbox{arg} (\,g^2) \leq \pi/2+\theta$.

If $ {\cal B}[E^{pert} ](t) $ has singularities along the direction $\mbox{arg} (t) = \theta$, this is called a Stokes direction.
For example if the original direction of integration (\ref{eq:borelresum}), $\mbox{arg} (t)= 0$, is a Stokes direction we can dodge the singularities by picking $\theta = \epsilon>0$ small and positive, in this way we avoid poles in the original integration cycle and we can calculate the well defined ${\cal S}_\epsilon [E^{pert} ](g)$. However we could equally choose $\theta = - \epsilon<0$ and calculate  ${\cal S}_{-\epsilon} [E^{pert} ](g)$. 

If $ {\cal B}[E^{pert} ](t) $ has singularities in the complex wedge $-\epsilon\leq \mbox{arg}( \,t)\leq \epsilon$, these two lateral summations, ${\cal S}_{\pm\epsilon} [E^{pert} ](g)$, are generically different but still yield the same asymptotic perturbative expansion (\ref{eq:Epert}) once expanded at weak coupling. 
This jumping as one deforms an integration cycle is known as the Stokes phenomenon.  

Generically, the original real positive coupling arg$(g^2)=0$ lies on a Stokes line -- the location at which such a jump takes place.  Thus we have traded our initial asymptotic series expression for the ground state energy for a finite but ambiguous result. Evidently this is still unsatisfactory; physical observables surely shouldn't be blighted with such ambiguity.   

We have of course missed the vital part of the physics; the non-perturbative sector. In these cases one should include in the path integral contributions from non-perturbative saddle points and the fluctuations around them.   These non-perturbative contributions quite certainly correct the real part of the energy as in quantum mechanics where instanton effects can lift perturbative degeneracies or give band structures.   These non-perturbative contributions are, however, by themselves ambiguous. As shown by Bogomolny \cite{Bogomolny:1980ur} and Zinn-Justin \cite{ZinnJustin:1981dx}, even the leading contribution to a path integral from integration over the quasi-zero-mode associated to separated instanton $I$ and anti-instanton ${\bar{I}}$ pairs has an ambiguous imaginary part.  Far from compounding our problems, this is just the medicine needed -- the $[I\,\bar{I}]$ ambiguity exactly cancels that of the perturbative sector.   This is remarkable: the divergent behaviour of late terms in the perturbative sector is compensated by early terms in a non-perturbative sector.   This is the first example of resurgent behaviour. 

The story runs deeper; the perturbative series around the one instanton sector $[I]$ will itself also have factorial growth and gives rise, in a similar fashion,  to an ambiguous imaginary part that will cancel against early terms in the $[I\,I\,\bar{I}]$ sector and so on.  
The resurgence conjecture is then that the unambiguous physical observable can be understood as an example of \'Ecalle's resurgent trans-series \cite{Ecalle:1981} of the form\footnote{For simplicity we are writing here only height-$0$, log-free transseries, see \cite{Edgar} for more details.},
\be\label{eq:TS}
E(g^2) = E^{pert} + \sum_{i} e^{-\frac{S_i}{g^2}} \sum_{k>0} c^{(i)}_k (g^2)^k   \, ,
\ee
in which in addition to the perturbative sector one should sum over all non-perturbative saddles labelled by $\{i \}$ with corresponding actions $S_i$.  

From a path-integral perspective the transseries representation of physical observables looks exactly like the semi-classical approximation. The perturbative expansion around the standard vacuum configuration gives rise to an asymptotic series without any exponentially suppressed factor sitting in front, while from each non-trivial saddle point we obtain an exponentially suppressed factor, coming from the classical action of the saddle, times its associated asymptotic perturbative expansion.

A not so widely appreciated fact about the semiclassical expansion is that the amplitudes associated with certain saddle points, for example the aforementioned instanton-anti-instanton events, are not well-defined precisely along the Stokes line $g^2$ real. This has a beautiful geometric origin, we can decompose \`{a} la Morse \cite{Witten:2010cx} the original contour of integration over the space of fields into the sum of certain privileged contours of integrations called Lefschetz thimbles. As we vary the argument of the coupling constant two things will happen: first the semi-classical expansion will generically receive contributions from all finite action \textit{complex} saddles living in a \textit{complexification of the space of fields} and secondly precisely at Stokes line this thimble decomposition will jump discontinuously, that
is, in addition to the jump in the resummation of the perturbative series, the sum over non-perturbative
saddle points will also jump and the two ``ambiguities'' will precisely cancel.

After we sum over all the finite action complexified saddles, the semiclassical expansion (\ref{eq:TS}) should not be
thought of as an approximation, a resurgent trans-series is an exact coded representation of the physical observable, i.e. it really is an analytic function in a certain wedge of the complex coupling plane.

\subsection{Resurgence in QFT and the PCM}

The general philosophy above is expected to hold in quantum field theories in which both perturbative and non-perturbative contributions play a role.  Factorially divergent series abound in $4$-d Yang-Mills or QCD  arising both from growth in diagrammatica {\em and} from the so-called renormalons \cite{Beneke:1998ui} coming from IR and UV contributions to loop momenta integration. This standard renormalons diagrams argument suggest that generically, in asymptotically free theory, we should expect singularities in the Borel transform along the real axis.
However, once we consider these theories on $\mathbb{R}^3 \times S^1$, due to the presence of non-trivial holonomies around the circle, we see \cite{Anber:2014sda} that renormalon type divergences are actually reproduced by the combinatorics of diagrams getting mixed with the discretized momenta along the $S^1$. 
We refer to IR/UV renormalons as singularities in the Borel transform along the positive/negative real axis that remain at finite locations once we take the approriate large-$N$ limits, without making any reference to their diagrammatic origins or otherwise\footnote{We thank Tin Sulejmanpasic for discussions on this issue.}.


As a first step towards full QCD, it is natural to study the ideas and implications of resurgence in a simpler setting of two-dimensions \cite{Cherman:2014ofa}.  The $SU(N)$ Principal Chiral Model (PCM), whose action is given by,
\be\label{eq:sPCM}
S_{PCM} = \frac{1}{2\pi t} \int \mathrm{d}^2 \sigma \,  \tr \left[ \partial_+g g^{-1} \partial_-g g^{-1}\right] \ , 
\ee
with $g\in SU(N)$, provides a rather effective toy model to QCD since it displays asymptotic freedom, a dynamically generated mass gap and confinement. Being an integrable theory in two-dimension \cite{Polyakov:1983tt,Wiegmann:1984ec} the PCM has the advantage of also being sufficiently simple to remain tractable; for instance the mass gap can be exactly determined \cite{Hasenfratz:1990zz,Balog:1992cm}.  Although the $SU(N)$ PCM does not possess instantons (in the sense that $\pi_2[SU(N)] = 0 $ and there is no topological charge),  it does have a non-perturbative sector whose constituents, known as {\em unitons} \cite{uhlenbeck1989harmonic}, are finite action field configurations that solve second order Euclidean equations of motion, i.e. they are not BPS objects but genuine saddles \cite{Piette:1987ia}.  

In \cite{Cherman:2013yfa,Cherman:2014ofa} the resurgent structure of the two-dimensional $SU(N)$ PCM was studied.  The technique employed was to study the spatially compactified  Euclidean theory with twisted boundary conditions corresponding to turning on a background gauge field for the diagonal $SU(N)\subset SU(N)_L \times SU(N)_R$  global symmetry of the theory.  This regime is adiabatically connected to the PCM defined on $\mathbb{R}^2$ and many features of the original theory remain visible in an effective one-dimensional quantum mechanics that comes when taking the radius of spatial compactification very small.  In particular,  the large order behaviour of this QM exhibits non-alternating factorial growth with poles along the positive real half-line of the Borel plane.  These poles can be matched precisely to certain instanton-anti-instanton configurations in the QM.  What is quite astonishing is that the origin of these non-perturbative saddles can be exactly identified in the two-dimensional quantum field theory -- they correspond to ``fractons''. When the PCM is put on   $\mathbb{R} \times S_1$  the unitons fractionalize and break up into these fractons when their size becomes comparable to the radius of compactification. In this way, the fractons lead to a semi-classical realisation of IR renormalons and the mass gap in the circle-compactified theory.

This is a generic feature of $2$-d (and perhaps $4$-d) asymptotically free QFTs as for example the $\mathbb{CP}^N$ model \cite{Dunne:2012zk,Dunne:2012ae}, the Grassmanian models and the $O(N)$ model \cite{Dunne:2015ywa}. In particular, the mass gap of the $O(6)$ sigma model directly relates to non-perturbative scales in AdS/CFT \cite{Basso:2009gh} and more specifically it translates to non-perturbative effects in the $4$-d $\mathcal{N}=4$ Super-Yang-Mills cusp anomalous dimension, whose resurgence properties were the subject of \cite{Aniceto:2015rua,Dorigoni:2015dha}.   We should emphasise resurgence is {\underline{not}} the statement that the perturbative sector encodes {\em all} non-perturbative information.  Instead resurgence can allow sectors of a theory of a given fixed topological charge to communicate when need to cancel ambiguities - this idea is called the resurgence triangle \cite{Dunne:2012zk}.  In some cases extra symmetries can mean certain sectors simply do not exist and the resurgence triangle can truncate  \cite{Dunne:2012zk}.  

  Whilst our main focus in the present context is in $2$-d QFTs, important progress has already been made in higher dimensional QFTs in the context of  3d Chern-Simons matter (ABJM) and ${\cal N}=2$ supersymmetric 4d Yang-Mills gauge theories \cite{Aniceto:2014hoa} building on \cite{Russo:2012kj}, where supersymmetric localisation allows for comparison to exact quantities. Recently this was further developed in \cite{Honda:2016mvg} and extended to 5d ${\cal N}=1$ supersymmetric theories.  The examples of 4d ${\cal N}=2$ \cite{Russo:2012kj,Aniceto:2014hoa} and 5d  ${\cal N}=1$  \cite{Honda:2016mvg} supersymmetric gauge theories show that the perturbative expansion is Borel summable within a given topological sector--presumably as a consequence of the extended supersymmetry--and the Borel resummation does not require communication between different sectors. 

\subsection{Resurgence Puzzles}
The study of resurgence has thus far proven to be very profitable but let us mention two puzzling features that the present study will help to shed light on:
\begin{enumerate}
\item {\em Complex Saddles and QFT}.   The situation described above in which the cuts in the Borel plane lie  on either the positive or the negative real axis is not at all generic.  In fact we should expect that in general the singularities might be scattered across the entire complex plane.  To understand how resurgence works in these cases will require a better understanding of the non-perturbative saddles that give rise to such singularities.  These will typically be associated to complex field configurations that will need to be understood in a field theory context. 
\item {\em Multiple couplings and Borel flow}.  The Lagrangian defining the QFT may have several bare parameters and complicated RG flows.  In this case it is not obvious how to understand physical observables as resurgent trans-series. Furthermore the semi-classical regime in which we will apply resurgence is only adiabatically connected to the original quantum field theory. The structure of the Borel plane does not remain invariant under this adiabatic flow and having multiple couplings, together with RG-invariant quantities, might tell us how this Borel flow takes place not only qualitatively but also quantitatively.
\end{enumerate}
 
 \subsection{The $\eta$-Deformed Principal Chiral Model}
To advance the study of resurgence in QFT and target the puzzling features, we would like to find a candidate theory that is rich enough to include a variety of non-perturbative sectors and multiple couplings but yet remain tractable.  A class of deformations of the PCM chiral model that has received substantial recent interest are the Yang-Baxter deformed $\sigma$-models introduced by Klimcik \cite{Klimcik:2002zj,Klimcik:2008eq}.   These theories, which exactly meet our requirements, are specified by a certain matrix ${\cal R}$, that we shall detail later, and a parameter $\eta$ (which {\em a priori} should be taken to be real) and are defined by the action,   
\be\label{eq:Seta}
S_{\eta} =  \frac{1}{2\pi t} \int \mathrm{d}^2 \sigma \,  \tr \left[ \partial_+g g^{-1}  \frac{1}{1 + \eta {\cal R}} \partial_-g g^{-1}\right] \, .
\ee
Whilst the deformation breaks the global symmetry down to $SU(N)_L \times U(1)_R$, the theory remains integrable (at least classically) and thus amenable to study.  The broken $SU(N)_R$ symmetry is not completely lost; its avatar is present in the structure of non-local charges whose  Poisson bracket relations give rise to  a classical version of a quantum group ${\cal U}_q (\frak{sl}(N))$ with parameter $q = \exp \left( \pi \eta t \right)$ \cite{Delduc:2013fga}.  This quantum group symmetry is expected to be manifest in the exact S-matrix of the theory.   

This Yang-Baxter deformation, after suitable modifications, can be equally well applied to the PCM on symmetric or semi-symmetric spaces (supercosets) \cite{Delduc:2013qra} including the well-known example of $PSU(2,2|4)/SO(4,1)\times SO(5)$ relevant to superstrings in $AdS_5 \times S^5$ \cite{Arutyunov:2013ega}. This offers intriguing prospects as a Lagrangian description for quantum group deformations of holography.  In the later context the Yang-Baxter $\sigma$-model is often called an $\eta$-deformation and we adopt this nomenclature. 

  There is much left to be understood about $\eta$-deformed $AdS_5\times S^5$ \cite{Arutyunov:2015qva,Arutyunov:2015mqj} and in particular the status of the deformed $\sigma$-model as a CFT is questionable but that concern is irrelevant to our current investigation.  It would be remiss not to mention that a related class of integrable theories \cite{Sfetsos:2013wia,Hollowood:2014rla,Hollowood:2014qma} known as $\lambda$-deformations has recently been developed that extend the idea of a current-current perturbation of a (gauged-)Wess-Zumino-Witten model.  These $\lambda$-deformations are classically related to \eqref{eq:Seta} via a Poisson-Lie T-duality together with some analytic continuations \cite{Hoare:2015gda,Sfetsos:2015nya,Klimcik:2015gba}.  This class of theories is expected to display a similar quantum group symmetry but with the parameter $q=\exp \left( \frac{\mathrm{i} \pi}{k} \right)$ a root of unity. There is by now good evidence that $\lambda$-deformed $\sigma$-models based on semi-symmetric spaces have target space times that solve the equations of type II supergravity and may define CFTs \cite{Sfetsos:2014cea,Demulder:2015lva,Borsato:2016zcf}. For a recent review see \cite{Thompson:2015lzd}.

\subsection{Summary of results}
In a fortuitous case of symbiosis, in this work we apply the ideas of resurgence to the $\eta$-deformed $\sigma$-model and draw concurrent insight about the technique of resurgence itself {\em and} the structure of the $\eta$-deformed theory.   In this work we focus on the simplest case of the $SU(2)$ $\eta$-deformed principal chiral model -- the extension to higher rank groups should be quite similar and will be pursued elsewhere.  Let us summarise some of the key findings:
\begin{itemize}
\item {\em Deformed unitons.}  We demonstrate explicitly that the known uniton solutions of the PCM persist in a  suitably deformed sense as classical saddles of the action \eqref{eq:Seta}.  These give rise to finite, and quantised, actions which are not classified by the usual topological instanton charge. 
\item  {\em Complex saddles.}   Intriguingly we find {\em new} complex saddles of \eqref{eq:Seta} whose actions are finite but as $\eta\rightarrow 0$ diverge and so do not contribute to the undeformed PCM.
\item {\em Dimensional reduction.}  After applying a twisted spatial compactification the $\eta$-deformed theory gives rise to a quantum mechanics with a two-term periodic  potential of the Whittaker-Hill type (known to be a quasi-integrable QM \cite{Hemery:2010:WHE,MagnusWinkler}).  This theory has both instanton configurations {\em and} complex field configurations whose actions match respectively those of the fractons constituents of the unitons and the complex saddles in the two-dimensional QFT. 
\item {\em Large order behaviour and resurgence.}  The perturbative expansion of the ground state energy gives an asymptotic series whose Borel transform has cuts in {\em both} the positive and the negative half line.  The leading and first sub-leading terms of the large order perturbation theory are precisely tied to the above non-perturbative saddles in line with expectations from resurgence. 
\item {\em Critical coupling $\eta_c$.}  For $\eta > \eta_c \approx 0.27$ the dominant contribution to the large order behaviour of the perturbative energy coefficients, and the singularity nearest the origin (but on the negative real axis) in the Borel plane, comes not from instanton-anti-instanton events but from the complex saddle.  The large order perturbation theory for $\eta>\eta_c$ is alternating in sign but, due to the sub-dominant contribution coming from the singularity in the positive half line associated with instanton-anti-instanton events, this series remains non-Borel summable along the positive real line. 
\item {\em Analytic continuation in $\eta$.}  We consider also the case of pure imaginary $\eta = \ii \eta_{\mathbb{R}}$, with $\eta_{\mathbb{R}}\in\mathbb{R}$.  The pole structure on the Borel plane now includes similar cuts in the positive real axis and also a complex conjugate pair of cuts (with positive real part) that correspond to complex saddles.  There are three regimes: $0\leq \eta_{\mathbb{R}} \lessapprox 0.4$  in which the singularity along the real axis dominate;  $ 0.4 \lessapprox \eta_{\mathbb{R}} \leq \frac{1}{\sqrt{2}} $ in which the dominating large order behaviour is due to the complex conjugate pairs of singularities;  $ \frac{1}{\sqrt{2}} < \eta_{\mathbb{R}}   \leq  1  $ in which the imaginary part of the complex conjugate pair vanishes and the relevant quantum mechanics has two different real types of instanton corresponding to tunnelling between vacua separated by large and small barriers. 
\item {\em Stokes jumps and Seiberg-Witten theory.}   A Stokes graph consists of trajectories with constant phase of a WKB functional\footnote{We will define this more precisely later}. As the coupling is analytically continued over a Stokes line the topology of this graph will change.  The same construction arises in Seiberg-Witten theory where the topology jumps are a manifestation of Wall Crossing.  We find that the QM of the $\eta$-deformed theory gives a Stokes graph for an $N_f=2$ mass deformation of an $SU(2)$ gauge theory with equal masses for the flavours.  See also the very recent papers \cite{Piatek:2016xhq,Ashok:2016yxz}.
\end{itemize}  
 
\subsection{Outline}
 We proceed in Section \ref{sec:etadef} where we shall first recapitulated the requisite details of the $\eta$-deformed Principal Chiral Model before discussing its non-perturbative semi-classical configurations of the deformed uniton and the complex saddle.  In Section \ref{sec:etareduction} we then consider the theory on $\mathbb{R}\times S^1$ and illustrate   the fractionalisation of the saddle point configurations and give details of the reduction to quantum mechanics.  In Section  \ref{sec:largeorder} we examine the large order behaviour of perturbation theory of this reduced theory paying attention to the location of the poles in the Borel plane.  In Section \ref{sec:Resurgence} we demonstrate the resurgent aspects of this system by comparison of the large order perturbative behaviour to low orders of perturbation theory around non-perturbative saddles. In Section \ref{sec:Stokesgraphs} we discuss the Stokes graphs and match to Seiberg-Witten theory.   We then return to a discussion and interpretation of our results. 
 
\section{The $\eta$-deformed PCM}\label{sec:etadef}

\subsection{PCM Background} 
 The Principal Chiral Model (PCM) for the group $G$  is given by,
\be\label{eq:SPCM}
S_{ \textrm{PCM}}  = \frac{1}{2\pi t} \int_{\Sigma} \mathrm{d}^{2}\s  \, \sum_{a=1}^{\vert G\vert} R_{+}^a R^a_{-} \ ,
\ee
in which $\Sigma$ is the world sheet taken to be $\mathbb{R}^{(1,1)}$ and the right-invariant forms  are $R^a_\pm =  R^a_\m \del_\pm X^\m = {\rm Tr}(T_a  \del_\pm g g^{-1} )$ for generators $T_a$ of the algebra\footnote{See the Appendix for conventions} $\frak{g}$ whose dimension is $\vert G\vert$.   The PCM is asymptotically free and strongly coupled in the IR with a mass scale generated through dimensional transmutation.  As is well known, this theory has a $G_L \otimes G_R$ symmetry under which $g$ transform as $g\to U_L \,g \,U_R$ with $U_{L/R}\in G_{L/R}$ two constant matrices, and is both classically and quantum mechanically integrable  \cite{Polyakov:1983tt,Wiegmann:1984ec}. The S-matrix is two-particle factorisable and hence multi-particle scattering processes can be obtained as a succession of two-particle scattering events.  The S-matrix can be determined from the factorised bootstrap and the theory solved using Bethe-ansatz. \cite{Bogomolny:1980ur,ZinnJustin:1981dx}

Although these theories, after analytic continuation to an Euclidean signature, have no topological sector, $\pi_2[SU(N)] = 0$, they do posses non-perturbative saddle configurations known as unitons introduced in the seminal work of Uhlenbeck \cite{uhlenbeck1989harmonic}.   A uniton is a finite action field configuration solving the Euclidean equations of motion which are a harmonic condition, 
\be
\partial( U^{-1}\bar\partial U ) + \bar\partial( U^{-1} \partial U )  = 0 \ ,
\ee
and obey, modulo a constant rotation,
\be
U\cdot U = - \mathbb{1} \ . 
\ee

Let us specialise to the case of the $G=SU(2)$.  We adopt a Hopf parametrisation for the group element,
\begin{equation}
g = \left(\begin{array}{cc} z_1 & \ii z_2 \\ \ii \bar{z}_2 &\bar{z}_1  \end{array} \right) \ , \quad z_1 = \cos\th  e^{\ii  \phi_1} \ , \quad  z_2 = \sin\th  e^{\ii \phi_2} \ ,
\end{equation}
with ranges $\theta \in [0,\pi]$, $\phi_1\in[0,\pi]$ and $\phi_2\in[0,2\pi]$, such that the PCM \eqref{eq:SPCM} defines a $\sigma$-model whose target space is a round $S^3$ with metric, 
\be
\mathrm{d}s^2 =\frac{1}{t} \left(  \mathrm{d}\theta^2 + \cos^2\theta \mathrm{d}\phi_1^2 + \sin^2\theta \mathrm{d}\phi_2^2 \right) \ .
\ee
The uniton $U: \mathbb{R}^{2} \rightarrow SU(2)$  is specified by a  holomorphic function $f(z)$ ($z= t + i x $)\ , 
\begin{equation}\label{eq:uniton1}
U = \frac{-\ii}{1 + |f|^{2}}  \left( \begin{array}{cc}
1- |f|^{2} & 2  \bar{f} \\ 2 f  &  |f|^{2} -1  
\end{array}\right) \ . 
\end{equation} 
When the function $f(z)$ is a polynomial of degree $k$ we speak of the $k$-uniton and we find that the uniton action is, 
\be
S_{PCM}[U] = \frac{1}{2\pi t} 8\pi k \ , \quad k\in \mathbb{Z}  \ . \label{eq:UnitonS}
\ee
It may seem counter intuitive to have such a quantised action but if we write the uniton in Hopf coordinates, 
\be\label{eq:unitonHopf}
\phi_1 =  \frac{\pi}{2} \ , \quad \phi_2 = \pi + \frac{\ii}{2}\log \frac{f(z)}{\bar{f}(\bar{z})} \ , \quad \cos\theta =  \frac{-1+ |f|^2}{1+|f|^2}  \ , 
\ee
it becomes evident that we are dealing not with general maps into the whole of $S^3$ but rather with maps that are varying over an $S^2$ since the uniton can be seen as the embedding of a $\mathbb{CP}^1$ lump into $SU(2)$.  More precisely the quantisation of the action can be related to a non-contractible loop in field space, i.e. $F= \{ g: S^2 \rightarrow S^3\}$ has $\pi_1(F) = \pi_3(SU(2)) =\mathbb{Z}$, see the beautiful orange book by Manton and Sutcliffe \cite{Manton:2004tk}.

\subsection{The $\eta$-deformed PCM Generalities } 
An interesting class of theories called variously as Yang-Baxter $\sigma$-models and $\eta$-deformations (we prefer the later term) has been obtained by Klimcik \cite{Klimcik:2002zj,Klimcik:2008eq} as integrable deformations of the PCM.  To introduce these we require an operator  ${\cal R}$  acting on the algebra  $\frak{g}$ of a compact bosonic group $G$, that solves the modified Yang-Baxter equation\footnote{Further algebraic details may be found in the Appendix.}, 
\be\label{eq:mYB}
[\R A, \R B] - \R\left( [ \R A,B] + [A , \R B]\right) =  [A, B]  \ ,  \quad \forall  A,B \in \frak{g} \,.  
\ee

Given a solution ${\cal R}$ of the modified YB equation we can consider the general action, 
 \begin{equation}
  \label{eq:YBaction}
  S_{\eta} = \frac{1}{2\pi t} \int  \mathrm{d}^2 \sigma\,  R_+^T (\mathbb{I} - \eta \R )^{-1} R_- \, .
\end{equation}

This is integrable -- at least classically.   Lets look at the equations of motion which, with the introduction of, 
\be
{\cal J}_{\pm} =  {\cal O}_{\pm}^{-1} R_{\pm} \ , \quad  {\cal O}_{\pm} =  \mathbb{I} \pm \eta \R \ , 
\ee 
can be written as,  
 \begin{equation} \label{eq:eqm}
  \begin{aligned}
0 &= \partial_{+} {\cal J}_{-} + \partial_{-} {\cal J}_{+} - [R_{+},{\cal J}_{-}]- [R_{-},{\cal J}_{+}] \ .
\end{aligned}
\end{equation} 
 In addition we have the Bianchi identity, 
\be
0 = \del_+R_--\del_-R_+ - [R_+,R_-]\, . 
\ee
  The classical integrability is then established \cite{Klimcik:2008eq} since  the equations of motion and Bianchi identity follow as a flatness conditions for a spectral  parameter dependent Lax connection ${\cal L}(\mu)$, 
\be\label{eq:Lax}
 {\cal L}_\pm(\mu)=\left[ \left( - \eta^{2} + \frac{1+\eta^{2}}{1\pm \mu }\right)  \mathbb{I}\pm\eta\cR\right] {\cal J}_{\pm }\, , \quad 
 \del_+{\cal L}_--\del_-{\cal L}_+=[{\cal L}_+,{\cal L}_-]\, .  
\ee

In the simplest rank $1$ example, $G=SU(2)$, there is a unique ${\cal R}$ that acts on generators normalised as $T_{i} = \frac{1}{\sqrt{2} }\sigma_{i}$ by,
 \be
 {\cal R}: \{ T_{1},  T_{2}, T_{3} \} \rightarrow \{ -T_{2}, T_{1}, 0 \} \ , 
 \ee 
 and the matrix entering the deformed PCM is,
\be
{\cal O}_{-}^{-1} =   (\mathbb{I} - \eta \R )^{-1}= \frac{1}{1+\eta^{2}} \left( \begin{array}{ccc}
     1& -\eta  & 0 \\ 
    \eta & 1& 0\\ 
    0 &0 & 1+\eta^{2} \\ 
  \end{array}\right) \ . 
\ee
 Then the deformed action is simply, 
\begin{equation} \label{eq:SetaR}
  \begin{aligned}
  S_{\eta} & = \frac{1}{2\pi t}\frac{1}{1+\eta^{2}} \int \mathrm{d}^2 \sigma\,\left[ R^{1}_{+}R^{1}_{-}+ R^{2}_{+}R^{2}_{-} + (1+\eta^{2})R^{3}_{+}R^{3}_{-}  - \eta ( R^{1}_{+}R^{2}_{-} - R^{2}_{+}R^{1}_{-}  ) \right]\ . 
  \end{aligned}
\end{equation}
As a $\s$-model we have a target space whose metric is a squashed three-sphere equipped with a pure gauge NS two-form,   
\be
\mathrm ds^{2} = \frac{1}{t(1+\eta^{2})} \left(R_{1}^{2} +  R_{2 }^{2} + (1+\eta^{2})R_{3}^{2} \right)\ , \quad B_{2} = \frac{\eta}{t (1+\eta^{2})} \mathrm d(\cos \theta  \mathrm d\psi) \ . 
\ee
Since the B-field is pure gauge in this case there is no danger in allowing $\eta$ to become pure imaginary.  In this case one see that $\eta^2 = -1$ corresponds to the $O(3)$ sigma model (the action will develop a gauge invariance allowing one to reduce the degrees of freedom to just an $S^2$'s worth), whilst $\eta=0$ is the $SU(2)$ PCM, or equivalently, the $O(4)$ sigma-model.  This interpolating theory was used in the classic work of Polyakov and Wiegmann \cite{Polyakov:1983tt}. Using TBA, the work \cite{Balog:1996im}  established the exact mass gap of the theory. The classical integrability of the $\sigma$-model on the squashed three-sphere was established long ago \cite{Cherednik:1981df} and an understanding of the symmetries as $q$-deformations was developed in \cite{Kawaguchi:2011pf}. 

\begin{figure}[h!]
   \begin{center}
    \includegraphics[width=6.5cm]{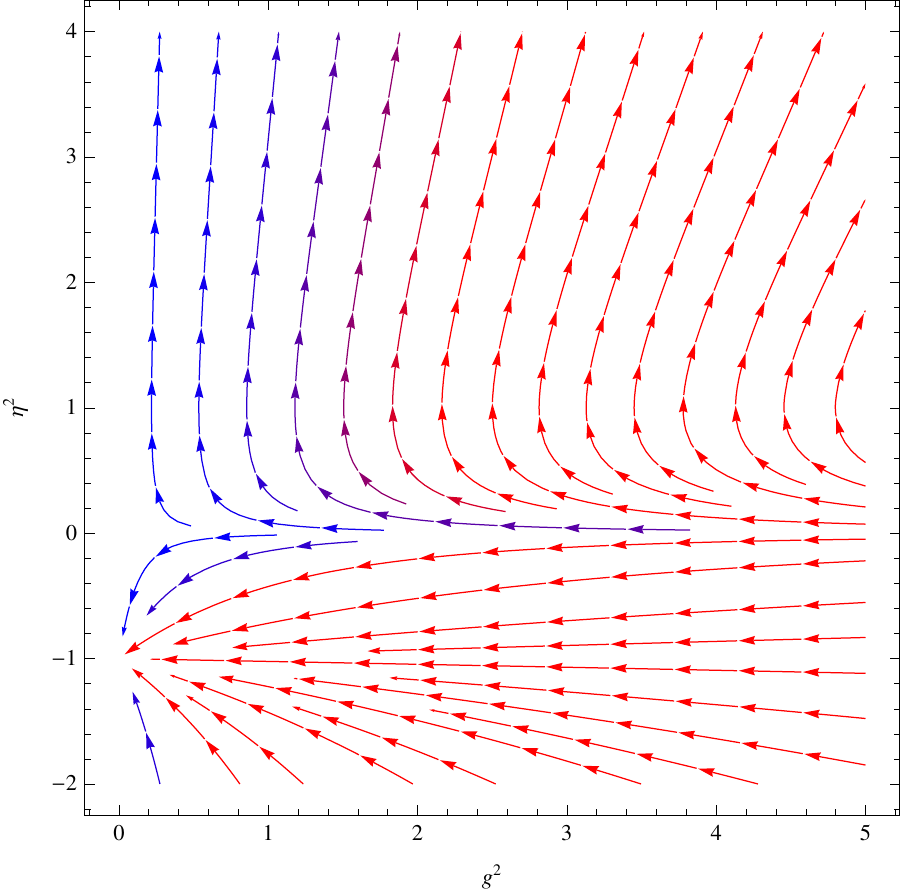}    
 \end{center}
  \caption{\footnotesize{ A phase portrait of the RG evolution of the couplings $g^2$ (horizontal) and $\eta^2$ (vertical).  The arrows point towards the UV and the colouring is such that from blue to red the absolute value of the RG invariant parameter $\zeta^2$ increases in magnitude ($\zeta^2>0  $ for $\eta^2>0$). }}  \label{fig:Rgstreamplots}
\end{figure}

Quantum mechanically, of the two couplings $\eta$ and $t$, there is an RG invariant parameter  $\zeta= t\, \eta$ and  a single RG flow equation chosen between, 
\begin{align}
 \frac{\mathrm d \,t}{\mathrm d \log \mu } & \label{eq:RGYB} =-\frac{c_G }{4}\,t^2\,\left(1+ \eta^2\right)^{2} \,,\\
\frac{\mathrm d\,\eta }{\mathrm d \log \mu } &=\frac{c_G\, \zeta }{4}\,\left(1+ \eta^2\right)^{2} \ , 
\end{align}
 where $c_{G}$ is the quadratic Casimir in the adjoint. A more convenient combination, that is closer to an overall coupling that sits outside the action  eq.~\eqref{eq:YBaction}, is given by, 
 \be
  g^2 = t (1+\eta^2)   \ , \quad  \frac{\mathrm d\,g^2}{\mathrm d \log \mu }=  \frac{c_G}{4} g^4 ( -1 + \eta^2) \, , 
 \ee
 in terms of which   $\zeta$ is given as,
 \be
 \label{eq:RGinv}
 \zeta^2 = g^4\frac{\eta^2}{(1+\eta^2)^2}\, . 
 \ee 
 The solution in parametric form (for $\zeta>0$) is, 
 \be
 \log \mu/\mu_0 = -\frac{1}{c_G \zeta} \arctan \frac{2 \zeta }{ \sqrt{g^4 -4 \zeta^2}  } + \frac{2}{c_G g^2} \,. 
 \ee
In Figure~\ref{fig:Rgstreamplots}  we plot the phase portrait of the RG flow of the couplings $g^2$ and $\eta^2$ allowing $\eta^2$ to take both positive and negative values. For $\eta^2 \leq 0$ the RG invariant quantity $\zeta^2\leq 0$ and the theory is asymptotically free \cite{Balog:1996im}
in the coupling $g^2$ with the coupling $\eta^2$ flowing to $-1$ in the UV. 
There are two fixed points, one at $\eta^2=-1$ which corresponds to the $O(3)$ $\sigma$-model, and the second one for $\eta^2=0$ which is the undeformed PCM. Note that the $\eta=0$ fixed line is ''unstable`` under this deformation, any non-zero $\zeta$ will automatically make us flow away from it.
Although the figure illustrates also the flow for $\eta^2<-1$ one should keep in mind that the target space metric will not have a positive definite signature.  

For the case of $\eta^2 >0$ we have that the RG invariant quantity $\zeta^2> 0$ and the situation is different; the theory in the IR is strongly coupled in $g^2$ as before however as we approach the UV this coupling will decrease to a minimum value $g^2_{min} = 2 \zeta $  which occurs when $\eta =1$ at a scale $\log \mu/\mu_0  \approx 1/\zeta$.   This minimum value of $g^2$ is not a fixed point of the RG flow since the running of $\eta^2$ continues at this point and then for higher energies the theory becomes once again strongly coupled in $g^2$. 
At large energies, for $\zeta>0$ target space of deformed $\sigma$-model will again become extremely curved rendering the $\sigma$-model description incomplete.  Note however that the ratio between the two strong coupling scales $\Lambda_{UV}$ and $\Lambda_{IR}$ is parametrically large $\Lambda_{UV}^2/ \Lambda_{IR}^2 = e^{\pi/(2\zeta)}$, so by picking a very small RG invariant $\zeta$ we can obtain an RG flow for which the coupling $g^2\sim 2\zeta$  remains small and $\eta^2 \sim O(1)$ for a long RG flow time thus justifying later on our weak coupling expansion $g\ll1$ and $\eta$ fixed.

Our philosophy here is that we will view the deformed $\sigma$-model being an effective theory up to some UV scale.  For all values of $\zeta$,  the Ricci scalar of the target space of deformed $\sigma$-model is diverging in the IR indicating that non-perturbative effects will take hold.   In general our goal in the resurgence analysis that follows is to understand this IR behaviour  and thus the issue of the UV behaviour, and its resolution, will not have an impact on our analysis.
We will use resurgence for the weak coupling expansion in $g\ll1$ while $\eta$ will be a fixed, possibly complex, number.
It is very likely that, to make a connection (i.e. the Borel flow mentioned in the Introduction) between the Borel plane of the QFT and that one of the reduced model that we will shortly introduce, we will have to perform some appropriate double scaling limit in $g$ and $\eta$ or possibly fix $\eta$ to be close to $1$ or $\ii$ modulo some small, order $g$ corrections. We will study this precise connection in future works.

 \subsection{Deformed-Uniton Solution} 
We would like to find uniton solutions, $U_\eta$ of the $SU(2)$ $\eta$-deformed PCM\footnote{ In \cite{Klimcik:2008eq} a bijection was set up between classical solutions of the undeformed and $\eta$-deformed PCM. However, this argument relied on Minkowski signature and the solutions we present in Euclidean do not immediately follow. We thank C. Klimcik for correspondence on this.} i.e. maps that obey the Euclidean equations of motion  and obey,
\be\label{eq:unitonconditions}
U_\eta.U_\eta =  - \mathbb{1}  \ ,  \quad  \lim_{\eta \rightarrow 0} U_\eta  = U  \ .
\ee
To ensure that the first of these conditions is solved let us make an assumption that in Hopf coordinates the solution for $\phi_1$ and $\phi_2$ is the same as in the undeformed case given in eq.~\eqref{eq:unitonHopf}.  This is consistent; upon plugging this into the full second order equation of motions in the deformed theory one finds that the equations of motion for $\phi_i$ reduce to an $\eta$-independent condition on $\theta$,
\be
f  \bar{f}' \partial \theta - \bar{f} f' \bar{\partial}\theta  = 0  \ , 
\ee
which is simply solved by $\theta = \theta( |f|^2)$. The second order equations of motion for $\theta(|f|^2)$ now dramatically simplify to, 
\be\label{eq:eqmtheta}
0= - \frac{1}{8}\left[1   + 2 \eta^2 \sin^2(\theta ) \right]\sin(2\theta) +  |f|^2 \theta'(|f|^2) + |f|^4 \theta''(|f|^2)  \ . 
\ee
 This has a solution that matches to the undeformed uniton in the limit $\eta\rightarrow 0$  given by, 
 \be\label{eq:thetasol}
 \cos \theta(|f|^2) =  \frac{\sqrt{1+\eta^{2}} (|f|^2-1) }{\sqrt{(1+|f|^2)^2 + \eta^{2}(|f|^2-1)^{2}}} \ . 
 \ee
 
 We can also rewrite equation (\ref{eq:eqmtheta}) by changing variables to $x = \log \vert f\vert ^2$, obtaining,
 \be\label{eq:eqmthetatime}
  \frac{\mathrm d^2\theta(x)}{\mathrm{d}x^2} = \frac{1}{8}\left[1   + 2 \eta^2 \sin^2(\theta ) \right]\sin(2\theta) \ ,
\ee
which is precisely the equation of motion of a particular quantum mechanical model that we will shortly study in detail.
 
\begin{figure}[h!]
   \begin{center}
    \includegraphics[width=4.5cm]{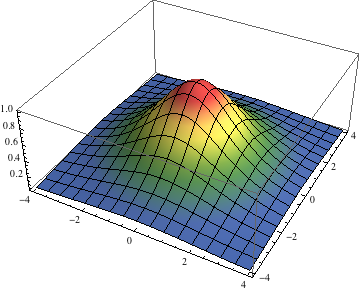}    
       \includegraphics[width=4.5cm]{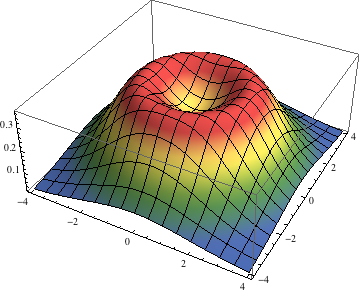}  
         \includegraphics[width=4.5cm]{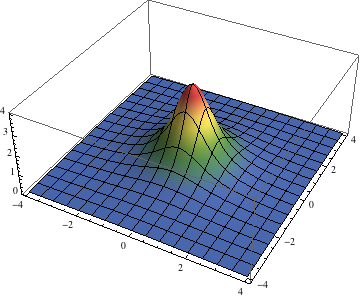}    
 \end{center}
  \caption{\footnotesize{The  (negative of the)  Lagrangian density corresponding to the real $SU(2)$ $\eta$-deformed uniton on $\mathbb{R}^2$. On the left we have no deformation $\eta=0$, in the center $\eta =1$ which has the effect of slightly spreading the support of the uniton and punching a hole in the middle of the uniton and on the right $\eta=\ii/\sqrt{2}$ in which case the deformation sharpens and reduces the support of the lump. (The moduli were fixed to $\lambda_1=0$ and $\lambda_2=0.5$ .) }}  \label{fig:R2uniton}
\end{figure}

 Although the full uniton is not a BPS configuration, note that the solution for the function $\theta$ obeys a rather suggestive $1^{st}$ order equation,  
  \be\label{eq:pseudoBPS}
 \frac{\mathrm d\theta(x)}{\mathrm  d x} = \mp\frac{1}{2} \left(\sin^{2}(\theta) + \eta^{2} \sin^{4}(\theta) \right)^{\frac{1}{2}} \qquad\mbox{for}\,\,x \gtrless 0\ .
 \ee 
 Indeed this equation arises from a Bogomolny rewriting of the action of the $\eta$-deformed PCM evaluated on the ansatz $\theta = \theta( |f|^2)$ and $\phi_{i}$ given by \eqref{eq:unitonHopf}.  In summary we have a deformed uniton solution, 
 \be\label{eq:defuniton}
 U_\eta = \frac{-\ii }{\sqrt{(1+|f|^2)^2 + \eta^2 (1- |f|^2)^2 }}   \left( \begin{array}{cc} \sqrt{1+\eta^2} (1 - |f|^2) & 2  \bar{f} \\
 2 f  & - \sqrt{1+\eta^2} (1- |f|^2  )   \end{array}  \right)  \ . 
\ee
Substitution of the uniton ansatz into the action (\ref{eq:SetaR}) making use of the pseudo-BPS condition  eq.~\eqref{eq:pseudoBPS} 
 shows that,
\be
\begin{aligned}
S_{\eta}[U_{\eta}] &= \frac{1}{2\pi t} \int {\textrm{d}}^{2}z |f|^{2}f'(z) \bar{f}'(\bar{z}) \theta'(|f|^{2})^{2}   \ ,
\end{aligned}
\ee 
from which it is easy to see, after changing variables $w=f(z)$ and making further use of  eq.~\eqref{eq:pseudoBPS}  that the action for a $k$-uniton (i.e. for $f(z)$ a polynomial of degree $k$) evaluates to,
\be\label{eq:Sunition}
\begin{aligned}
S_{\eta}[U_{\eta}] & = \frac{1}{2\pi t  (1+\eta^{2}) } 4 \pi k \left( 1 + (\eta + \eta^{-1}) \arctan(\eta) \right)   \ , \quad k\in \mathbb{Z} \\  
& =  \frac{k}{ t (1+\eta^{2}) }  \,2S_I \ ,
\end{aligned}
\ee
where, 
\be
\label{eq:SInst}
S_I = \left( 1 + (\eta + \eta^{-1}) \arctan(\eta) \right) \, ,
\ee
will correspond later to the action of a quantum mechanical instanton (i.e. the fracton).
Note that as $\eta \to 0$ the uniton action $S_{\eta}[U_{\eta}] $ reduces to the undeformed case (\ref{eq:UnitonS}) and the instanton action $S_I\to 2$.
The minimal uniton solution is obtained by plugging into eq.~\eqref{eq:defuniton},   
\be
f(z) = \lambda_1 + \lambda_2\,z \ ,
\ee
where $\lambda_i \in \mathbb{C}$ are the moduli of the solution, with $\lambda_2\neq0$. The modulus $\lambda_1$ can be thought as the centre of the uniton while $\lambda_2$ is a size modulus. 
The real uniton on $\mathbb{R}^2$ behaves like a lump and can be best visualised by plotting its Lagrangian density,
\be
{\cal L}_\eta[U_\eta] = \frac{-4 (1+|f|^2)^2 f' \bar{f}' }{\left[ (1+|f|^2)^2 +\eta^2 (1-|f|^2)^2  \right]^2} \ ,
\ee
as shown in Figure~\ref{fig:R2uniton}.
The effect of the deformation is rather mild, as we increase $\eta \in \mathbb{R}$ the Lagrangian density becomes slightly punched out.  
We can easily consider the case $\eta = \ii \eta_R$ with $\eta_R\in\mathbb{R}$ just by analytical continuation of the deformed uniton solution (\ref{eq:defuniton}) and the uniton action (\ref{eq:Sunition}) remains real. As we increase $\eta_R$, the action density for the deformed uniton in this case steepens and its support gets smaller as we can see in Figure \ref{fig:R2uniton}.
In both cases, $\eta$ real and purely imaginary, we can vary over the moduli space of $\lambda_i$ and no fractionalisation occurs, the lump remains coherent.

\subsection{Complex Uniton Solution} 
Conventionally one would consider only saddle points of PCM for which the field $g$ is indeed an element of $G$. However, there is by now some substantial evidence \cite{Witten:2010zr,Witten:2010cx,Harlow:2011ny,Basar:2013eka,Cherman:2014ofa,Behtash:2015loa} that suggests that complex field configurations could play an important role in determining the resurgent structure of the theory and that we should also look for {\em finite action} solutions to the equations of motion when $g \in G^{\mathbb{C}}$.   To this end, in the $\eta$-deformed theory we can search for solutions in which, as in the uniton, $\phi_{i}$ are real and obey \eqref{eq:unitonHopf} but $\theta(|f|^{2})$ is allowed to become complex but still satisfy the equation of motion eq.~\eqref{eq:eqmtheta}.  Remarkably there is indeed such a solution,
\be\label{eq:cthetasol} 
\theta(\vert f \vert^2)  = \frac{\pi}{2}+ \ii \, \textrm{arctanh} \left(\frac{1}{2} \sqrt{1+\eta^{2}} \left(\frac{1}{ |f| } + |f| \right) \right) \ , 
\ee
which corresponds to  a complex-uniton $U^{c}_{\eta} \in SL(2, \mathbb{C})$ given by, 
\be\label{eq:defcuniton}
U^{c}_{\eta} = \frac{1}{\sqrt{ 4 |f|^{2 }  - (1+\eta^{2} ) (1+|f|^{2})^{2}  }}  \left( \begin{array}{cc} \sqrt{1+\eta^2} (1 + |f|^2) & 2  \ii  \bar{f} \\
 2 \ii f  &  - \sqrt{1+\eta^2} (1+|f|^2  )   \end{array}  \right)  \ . 
\ee

We call this solution to the complexified $\eta$-deformed PCM equation of motion a \textit{complex-uniton} because the matrix $U^{c}_{\eta}$ lives in the natural complexification $SL(2,\mathbb{C})$ of the original target space $SU(2)$.

In this case, the action when $f(z)$ has degree $k$ is given by,
\be\label{eq:CUnitonS}
\begin{aligned}
S_{\eta}[U_{\eta}^{c}] &= \frac{1}{2\pi t(1+\eta^{2})}4 \pi k  \left( 1 - (\eta + \eta^{-1}) {\textrm{arccot}}(\eta) \right)  \ , \quad k\in \mathbb{Z} \ .\\  
& =  \frac{k}{ t (1+\eta^{2}) }  \,2S_{\mathbb{C}I} \ ,
\end{aligned}
\ee
where,
\be
\label{eq:SCInst}
S_{\mathbb{C}I} =  \left( 1 - (\eta + \eta^{-1}) {\textrm{arccot}}(\eta) \right), 
\ee
will correspond later to a complex-instanton solution in the quantum mechanically reduced deformed PCM.

The reason quite evidently that these solutions have not been, to the best of our knowledge, considered before in the literature is that in the limit $\eta\rightarrow0$ the action diverges and can not give any contributions to the PCM path integral. This is readily seen by looking at the Lagrangian density for $\eta\rightarrow0$ and $f(z)= \lambda z $ (the constant term in $f(z)$ just corresponds to a shift in the centre of the uniton and can be set to zero here) which reads, 
\be\label{eq:Szerosing}
{\cal L}_{\eta=0} = \frac{4 |\lambda|^{2}}{ (-1 + |z|^{2} |\lambda|^{2})^{2}}   \ ,
\ee
and evidently diverges along a circle in the complex plane where the denominator vanishes. The effect of the $\eta$-deformation is to smooth out this singularity giving a finite value to the action. 

The minimal complex-uniton is obtained from eq.~\eqref{eq:defcuniton} by choosing,
\be
f(z) = \lambda_1 + z \lambda_2 \ ,
\ee
where $\lambda_i \in \mathbb{C}$ are the moduli of the solution. 

\begin{figure}[h!]
   \begin{center}
    \includegraphics[width=4.5 cm]{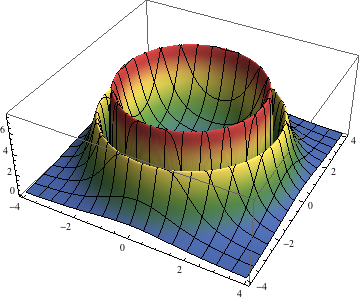}    
          \includegraphics[width=4.5 cm]{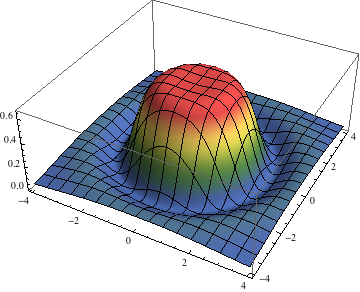}    
              \includegraphics[width=4.5 cm]{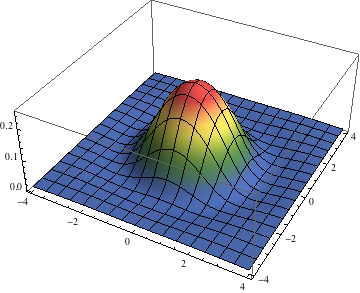}    
                  \includegraphics[width=4.5 cm]{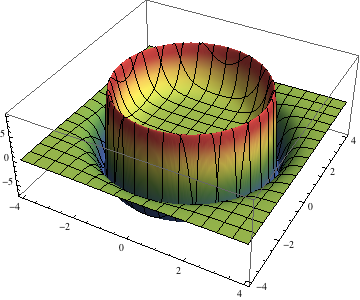}    
          \includegraphics[width=4.5 cm]{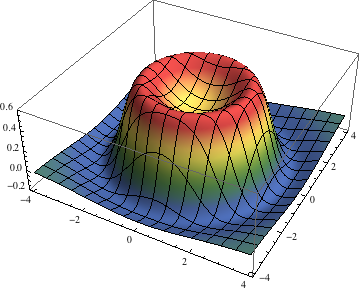}    
              \includegraphics[width=4.5 cm]{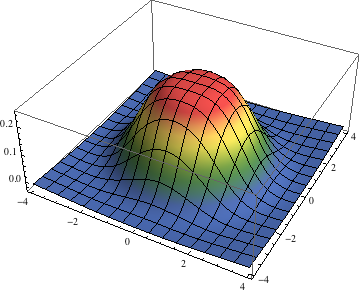}     
 \end{center}
  \caption{\footnotesize{The (negative of the) Lagrangian density corresponding to the   $n=2$ $SU(2)$ $\eta$-deformed complex-uniton on $\mathbb{R}^2$ for values of $\eta = \{0.1, 0.5, 1 \}$.  (The top row shows the real part, while the second row shows the imaginary part.)   We centre the uniton by setting $\lambda_1=0$ and fix the size as $\lambda_2 = 0.5$ .   The effect of the $\eta$-deformation is very pronounced and serves to resolve a singularity in the Lagrangian density.  }}  \label{fig:ComplexUnitononR2}
\end{figure}

In contrast, for the complex-uniton on  $\mathbb{R}^2$ the effect of the $\eta$-deformation is drastic, it has the effect of smoothing out a singular Lagrangian density to give a finite action configuration for $\eta>0$ (see Figure~\ref{fig:ComplexUnitononR2}). 

Note also that if we analytically continue the complex-uniton solution (\ref{eq:defcuniton}) to $\eta=\ii\eta_R$ with $\eta_R\in\mathbb{R}$ we obtain a Lagrangian density which has a non-integrable singularity. The reason is that our solution (\ref{eq:cthetasol}) becomes singular when analytically continued to purely imaginary $\eta$. The complex-uniton action in this regime can be obtained via analytic continuation of (\ref{eq:CUnitonS}) from $\mbox{arg}(\eta) = \pi/2\pm\epsilon$ (see later on Section \ref{sec:NPsaddles} as well as \cite{Behtash:2015loa}) obtaining two types of actions,
\be\label{eq:CUnitonSIM}
\begin{aligned}
S_{\eta}[U_{\eta}^{c}] & =  \frac{k}{ t (1+\eta^{2}) }  \,2S_{\mathbb{C}I} \,,\\
S_{\eta}[\widetilde{U_{\eta}^{c}}] & =  \frac{k}{ t (1+\eta^{2}) }  \,2S_{\widetilde{\mathbb{C}I}} \,,
\end{aligned}
\ee
where
\be
\label{eq:SCInstIm}
\begin{aligned}
 S_{\mathbb{C}I} &=  \left(1- (\eta_R-\eta_R^{-1})\,\mbox{arctanh}( \eta_R )\right)-\ii \,\frac{\pi}{2}(\eta_R-\eta_R^{-1})\,,\\
S_{\widetilde{ \mathbb{C} I } }  &=   \left(1- (\eta_R-\eta_R^{-1})\,\mbox{arctanh}( \eta_R )\right)+\ii \,\frac{\pi}{2}(\eta_R-\eta_R^{-1})\,.
 \end{aligned}
 \ee
 
There is another way of smoothening these field configuration for $\eta$ purely imaginary. If we allow, as we should, all the field configuration to become complex, then we can replace in (\ref{eq:thetasol}-\ref{eq:cthetasol}) $\vert f(z)\vert^2\to Z_0 \vert f(z)\vert^2$  with $Z_0\in\mathbb{C}$. This corresponds to the choice of a complex centre $x_0 = \log Z_0$ for the instanton or complex-instanton solution to (\ref{eq:eqmthetatime}). The modulus of $Z_0$ acts simply as a scale modulus but its phase allow us to move the singularities of the Lagrangian density away from the domain of the spatial integration. In particular the action density for the complex-uniton for $Z_0 = e^{\ii\alpha}$ in the undeformed theory becomes,
\be
{\cal L}_{\eta=0} = \frac{4 |\lambda|^{2} e^{\ii\alpha}}{ (-1 + |z|^{2} e^{\ii\alpha} |\lambda|^{2})^{2}}   \ ,
\ee
to compare against (\ref{eq:Szerosing}).
For $\alpha\in(0,2\pi)$ the above action density has no singularities over $z\in\mathbb{C}$ and it integrates precisely to the uniton action but we stress again that the corresponding complex-uniton is a genuine $SL(2,\mathbb{C})$ matrix even in the undeformed theory.
This issue of complex moduli is at the very core of the Morse-Picard-Lefschetz approach to the path integral of quantum field theories as discussed in Section \ref{sec:cSaddles}.


\section{Fractionalization and reduction to QM} \label{sec:etareduction}

\subsection{Twisted spatial compactification} 
Recent works \cite{Unsal:2010qh,Argyres:2012ka,Dunne:2012zk,Dunne:2012ae,Cherman:2016hcd} have emphasised the idea of {\em adiabatic continuity} as a useful technique in the study of the IR physics of QCD-like gauge theories on $\mathbb{R}^4$ (or $\mathbb{R}^2$).  The idea is to place the theory on $\mathbb{R}^3\times S_L^1$ (or $\mathbb{R}\times S_L^1$) such that when the radius $L$ is sufficiently small the theory enters a weakly coupled domain. 

Of course, the physics in this regime may be quite different to that of the theory on $\mathbb{R}^4$; for a thermal compactification we have a deconfined phase at small $L$ and a confined phase at large $L$. In such a case calculations performed at small $L$ will not tell us a great deal about the long distance physics relevant for the theory at large $L$. If instead one takes a spatial compactification, the small $L$ behaviour can be adiabatically connected to the large $L$-regime, note that the fermions need a spatial compactification as well to prevent any Hagedorn instability  \cite{Basar:2013sza}.
For $2$-d theories on $\mathbb{R}\times S_L^1$ a thermal compactification would only give rise to a cross-over between the deconfined regime at small $L$ and the confined one at large $L$, it is only in the $N= \infty$ limit that we obtain a true phase transition as we dial the radius of the $S^1$.

This idea has been applied to the PCM eq.~\eqref{eq:sPCM}, which as discussed in the introduction, can serve in many ways as a good prototype for QCD but in the easier scenario of two-dimensions. In \cite{Cherman:2014ofa} it was studied in detail how the properties of compactification of the $SU(N)$ PCM on $\mathbb{R}\times S^1_L$  are related to a choice of boundary conditions,
\be
g(t, x+L) = e^{\ii H_L} g(t ,x )e^{-\ii  H_R} \ , \quad  \psi(t, x+L) = \pm e^{i H_L} \psi(t ,x )e^{-\ii  H_R}  \ , 
\ee
in which $\psi$ denote any fermions in the game and the $-$ sign corresponds to anti-periodic thermal boundary conditions and the $+$ sign to periodic spatial boundary conditions. 
These boundary conditions can equally well be thought of as turning on background gauge fields for $H_L\times H_R$ inside the global $SU(N)_L \times SU(N)_R$ symmetry. 
Two conditions are required of the choice of $H_L$ and $H_R$ for adiabaticity.  First is that the free energy ${\cal F}$ should be at a stationary point as we vary the boundary conditions and second that   ${\cal F}/N^2 \rightarrow 0$ for any $L$.  The eigenvalues of the axial $H_A = \frac{1}{L} (H_L- H_R)$ obey a tree level effective potential which is extremised only when $H_A=0$ and thus attention can be restricted to vectorial gauge field for $H_V= \frac{1}{L}(H_L+ H_R)$.   By integrating out the KK-modes one finds an effective potential $V(\Omega)$ for the Wilson loop $\Omega = \exp \ii  \oint dx H_V  = \exp  i L H_V$ which depends crucially on the sign choice of fermionic boundary conditions.  For thermal boundary conditions this potential is minimised by, 
\be
\Omega_{thermal} = e^{\ii  \frac{2\pi k}{N}} \mathbb{1} \ , 
\ee 
leading to a free energy that does not satisfy the adiabaticity requirements as expected. 
For QCD-like theories on $\mathbb{R}^3\times S^1_L$, this holonomy would correspond to a VEV for the Polyakov line that breaks the $\mathbb{Z}_N$ centre symmetry. 

Conversely, the minimisation of the effective potential for spatial compactifications is achieved by, 
\be
\Omega_{thermal} = e^{\ii  \frac{\nu \pi }{N}}  \diag \left( 1, e^{\ii  \frac{2\pi}{N}} , \dots , e^{\ii  \frac{2\pi (N-1)}{N}}\right) \ , \quad  \nu = 0,1  {\textrm{ for }N} = {\textrm{ odd, even}}. 
\ee
For QCD-like theories on $\mathbb{R}^3\times S^1_L$ this would correspond to a VEV for the Polyakov line that preserves the $\mathbb{Z}_N$ centre symmetry since $\tr (\Omega) =0$.

For $SU(2)$ this has the simple interpretation of turning on a background field for the $U(1)_V$, 
\be\label{eq:Hbc}
H_V = \frac{ \pi}{2 } \left(\begin{array}{cc} 1 & 0 \\ 0 & -1 \end{array} \right)\ . 
\ee

In the case at hand, the $\eta$-deformations the situation is slightly subtle.  For $-1 < \eta^2 <0$, as discussed previously, the theory is asymptotically free and the story outlined above transfers directly.  Indeed one can calculate the effective potential in this case and establish that the same boundary conditions, eq.~\eqref{eq:Hbc}, are required.  
For the case of $\eta^2>0$ the theory on $\mathbb{R}^2$ is both IR and UV strongly coupled.  We can, and shall, of course still study the spatial compactification but should be appropriately more cautious in interpreting the implications of the results of the reduced theory to that on $\mathbb{R}^2$.   Nonetheless what we shall find is a rather consistent story that is valid for both  $-1 < \eta^2 <0$ and $\eta^2> 0$. 

Furthermore, since we are not working at $N=\infty$, our choice of twisted boundary condition on $\mathbb{R}\times S_L^1$ is by no means a guarantee of adiabatic continuity. There is no phase transition as we shrink the radius of $S^1$, nonetheless, as noted in \cite{Leurent:2015wzw}, this choice of maximal twist is very particular even at finite $N$ hinting towards volume independence also when $N=2$. These caveats aside, we will study the compactification of the $\eta$-deformed $SU(2)$ PCM in the presence of the maximal twist (\ref{eq:Hbc}) and leave for further studies the precise connection with the theory on $\mathbb{R}^2$.

Rather than working with twisted boundary conditions it is expedient to work instead with periodic fields, 
\be
\tilde{g}(t,x)= e^{-\ii  H_L \frac{x}{L} } g e^{\ii  H_R \frac{x}{L}  }  \ , \quad \tilde{g}(t,x+L) = \tilde{g}(t,x) \ . 
\ee
Following the discussion above, we will turn on only a vectorial twist,
\be
H_L= H_R = \left(
\begin{matrix}
\xi && 0\\
0 && -\xi
\end{matrix}\right)\,,
\ee
where $\xi = \pi/2$ will correspond to the maximal twist (\ref{eq:Hbc}).

We calculate the Lagrangian for this periodic field given in terms of right-invariant forms as,
\be 
{\cal L}_{\eta}[\tilde{g}]  = \frac{1}{t} \sum_{a=1}^3 R^a_-[\tilde g]( {\cal O}_{-}^{-1})_{ab}R^b_+[\tilde g]  \ , \\ 
\ee
using the coordinates $x^\pm = \frac{1}{2} (t \pm i x) $ resulting in, 

\be
\begin{aligned}
g^2 {\cal L}_{\eta}[\tilde{g}] &= \partial_{i}\theta \partial_{i}\theta + f_{+}(\theta) \cos^{2}\theta \partial_{i}\phi_{1}\partial_{i}\phi_{1}+  f_{-}  (\theta)  \sin^{2}\theta \partial_{i}\phi_{2}\partial_{i}\phi_{2}+  \frac{1}{2} \eta^{2} \sin^{2}(2\theta)\partial_{i} \phi_{1} \partial_{i}\phi_{2}  \\
& \qquad - 2 \xi \sin^2\theta \left( \eta^2 \cos^2 \theta \partial_x \phi_1 + f_-(\theta) \partial_x \phi_2 \right)  + \xi^2 f_-(\theta) \sin^2 \theta     + \ii   \eta \xi  \sin 2\theta \dot{\theta} \ .
\end{aligned}
\ee
where we defined,
\[
2f_{\pm}(\theta) = 2 + \eta^{2} (1\pm \cos(2\theta)) \ .
\]
In the above, we have introduced the coupling $g^2 = t(1+\eta^2)$ and  we have dropped the pure gauge B-field of the $\eta$-deformation (the term in the action is a total derivative and imaginary in Euclidean signature) and note  that the final imaginary term coming from the twisted boundary conditions  is also a total derivative and will be henceforth discarded.  

The $U(1)_L$- and $U(1)_R$-currents, corresponding to the symmetry $g\rightarrow e^{i \epsilon_L \sigma_3  }  g e^{-i \epsilon_R \sigma_3}$ of the untwisted theory, are given by,  
\begin{equation}
\begin{aligned}
 {\cal J}_\mu^L&=  - \frac{2}{t} \left( \cos^2 (\theta) \partial_\mu \phi_1 + \sin^2 (\theta) \partial_\mu \phi_2  \right) \ , \\
{\cal J}_\mu^R &=    \frac{2}{t(1+\eta^2)} \left( \cos^2 (\theta) \left( 1+ \eta^2 \cos(2\theta) \right) \partial_\mu \phi_1 + \sin^2 (\theta)\left( -1+ \eta^2 \cos(2\theta) \right) \partial_\mu \phi_2 \right) \ .
\end{aligned}  
\end{equation}
Reflecting the isometries of the squashed sphere target space, we note that the left acting symmetry is insensitive to the deformation while the right acting symmetry is modified.  We further see that the twisting used is equivalent to turning on a spatial component of the background gauge field of the vectorial action since,
\be
\begin{aligned}
{\cal L}_{\eta}[\tilde{g}] &= {\cal L}_{\eta}[g]   + \frac{\xi}{2} \left( {\cal J}_x^L + {\cal J}_x^R \right)  +  \frac{1}{g^2}   \xi^2 f_-(\theta) \sin^2 \theta \ .
\end{aligned}
\ee 
We will use now a small-$L$ effective quantum mechanics by performing a KK-reduction and dropping all the $x$ dependence from the fields.
Clearly such a reduction will be \textit{forgetful} regarding the UV-renormalon singularities, which can be only extracted from the microscopic theory.

An important subtlety in performing the KK-reduction is that it is not quite true that all
KK-momentum carrying states can be discarded. 
The issue is that states that carry non-zero winding number can contribute to the low-energy dynamics on the same footing as states that carry zero winding number, as already noted in \cite{Cherman:2014ofa}.

In particular we have to consider when $\phi_2$ carries a non-trivially winding around the $S^1$, 
\be
\phi_2(t,x) = 2\pi \,n\,\frac{x}{L}+\tilde{\phi}_2(t)\,,
\ee
with $n\in\mathbb{Z}$.

Taking into account the factor of $L$ that comes from performing the integral around the $S^1$ one finds, after rotating back to Lorentzian signature, an Hamiltonian of the form, 
 \be\label{eq:QM1}
 {\cal H} = \frac{g^2}{4 L} P_I {\cal G}^{IJ} P_J +  \frac{L (\pi n +\xi)^2}{g^2} \sin^2(\theta)\left(1+ \eta^2 \sin^2(\theta) \right) \ ,
 \ee
 where the momenta $P_I = \{ p_\theta, p_{\phi_1} , p_{\phi_2}\}$ are coupled via the inverse metric on the squashed sphere namely, 
 \be
 {\cal G}^{IJ}  = \left( \begin{array}{ccc} 1 & 0& 0 \\ 0 & \frac{f_-(\theta)\sec^2\theta}{1+\eta^2 } &- \frac{\eta}{1+\eta^2}  \\ 0 &- \frac{\eta}{1+\eta^2} &   \frac{f_+(\theta)\csc^2\theta}{1+\eta^2 }  \end{array}\right) \ . 
 \ee
 As explained above we see that, after picking the maximal twist $\xi=\pi/2$, the states with zero winding number for $\phi_2$ will have exactly the same energy as the states with winding number $-1$.

We are interested in the ground state of this system and to this end we employ the Born-Oppenheimer approximation which separates the dynamics of the light degrees of freedom $p_\theta$ with the heavy ones $p_{\phi_i}$ as  in \cite{Cherman:2014ofa}. Remembering the subtleties about winding modes we can also set $n=0$ from now on and choose the maximal twist $\xi = \pi/2$.
Thus we arrive at a one-dimensional quantum mechanics of interest given by eq.~\eqref{eq:QM1} with $p_{\phi_i}=0,n=0$ and $\xi=\pi/2$,
\be
\mathcal H = \frac{g^2}{4L} p_\theta^2+ \frac{L\pi^2}{4 g^2}\sin^2(\theta)\left(1+ \eta^2 \sin^2(\theta) \right) \,.
\ee

We can rescale the time variable to remove the energy scale $1/L$ and some numerical factor to put the above Hamiltonian in the form,
\be\label{eq:QM}
\mathcal H_{QM} = \frac{g^2}{2} p_\theta^2+ \frac{1}{2 g^2}\sin^2(\theta)\left(1+ \eta^2 \sin^2(\theta) \right) \,,
\ee
with the corresponding Lagrangian given by,
\be
\mathcal L_{QM} = \frac{1}{2g^2}\left[\left( \frac{\mathrm d\theta(t)}{\mathrm{d}t}\right)^2- \sin^2(\theta)\left(1+ \eta^2 \sin^2(\theta) \right)\right]\,.
\ee
The corresponding Schr\"odinger equation can be put in the form, 
\be\label{eq:WH}
\Psi''(\theta) + (a -2 q  \cos(2\theta) -2p \cos(4\theta) ) \Psi(\theta) = 0 \ , 
\ee
with the parameters,
\be\label{eq:WH2}
p = \frac{\eta^2}{16 g^4} \ , \quad q= -\frac{(1+\eta^2)}{4g^4}  \ , \quad a= \frac{1}{8 g^2} \left( 16 E - 4-3\eta^2  \right) \ .
\ee
This differential equation is called the three-term Whittaker-Hill equation and some of its properties are known\footnote{See e.g. \href{http://dlmf.nist.gov/28.31}{http://dlmf.nist.gov/28.31} and the books   \cite{MagnusWinkler,WhittakerWatson}, we thank Gerald Dunne and Gleb Arutyunov for related discussions. }.   Although it will not be directly relevant for what follows, it is notable that whilst this is not an integrable QM (in the sense that its spectrum can't be given in closed form in terms of known functions), it is an example of a quasi-exactly-solvable QM (for a recent comprehensive treatment \cite{Turbiner:2016aum}) in which some--but not all--eigenstates for particular values of the parameters can be found exactly.  This Whittaker-Hill equation also emerges in certain inflationary scenarios as describing entropy fluctuations during reheating   \cite{Lachapelle:2008sy} (this paper also contains a useful exploration of stability properties of the equation). 

We notice that our Hamiltonian can be directly related (for $\eta^2<0$) to the supersymmetric quantum mechanics studied in \cite{Behtash:2015zha,Behtash:2015loa} with a superpotential of the form $W(\theta) = \cos(\theta)$ after the fermions are integrated out. 

That our deformed bosonic theory reduces in one-dimension to the supersymmetric completion of the undeformed theory seems, at first sight, rather peculiar.  One anticipates that the twisted reduction of the two-dimensional supersymmetric Principal Chiral Model would be the natural origin for the supersymmetric quantum mechanics studied in \cite{Behtash:2015zha,Behtash:2015loa} and that this would be a rather different theory from the bosonic $\eta$-deformed PCM.  However there is one sense in which the $\eta$-deformation and the supersymmetric deformation are similar; they both modify the S-matrix of the theory by means of a quantum group.  Indeed the supersymmetric part of the S-matrix of the SUSY PCM for spinors in the $O(4)$ model is identical to the soliton S-matrix of the supersymmetric sine-Gordon theory evaluated at a value of coupling such that scattering is reflectionless \cite{Ahn:1990uq,Hollowood:1996ex} .  This in turn is related to an affine $SU(2)$ quantum group with parameter $q$ a root of unity.    This was extended in \cite{Evans:1996ah} to the $SU(N)$ PCM wherein the supersymmetric part of the S-matrix is a   $q$ root unity affine quantum group deformation of $SU(N)$.   It would be interesting to try and clarify this possible relation at the level of Lagrangians perhaps by directly integrating out the fermions from the PCM.

As expected, for $\eta \rightarrow 0$ the QM defined by eq.~\eqref{eq:WH} reduces to the well known Mathieu equation studied in this PCM  context   in   \cite{Cherman:2013yfa,Cherman:2014ofa}.  More generally the Mathieu equation has been carefully studied in a number of works as an example of a QM where degenerate minimum give rise to a banded spectrum \cite{Dunne:2014bca,Misumi:2015dua}.  This system also plays a key role in the study of $4$-dimensional $N=2$ supersymmetric pure $SU(2)$ gauge theory and its wall crossing phenomenon \cite{Basar:2015xna,Kashani-Poor:2015pca}.  
     
  Before proceeding, let us make a small digression concerning the reduced quantum mechanics  eq.~\eqref{eq:QM1}.  If instead of adopting Euler angles we use $\mathbb{R}^{4}$  embedding coordinates to parametrise the group element as, 
  \be
  g = \left( \begin{array}{cc} x_{1}  + \ii x_{2 } & x_{3} + \ii x_{4} \\ -x_{3} - \ii x_{4} & x_{1} - \ii x_{2}\end{array}\right) \ ,  \quad |\vec{x}|^{2}= 1 \ ,
  \ee
 then the Lagrangian corresponding to eq.~\eqref{eq:QM1} can be expressed compactly as, 
  \be\label{eq:QM2}
\frac{ g^{2}}{L}  {\cal L}  = \left( \delta_{ij} +\eta^{2} \tilde{x}_{i} \tilde{x}_{j}\right)\dot{x}_{i} \dot{x}_{j}  - \xi^{2} (x_{3}^{2}+x_{4}^{2})\left(1+ \eta^{2} (x_{3}^{2}+x_{4}^{2}) \right) \ , 
  \ee
  in which $\tilde{x}_{i}= \{ x_{2},-x_{1},x_{4},-x_{3}\}$.   For $\eta^{2}=0$ this quantum mechanics on an $S^{3}$ with a quadratic potential in the embedding coordinates has been well studied under the name of the C. Neumann model \cite{Neu} and is known to be  integrable both classically \cite{uhlenbeck1989harmonic,Mo} and quantum mechanically \cite{Avan:1991ib}.\footnote{We thank Oleg Evnin for drawing our attention to this.}  It seems quite likely then, given the integrable 2-d origin of eq.~\eqref{eq:QM2}  that the full $\eta$-deformed quantum mechanics is also integrable and will be pursued elsewhere \cite{OlegDan}.  Intriguingly an integrable deformation of the Neumann model has already arisen in relation to spinning strings in $\eta$-deformed $AdS_{5}\times S^{5}$ in \cite{Arutyunov:2014cda} -- though the structure there seem more involved than the simple Lagrangian of eq.~\eqref{eq:QM2}. 
  
\subsection{Unitons on $\mathbb{R}\times S^1$} 
To put the minimal unitons and the minimal complex-unitons on $\mathbb{R}\times S_L^1$, and to take care of the twisted boundary conditions we use the results of \cite{Bruckmann:2007zh,Brendel:2009mp} and we simply make the replacement,
\be
f(z) = e^{-\pi \frac{z}{L} } \left( \lambda_1 + \lambda_2 e^{2\pi \frac{z}{L} } \right) \ , 
\ee
with $z= t+\ii  x$ such that, 
\be
U_\eta(t, x+L) = e^{\ii  H_V} U_\eta(t, x) e^{-\ii H_V} \ ,
\ee  
where $H_V$ is the maximal twist matrix of (\ref{eq:Hbc}).

\begin{figure}[h!]
   \begin{center}
    \includegraphics[width=4.5cm]{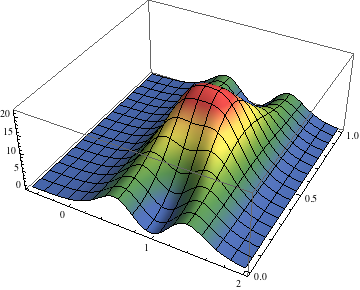}    
       \includegraphics[width=4.5cm]{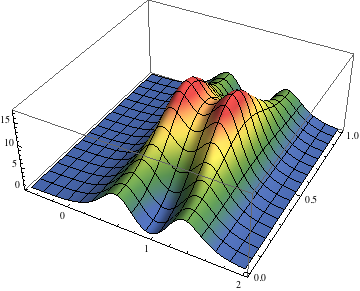}    
          \includegraphics[width=4.5cm]{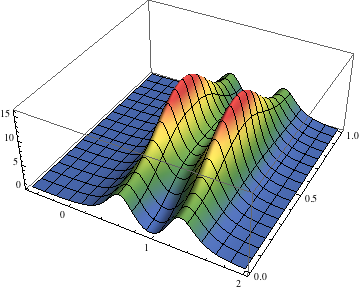}    
              \includegraphics[width=4.5cm]{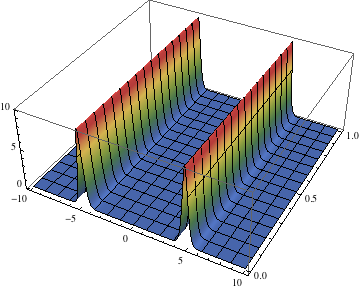}    
       \includegraphics[width=4.5cm]{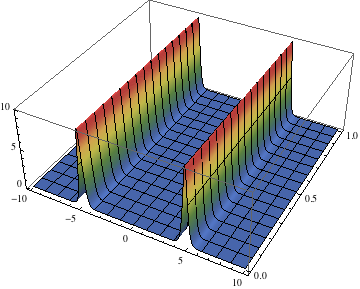}    
          \includegraphics[width=4.5cm]{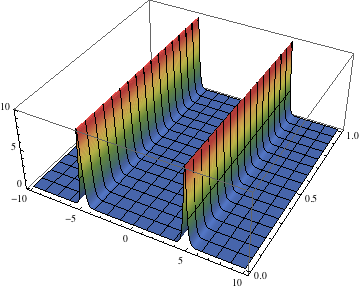}   
 \end{center}
  \caption{\footnotesize{The (negative of the)  Lagrangian density corresponding to the real $SU(2)$ $\eta$-deformed uniton on $\mathbb{R}\times S^1$ with twisted boundary conditions for real moduli.  The radius of the $S^1$ is set to one.  The top line shows for $\eta = \{ 0 , 0.5 , 1 \}$  (left to right)   a regime of uniton moduli space   $\lambda_1\gg 1 \gg  \lambda_2   $ (the plot shows $\lambda_1=e^2$ and $\lambda_2=e^{-4}$) in which there is no fractionalisation.  Here the effect of the $\eta$-deformation creates some partial splitting in the uniton but not a complete fractionalisation.  The lower lines shows for $\eta = \{ 0 , 0.5 , 1 \}$  (left to right)   when $\lambda_1 \sim \lambda_2 \ll 1 $ (the plot shows $\lambda_1=\lambda_2=e^{-15}$) and regardless of the value of $\eta$ the uniton has fully broken up into a fracton-anti-fracton pair.  }}  \label{fig:fractionalization}
\end{figure}

\begin{figure}[h!]
   \begin{center}
    \includegraphics[width=4.5cm]{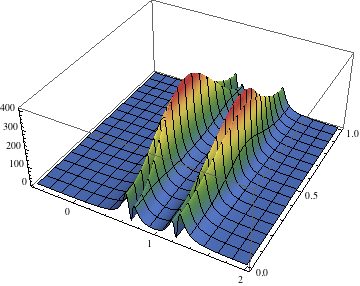}    
       \includegraphics[width=4.5cm]{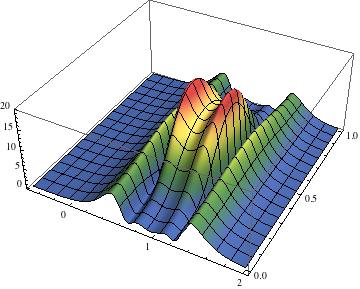}    
          \includegraphics[width=4.5cm]{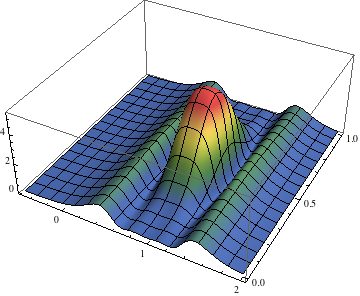}    
              \includegraphics[width=4.5cm]{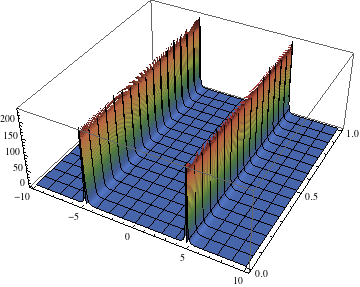}    
       \includegraphics[width=4.5cm]{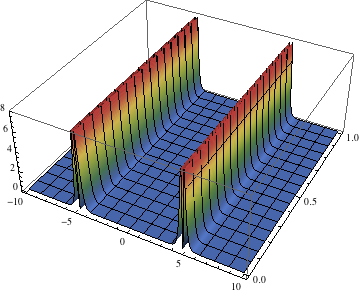}    
          \includegraphics[width=4.5cm]{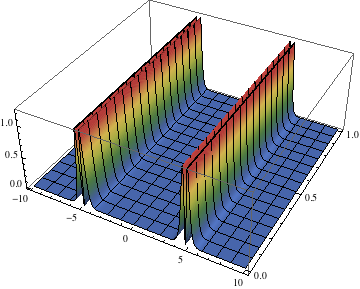}   
 \end{center}
  \caption{\footnotesize{The real part of the Lagrangian density corresponding to the $SU(2)$ $\eta$-deformed complex-uniton on $\mathbb{R}\times S^1$ with twisted boundary conditions for real moduli.  The radius of the $S^1$ is set to one.  The top line shows for $\eta = \{ 0.1 , 0.5 , 1 \}$  (left to right)   a regime of uniton moduli space   $\lambda_1\gg 1 \gg  \lambda_2   $ (the plot shows $\lambda_1=e^2$ and $\lambda_2=e^{-4}$) in which there is no fractionalisation.  The lower lines shows for $\eta = \{ 0.1 , 0.5 , 1 \}$  (left to right)   when $\lambda_1 \sim \lambda_2 \ll 1 $ (the plot shows $\lambda_1=\lambda_2=e^{-15}$) and regardless of the value of $\eta$ the complex-uniton has fully broken up.    Note the vertical scale changes between plots since the densities are divergent as $\eta\rightarrow 0$.  }}  \label{fig:fractionalizationcomplexunitonRe}
\end{figure}

\begin{figure}[h!]
   \begin{center}
    \includegraphics[width=4.5cm]{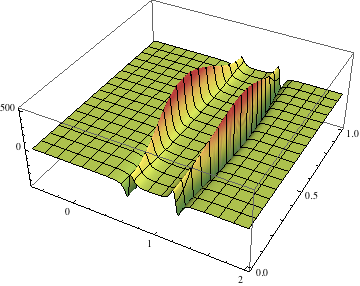}    
       \includegraphics[width=4.5cm]{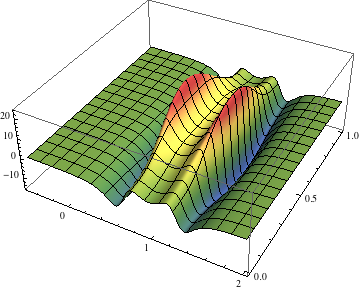}    
          \includegraphics[width=4.5cm]{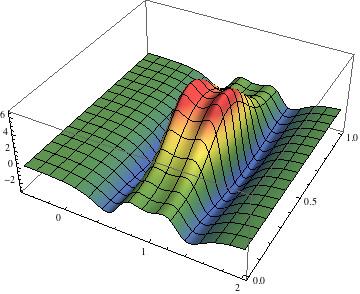}    
              \includegraphics[width=4.5cm]{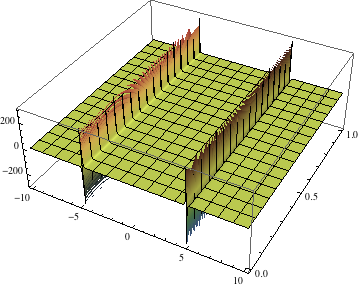}    
       \includegraphics[width=4.5cm]{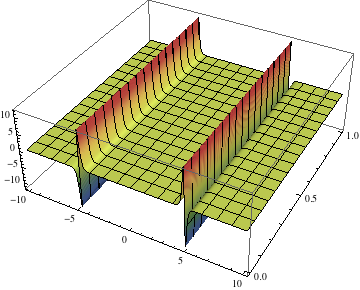}    
          \includegraphics[width=4.5cm]{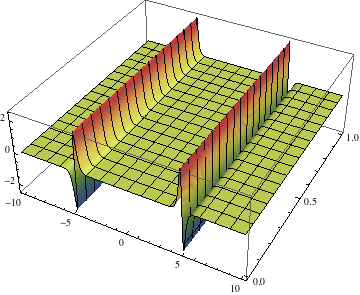}     
 \end{center}
  \caption{\footnotesize{The imaginary part of the Lagrangian density corresponding to the $SU(2)$ $\eta$-deformed complex-uniton on $\mathbb{R}\times S^1$ with twisted boundary conditions for real moduli.  Plots ordered as in fig.~\ref{fig:fractionalizationcomplexunitonRe} (for the same values of $\lambda_{1,2}$ respectively)}}  \label{fig:fractionalizationcomplexunitonIm}
\end{figure}

Here something quite different happens as we vary over the moduli-space, both the real and complex-uniton solutions break-up or fractionalize. In the regime $\lambda_1\gg 1 \gg  \lambda_2   $ there is no fractionalisation however  when $\lambda_1 \sim \lambda_2 \ll 1$, regardless of the value of $\eta$ these semi-classical solutions break up into widely separated lumps of Lagrangian density. This is illustrated in Figure~\ref{fig:fractionalization} for the real uniton and Figures~\ref{fig:fractionalizationcomplexunitonRe} and \ref{fig:fractionalizationcomplexunitonIm} for the real and imaginary contributions to the Lagrangian density of the complex-uniton.

As well as seeing the fractionalisation in the Lagrangian density, we can see it directly at the level of field space.  Let us focus on the field $\theta(x,t)$ defined via eq.~\eqref{eq:thetasol} for the real uniton and   \eqref{eq:cthetasol}  for the complex-uniton but with the function $f(z)$ appropriate for $\mathbb{R}\times S^1_L$ with for concreteness real moduli.   For $\lambda_1 \lambda_2 \ll 1 $, which holds  in the fractionalized regime, the derivative with respect to $x$--i.e. the dependence in the $S^1_L$-- is suppressed and $\theta(x,t)\sim \theta(t)$ effectively truncates to the zero-KK mode sector. The field $\phi_2$ given by equation (\ref{eq:unitonHopf}) does instead depend on $x$ and interpolates between a configuration with $0$ winding and a configuration with winding $-1$.
For the real uniton this gives a correlated fracton--anti-KK-fracton profile so schematically we write the fractionalisation as $U \rightarrow [ {\cal F}_0 \,\overline{{\cal{F}}}_{-1}] $ where the subscript are to remind that the field $\phi_2$ has different winding numbers. 

As we will see later on, the action of each fracton event can be computed from our quantum mechanic model and it is precisely equal to (\ref{eq:SInst}) so that the uniton, even in the deformed theory, is indeed composed by a fracton and a KK-anti-fracton and its action is given by (\ref{eq:Sunition}).

For the complex-uniton, we find that the profile of $\theta(t)$ is such that the real part remains constant however the imaginary part is non-constant and it fractionalizes into two equal constituents.
As for the uniton, even in this case the field $\phi_2$ is given by equation (\ref{eq:unitonHopf}) so it does depend on $x$ and interpolates between a configuration with $0$ winding and a configuration with winding $-1$.
For these reasons we denote the fractionalisation of the complex-uniton as $U^C \rightarrow [\mathbb{C} {\cal F}_0\, \mathbb{C} {\cal F}_{-1}]$, i.e. the complex-uniton is formed by a complex-fracton and a KK-complex-fracton.
We will shortly show that the action of a complex-fracton can be computed from our quantum mechanical model and it is precisely equal to (\ref{eq:SCInst}) explaining why the complex-uniton action is given by (\ref{eq:Sunition}).

\begin{figure}[h!]
   \begin{center}
    \includegraphics[width=6cm]{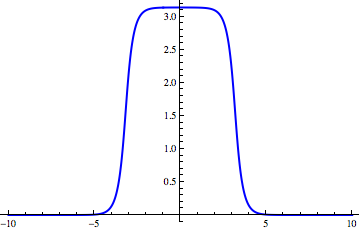}   
       \includegraphics[width=6cm]{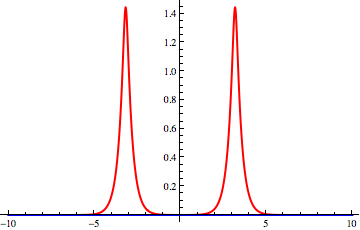}    
 \end{center}
  \caption{\footnotesize{ The profile of $\theta(t)$ in the fractionalised regime for the uniton (left) and complex-uniton (right) with real parts in blue and imaginary in red.  }}  \label{fig:thetaprofiles}
\end{figure}

The profiles of $\theta$ for both these cases are plotted in Figure.~\ref{fig:thetaprofiles}.  

\subsection{Non-perturbative saddles in the reduced quantum mechanics}\label{sec:NPsaddles}
We can understand the origin of the fracton and the complex-fracton from the reduced quantum mechanics.
Let us rewrite here the Hamiltonian of interest after a rescaling\footnote{We rescaled the Lagrangian so that the factor $1/(2g^2)$ sits outside of the action and $g^2$ plays the role of $\hbar$.} becomes
\be\label{eq:QMWKB}
\mathcal H =  p_\theta^2+ \sin^2(\theta)\left(1+ \eta^2 \sin^2(\theta) \right)= p_\theta^2+V \,,
\ee
and the corresponding Euclidean Lagrangian is obviously
\be\label{eq:LQM}
\mathcal L_E =  \left(\frac{\mathrm d\theta(t)}{\mathrm{d}t}\right)^2+ \sin^2(\theta)\left(1+ \eta^2 \sin^2(\theta) \right)=\dot{\theta}^2 +V \,,
\ee
whose variation gives the second order equation of motion
\be\label{eq:EOM}
\ddot{\theta}(t) = \frac{1}{2} \left( 1+2\eta^2 \sin^2(\theta)\right) \sin(2\theta)\,.
\ee

To classify all the non-perturbative solutions to the equation of motion we first have to find the \textit{complex} critical points of the potential 
\be\label{eq:Pot}
V(\theta)=\sin^2(\theta)\,(1+ \eta^2 \sin^2(\theta))
\ee in the strip $\mbox{Re}\,\theta\in [0,\pi]$ of the complex plane $\theta\in\mathbb{C}$.
We can easily solve for $V'(\theta) =0$ for arbitrary $\eta\in\mathbb{C}$ with $\vert \eta \vert \leq 1$, finding generically $7$ critical points:
\be
\begin{aligned}
&\theta = 0\,\qquad\theta=\frac{\pi}{2}\,,\qquad\theta = \pi\,,\\
&\theta = \theta_{cr}\,,\qquad\qquad\qquad\,\theta=\bar{\theta}_{cr}\,,\\
&\theta = \pi-\theta_{cr}\,\qquad\qquad\,\,\,\theta= \pi-\bar{\theta}_{cr}\,,
\end{aligned}
\ee
where $\theta_{cr}\in\mathbb{C}$ is given by,
\be\label{eq:ThetaCr}
\theta_{cr}=\frac{\pi}{2} - \frac{1}{2} \mbox{Im}\left[ \log \left( \frac{1 + \sqrt{1 + 2 \eta^2} }{1 - \sqrt{1 + 2\eta^2}}\right)\right]+\frac{\ii}{2}\mbox{Re}\left[ \log \left( \frac{1 + \sqrt{1 + 2 \eta^2} }{1 - \sqrt{1 + 2\eta^2}}\right)\right]\,.
\ee

\begin{figure}[tb]
\begin{center}
\begin{tabular}{cc}
\resizebox{65mm}{!}{\includegraphics{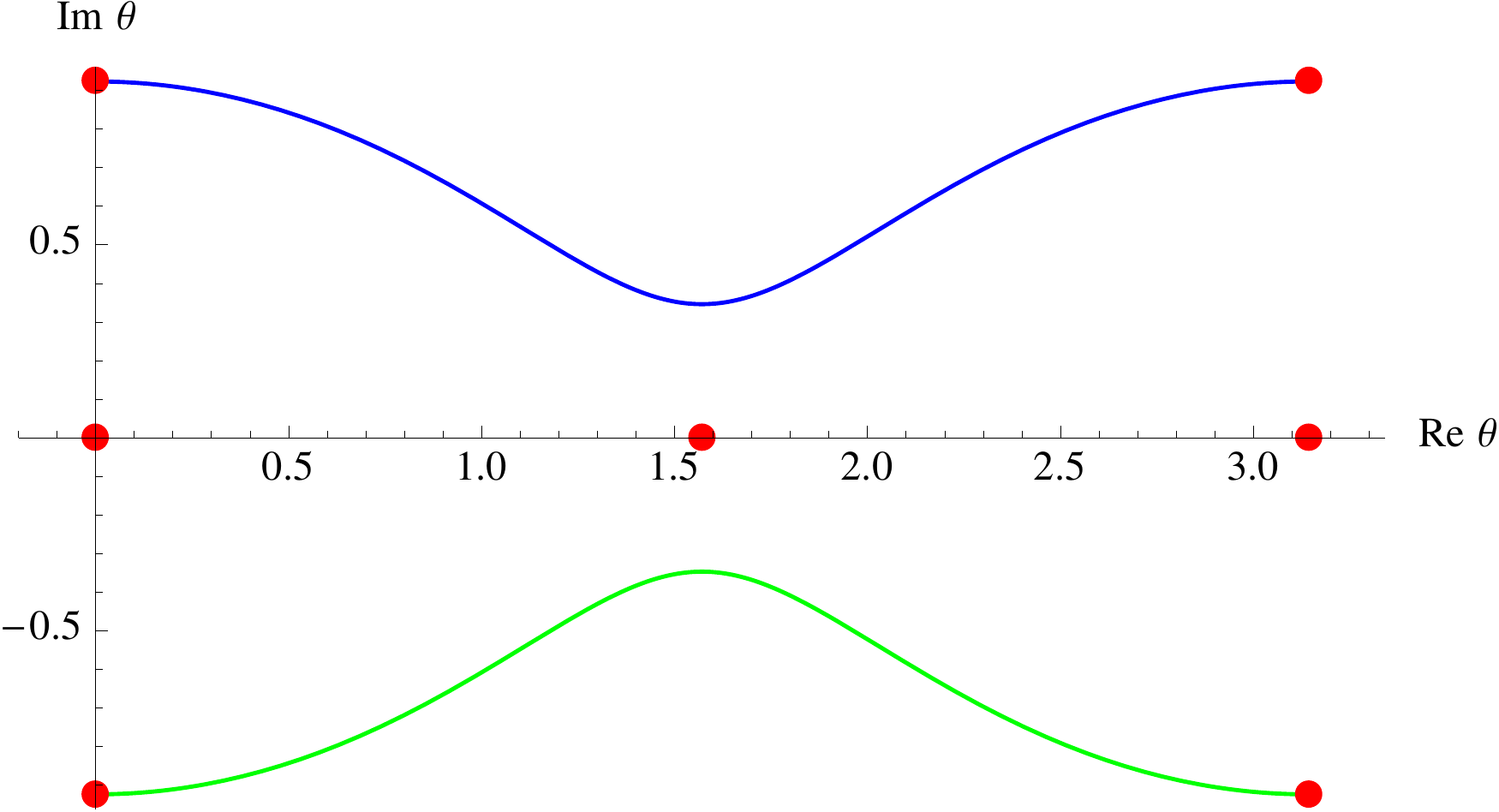}}
\hspace{2mm}
&
\resizebox{65mm}{!}{\includegraphics{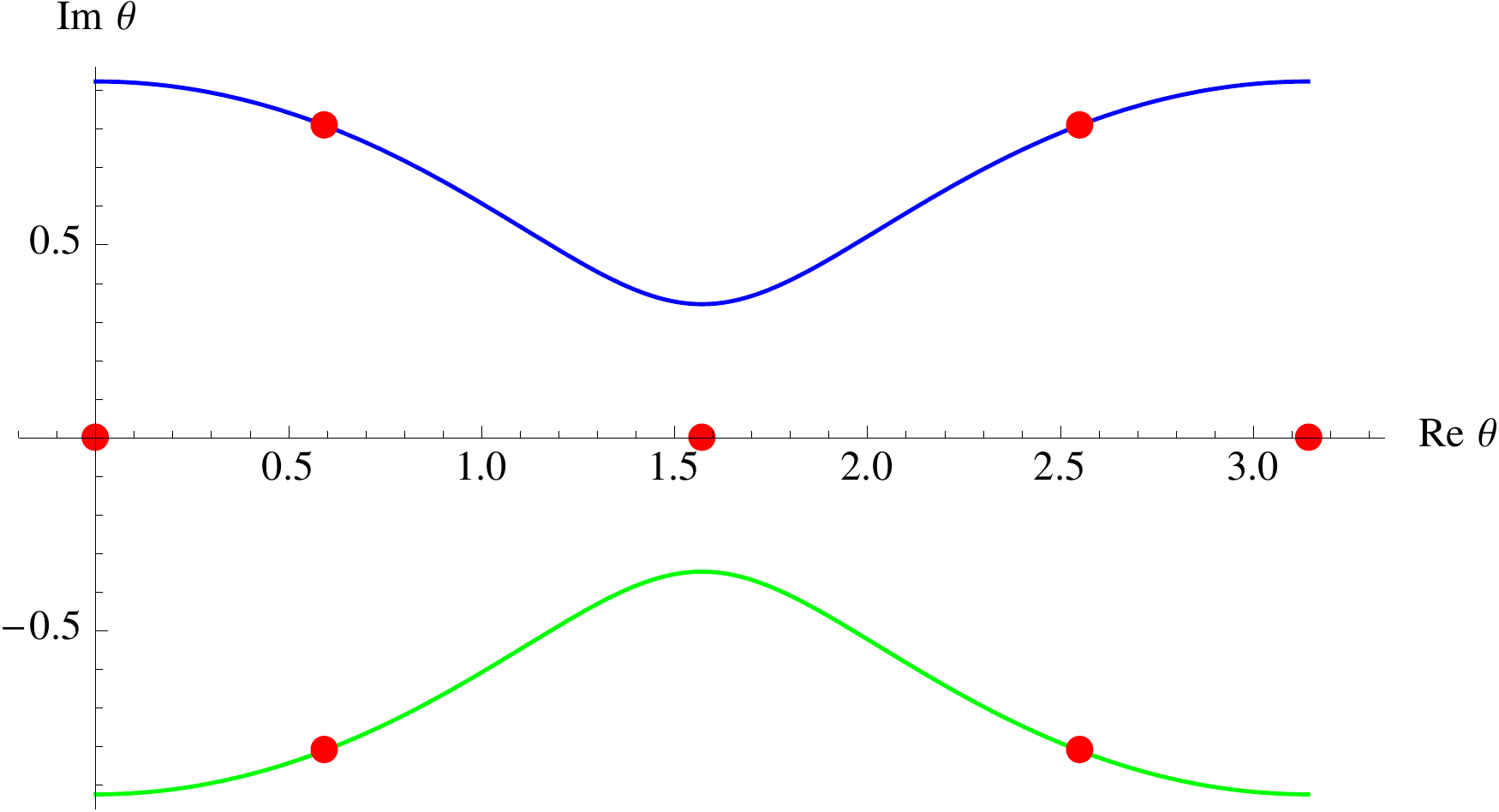}}\\
{\small $\eta=\frac{2}{3}$} & {\small $\eta=\frac{2}{3} \,e^{i \pi/4}$}
\vspace{5mm}
\end{tabular}
\begin{tabular}{c}
\resizebox{65mm}{!}{\includegraphics{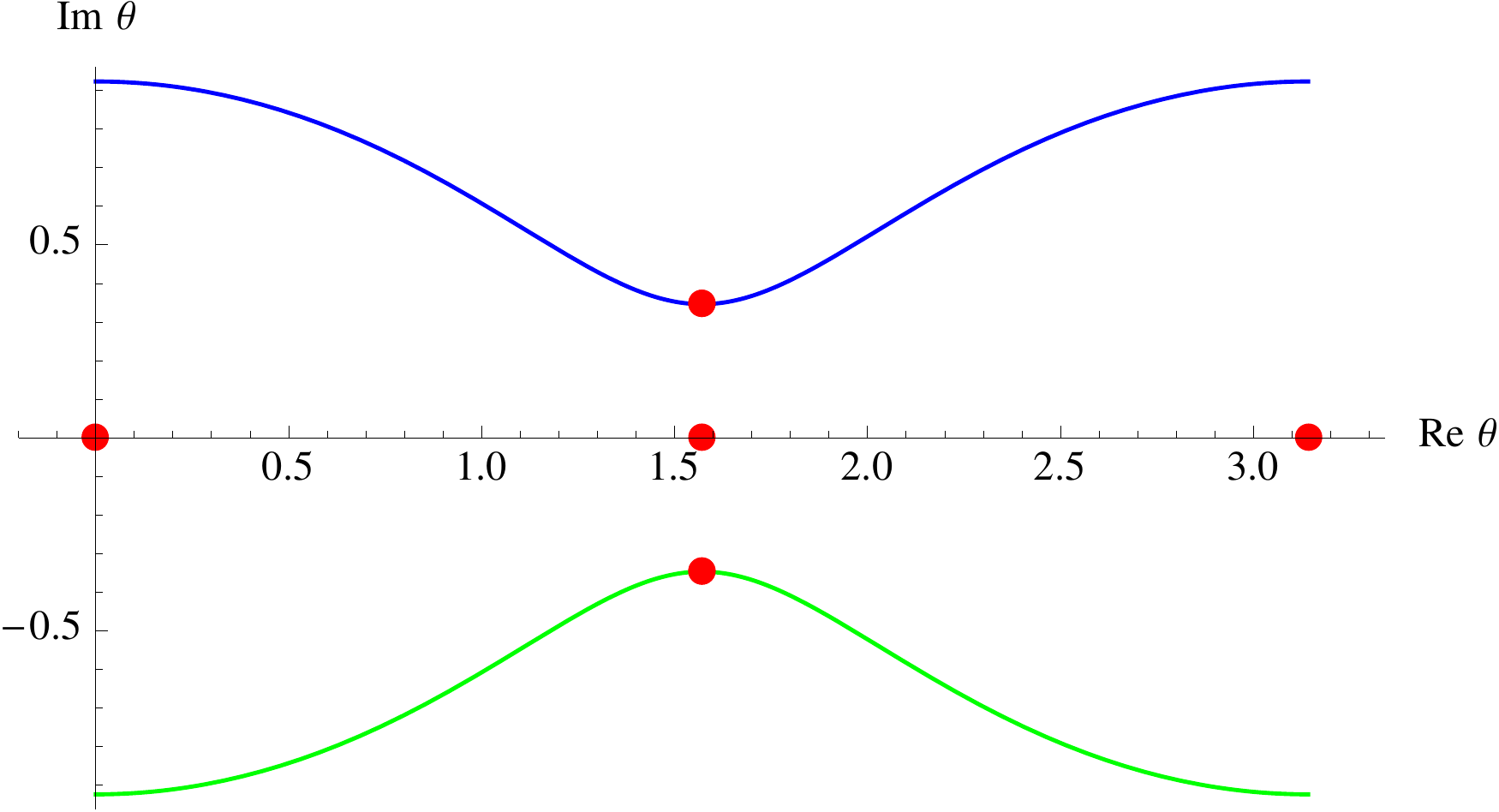}} \\
{\small $\eta=\frac{2}{3} \,e^{i \pi/2}$}
\vspace{-3mm}
\end{tabular}
\end{center}
  \caption{The red dots correspond to the critical points of the potential (\ref{eq:Pot}) for different values of $\eta$. As we increase the argument of $\eta$ while keeping its modulus fixed, the two critical points $\theta_{cr}$ and $\pi-\bar{\theta}_{cr}$ come closer together moving on the blue curve, while $\bar{\theta}_{cr}$ and $\pi-\theta_{cr}$ move closer along the green curve. For $\mbox{arg}(\eta)=\pi/2$ we are left with $5$ critical points.
}
\label{fig:CritPts1}
\end{figure}

If we start with $\eta=0$ we immediately see that $\theta_{cr}\to\infty$ so we are only left with the three critical points $\{0,\pi/2,\pi\}$ of the undeformed Mathieu case, as discussed in \cite{Cherman:2014ofa}.
As we turn on $\eta\neq0$ with $\mbox{arg}(\eta)=0$ we see that $\theta_{cr}$ becomes finite and purely imaginary, of the $7$ critical points $3$ are real while $4$ are complex, see Figure \ref{fig:CritPts1}.

\begin{figure}[tb]
\begin{center}
\begin{tabular}{cc}
\resizebox{65mm}{!}{\includegraphics{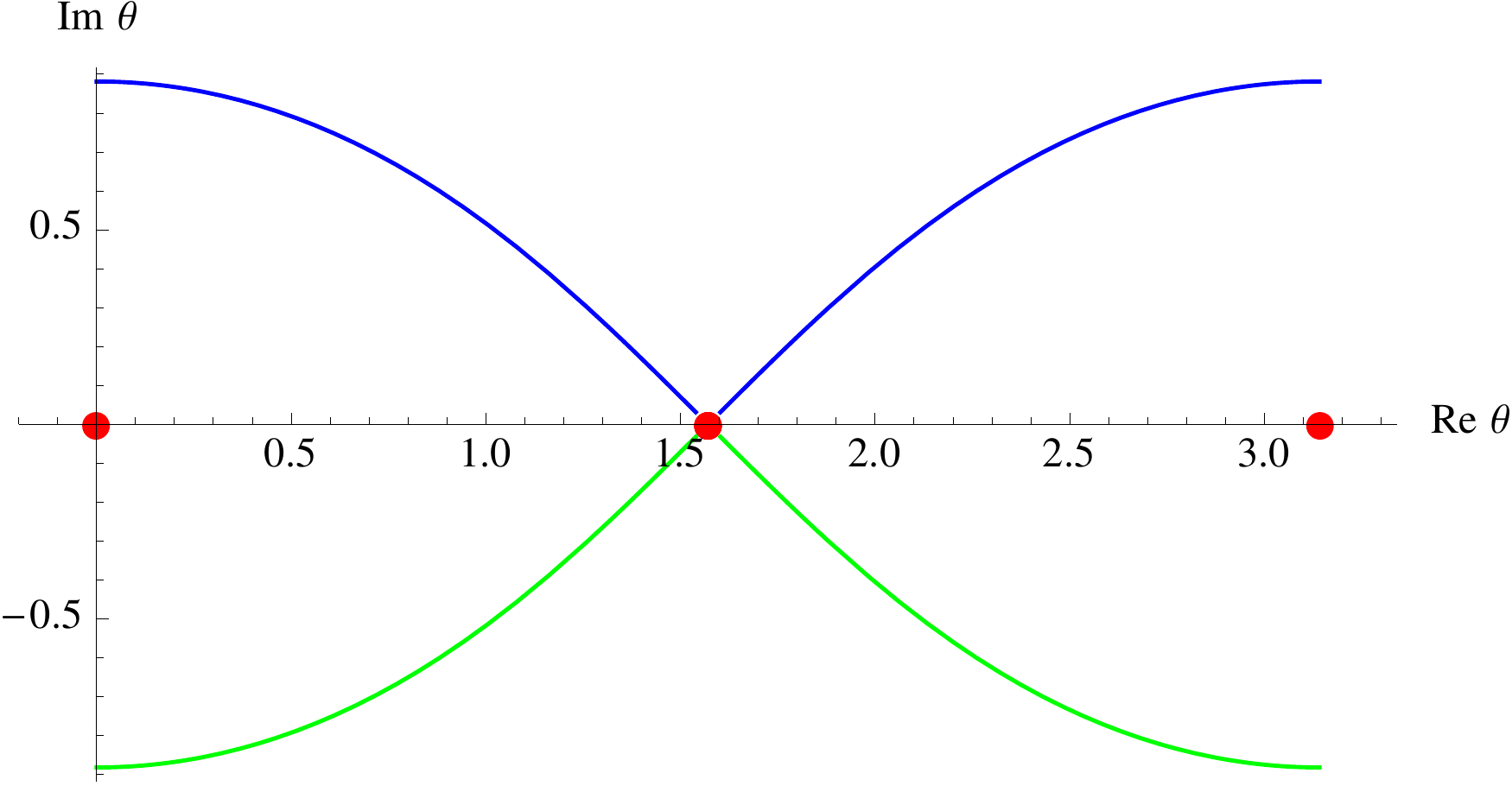}}
\hspace{2mm}
&
\resizebox{65mm}{!}{\includegraphics{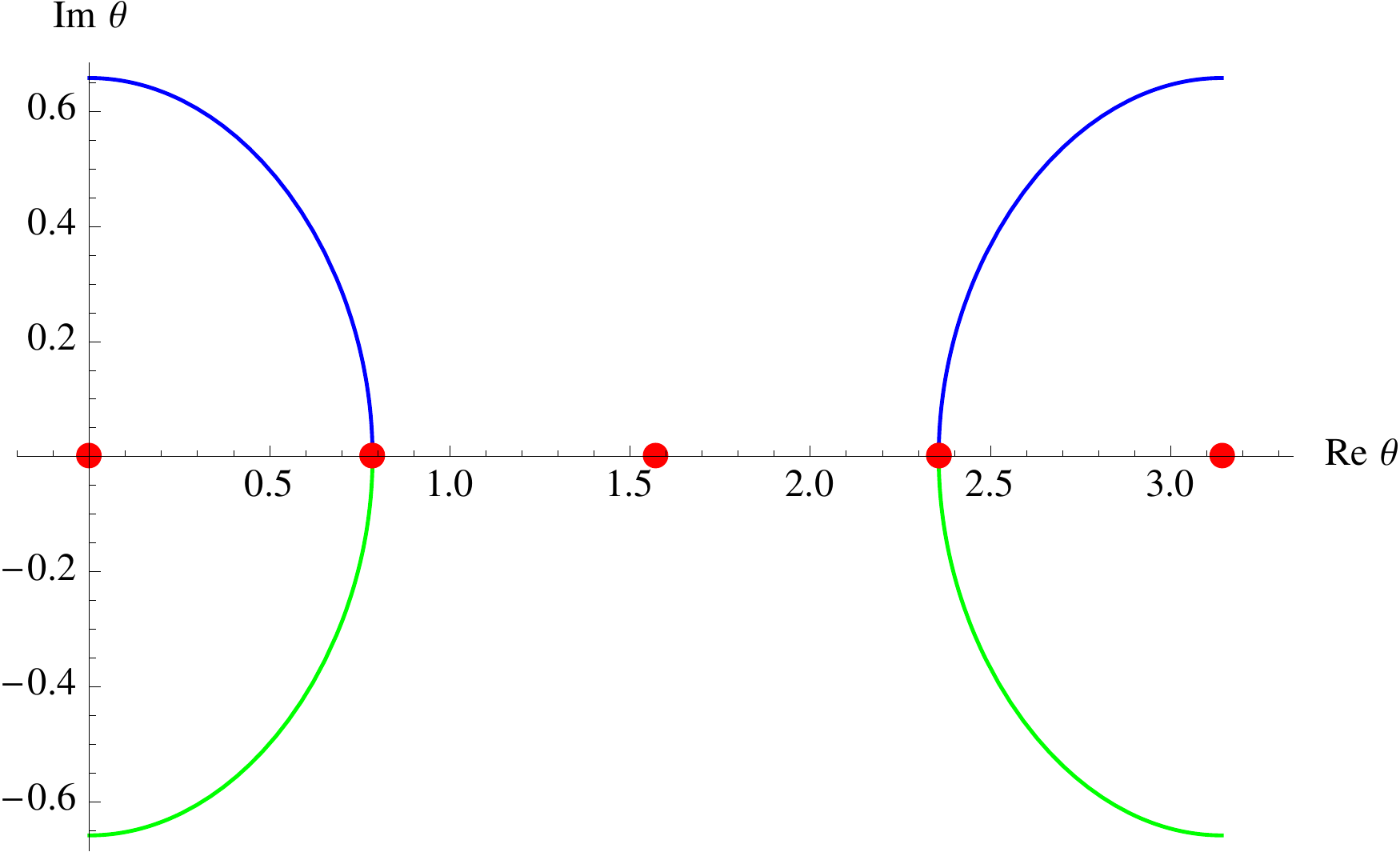}}\\
{\small $\eta=\frac{1}{\sqrt{2}} \,e^{\ii\pi/2}$} & {\small $\eta=e^{\ii\pi/2}$}
\vspace{5mm}
\end{tabular}
\end{center}
  \caption{The red dots correspond to the critical points of the potential (\ref{eq:Pot}) for different values of $\eta$. For $\eta=\ii/\sqrt{2}$ the critical point $\theta_{cr}$ coincide with $\pi/2$ and we only have $3$ real critical points. For $\vert\eta\vert>1/\sqrt{2}$, while keeping $\mbox{arg}(\eta)=\pi/2$, $\theta_{cr}$ moves along the real axis and we have $5$ real critical points.
}
\label{fig:CritPts2}
\end{figure}

If we increase the argument of $\eta$, $\theta_{cr}$ acquires a real part that becomes exactly equal to $\pi/2$ for $\eta = \ii\eta_R$ with $0<\eta_R<1/\sqrt{2}$ so generically we have $7$ critical points, $3$ real and $4$ complex. When $\eta$ is purely imaginary and in modulus less than $1/\sqrt{2}$ we have that $\theta_{cr} = \pi-\bar{\theta}_{cr}$ so we are left with only $5$ critical points, $3$ real and $2$ imaginary, see Figure \ref{fig:CritPts1}.
As $\eta$ approaches $\ii/\sqrt{2}$ the two critical points $\theta_{cr}$ and $\bar{\theta}_{cr}$ approach one another, and for $\eta=\ii/\sqrt{2}$ we have that $\theta_{cr}=\pi/2$ so we are left with only $3$ real critical points, see Figure \ref{fig:CritPts2}.

\begin{figure}[tb]
\begin{center}
\resizebox{65mm}{!}{\includegraphics{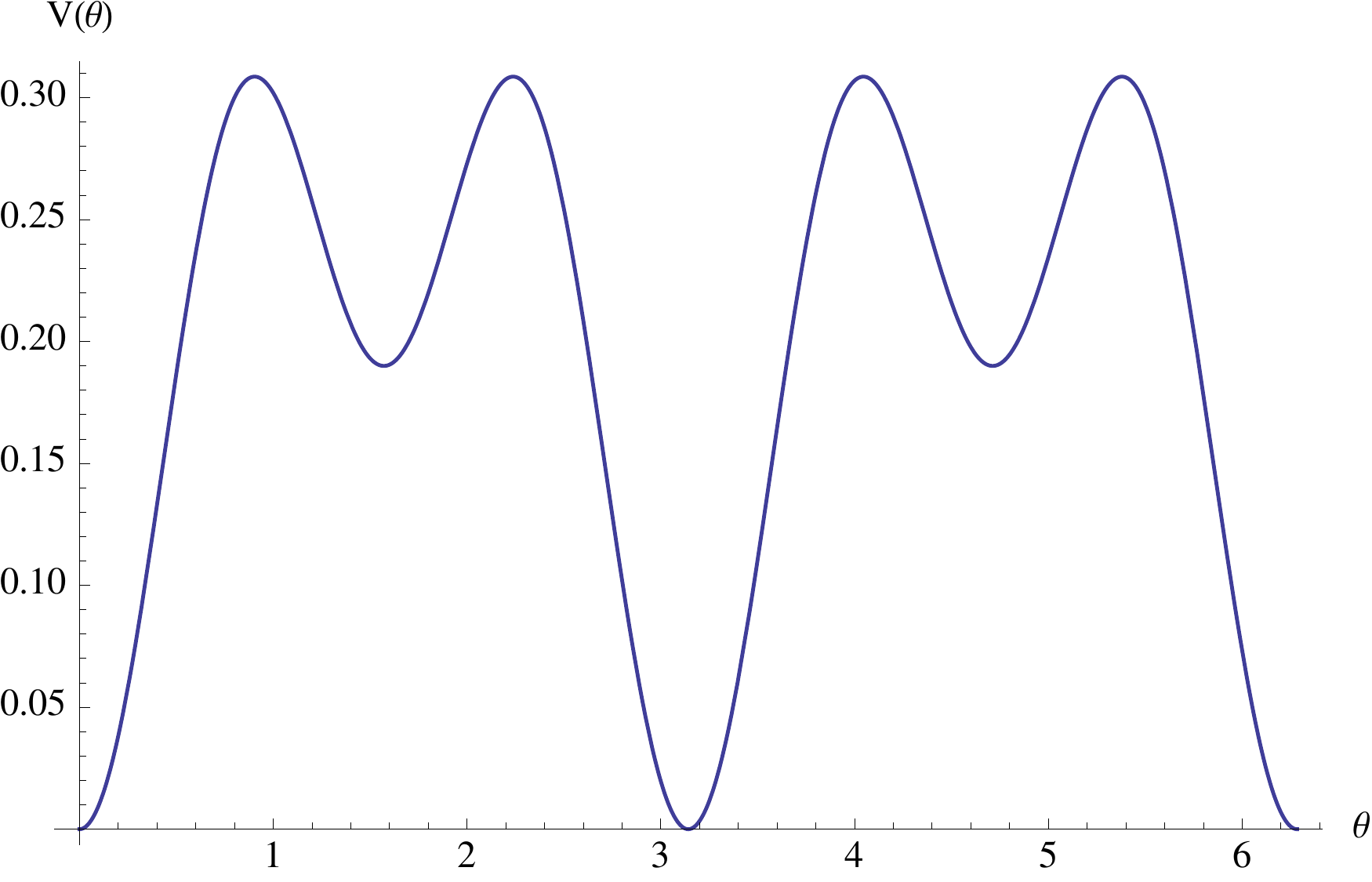}}
\end{center}
  \caption{Profile of the potential (\ref{eq:Pot}) for $\theta$ real and $\eta = \frac{9}{10}\ii$. For $\vert \eta \vert >1/ \sqrt{2}$ with $\mbox{arg}(\eta)=\pi/2$ the critical points are all real.
}
\label{fig:Potential}
\end{figure}

If we keep increasing the modulus of $\eta$ while keeping its argument to $\pi/2$ we see that $\theta_{cr}$ remains real and moves on the real axis towards the origin, while $\theta-\theta_{cr}$ moves on the real axis towards $\pi$. In this regime we have once again $5$ critical points all reals, see Figure \ref{fig:CritPts2}.
For $\eta$ purely imaginary and in modulus larger than $1/\sqrt{2}$, the critical points can be easily understood by plotting the potential $V(\theta)$ for real $\theta$ as in Figure \ref{fig:Potential}. In this regime there are two different types of instanton events corresponding to tunnelling between the tall barrier or the short one. 

The most general solution to the equation of motion (\ref{eq:EOM}) can be obtained by making use of the Weierstrass elliptic functions \cite{Behtash:2015loa} but we will only be interested in finite action solutions to (\ref{eq:EOM}).
The first of such solutions is a typical instanton event interpolating between $\theta\to0$ as $ t\to-\infty$ and $\theta\to \pi$ as $t\to+\infty$, with profile given by,
\be\label{eq:Inst}
\theta_I(t)= \pi - \arccos\left[ \frac{\sqrt{1 +\eta^2} \tanh(t-t_0)}{\sqrt{
  1 + \eta^2 \tanh(t -t_0)^2} }\right]\,,
  \ee
  where $t_0\in\mathbb{R}$ denotes the instanton centre zero-mode.
  The action can be computed,
  \be
  \label{eq:InstAct}
  S_I = 1+(\eta+\eta^{-1})\arctan \eta\,,
  \ee
  and it is precisely equal to the fracton action (\ref{eq:SInst}).
  In the $\eta\to0$ limit the instanton reduces to the standard instanton of the reduced PCM case of \cite{Cherman:2014ofa} and its action stays finite $S_I\to2$.
  
  Clearly an anti-instanton solution can also be found and it has exactly the same action while its profile is given by,
  \be\label{eq:aInst}
\theta_{\bar{I}}(t)=  \arccos\left[ \frac{\sqrt{1 +\eta^2} \tanh(t-t_0)}{\sqrt{
  1 + \eta^2 \tanh(t -t_0)^2} }\right]\,.
  \ee
  
  For $\eta$ real the instanton and the anti-instanton are real solutions (if the modulus $t_0$ is chosen to be real) while for generic complex $\eta$ the profile $\theta_I$ and the action $S_I$ have a non-zero imaginary part. As $\eta$ becomes purely imaginary, $\eta = \ii \eta_R$ with $\eta_R\in\mathbb{R}$, the function $\theta_I$ becomes real once again (for $t_0\in\mathbb{R}$) and the instanton action can be rewritten as,
  \be
  \label{eq:InstActIm}
 S_I = 1-(\eta_R-\eta_R^{-1})\,\arctanh \,\eta_R\,,
  \ee
  which is clearly real.
  
  Note that for any $\eta\in \mathbb{C}$ both the instanton and the anti-instanton interpolates between \textit{different} vacua, namely $\theta = 0$ and $\theta = \pi$. This tells us that only correlated event $I\bar{I}$ interpolating between the same vacuum $\theta= 0$ can (and will) communicate with standard perturbation theory around the vacuum $\theta= 0$.

  Perhaps more interesting are the complex-instanton solutions, called complex-bions in \cite{Behtash:2015loa},
  \be\label{eq:CInst}
\theta_{\mathbb{C}I}(t)=\frac{1}{2} \left[ -\pi + 
   2 \ii \,\mbox{arctanh}\left( \sqrt{1 + \eta^2} \cosh(t-t_0)\right)\right]\,,
  \ee
  together with $\{\pi-\theta_{\mathbb{C}I},\overline{\theta_{\mathbb{C}I}},\pi-\overline{\theta_{\mathbb{C}I}}\}$,
  all with action,
  \be
  \label{eq:SCI}
  S_{\mathbb{C}I} = 1-(\eta+\eta^{-1})\,\mbox{arccot}\, \eta\,.
  \ee
  
\begin{figure}[tb]
\begin{center}
\begin{tabular}{cc}
\resizebox{65mm}{!}{\includegraphics{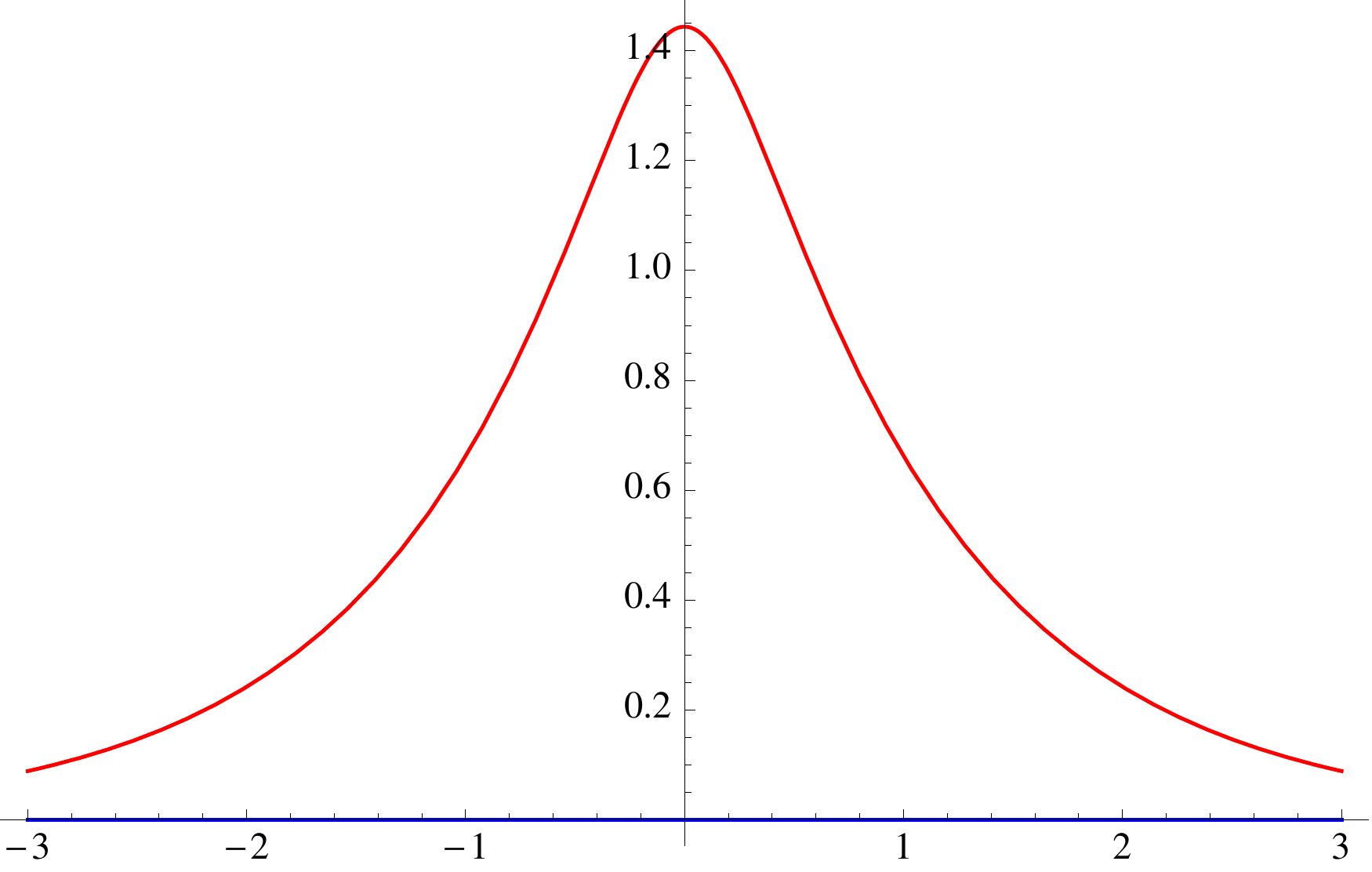}}
\hspace{2mm}
&
\resizebox{65mm}{!}{\includegraphics{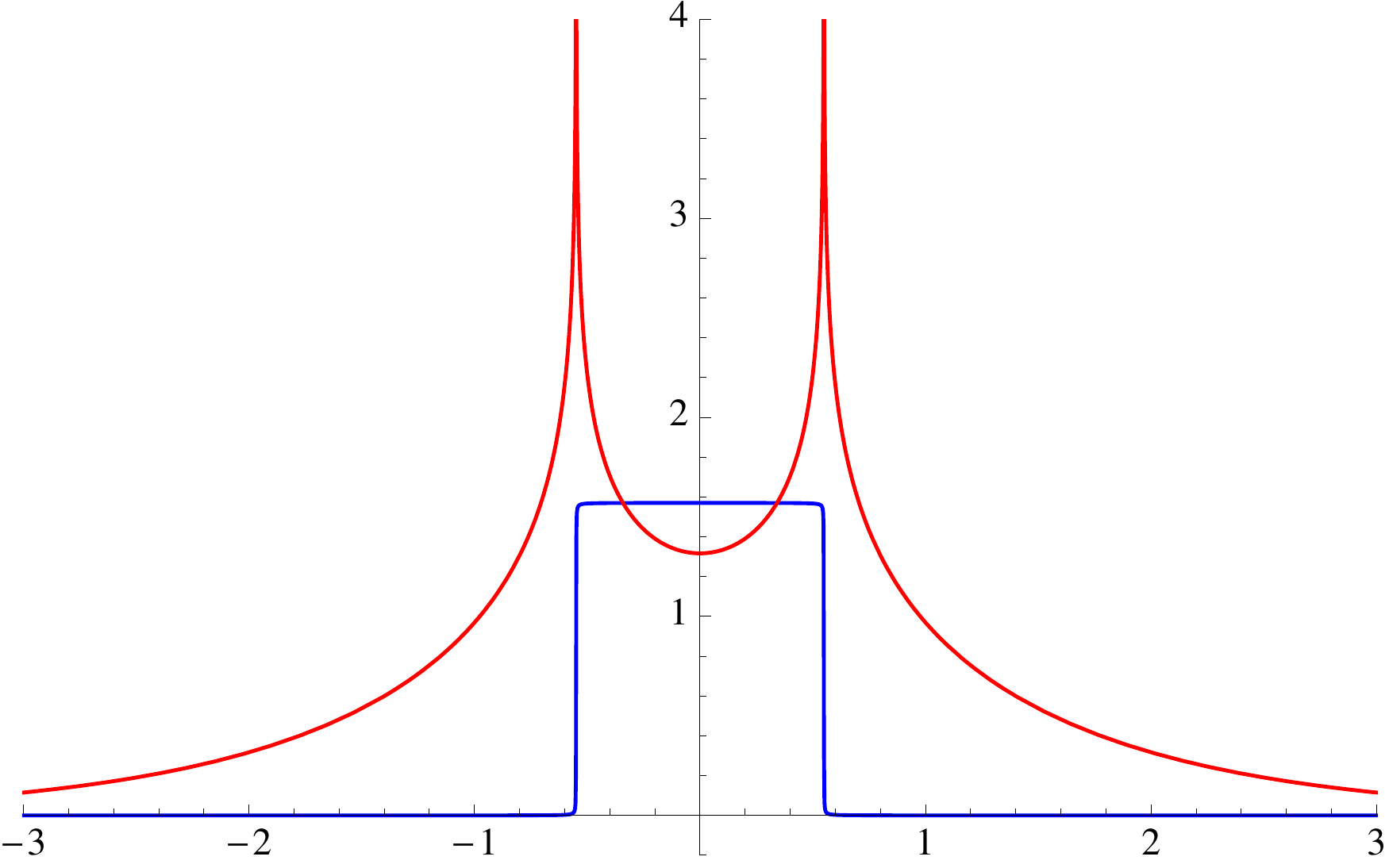}}\\
{\small $\eta=\frac{1}{2}$} & {\small $\eta=\frac{1}{2}\,e^{i\pi/2}$}
\vspace{5mm}
\end{tabular}
\end{center}
  \caption{Profile of the complex-instanton solution (\ref{eq:CInst}) for $\eta=\frac{1}{2}$ (right) $\eta=\frac{1}{2}\,e^{i\pi/2}$ (left), with real part in blue and imaginary part in red.
}
\label{fig:ComplexInstProfile}
\end{figure}

  As we see in Figure \ref{fig:ComplexInstProfile}, for $\eta$ real $\theta_{\mathbb{C}I}(t)$ is purely imaginary (for $t_0\in\mathbb{R}$). The complex-instanton solution tunnels from $\theta \to 0 $ as $t\to -\infty$ to $\theta\to\theta_{cr}$ given in (\ref{eq:ThetaCr}) and then tunnels back to the \textit{same} vacuum $\theta \to 0$ as $t\to +\infty$, and similarly for all the other complex-instanton solutions written above. Note that despite the complex-instanton being a solution living in the complexification of the field space its action is real. 
  In the limit $\eta \to 0$ the complex-instanton $\theta_{\mathbb{C}I}(t)$ becomes singular for $t=t_0$ and its action (\ref{eq:SCI}) diverges but this only happens if we keep the centre modulus real! If we allow ourself to consider $t_0\in\mathbb{C}$, and we should since the field space should be complexified in its entirety, we can see the presence of this additional saddle even in the $\eta=0$ case (see the discussion on complex-unitons of Section \ref{sec:etadef}).
  
If we increase the argument of $\eta$ the solution $\theta_{\mathbb{C}I}(t)$ will stop being purely imaginary and the action \ref{eq:SCI} will become complex, but the complex-instanton will remain a configuration interpolating between the vacuum $\theta\to0$ for $t\to-\infty$, then off to the complex critical point $\theta_{cr}$ and back again for $t\to+\infty$ to the \textit{same} vacuum $\theta\to0$.

When $\eta = \ii \eta_R$ with $0<\eta_R<1/\sqrt{2}$ the complex-instanton (\ref{eq:CInst}) develops two non-integrable singularities as displayed in Figure \ref{fig:ComplexInstProfile}. To obtain a smooth profile function we can either reach $\eta$ purely imaginary from $\mbox{arg}(\eta) = \pi/2\pm\epsilon$ or we can consider a complex centre $t_0$ modulus.
As explained in detail in \cite{Behtash:2015loa}, when $\eta = \ii \eta_R$ with $0<\eta_R<1/\sqrt{2}$ the branch cut of $\arctanh$ produces two types of complex-instanton solutions with actions:
\be\label{eq:SCIimag}
\begin{aligned}
  S_{\mathbb{C}I} &=  \left(1- (\eta_R-\eta_R^{-1})\,\mbox{arctanh}( \eta_R )\right)-\ii \,\frac{\pi}{2}(\eta_R-\eta_R^{-1})\,,\\
S_{\widetilde{ \mathbb{C} I } }  &=   \left(1- (\eta_R-\eta_R^{-1})\,\mbox{arctanh}( \eta_R )\right)+\ii \,\frac{\pi}{2}(\eta_R-\eta_R^{-1})\,,
\end{aligned}
\ee
one complex conjugate of the other.

  Note that for any $\eta\in \mathbb{C}$ the complex-instanton interpolates between the \textit{same} vacuum. This tells us that a single complex-instanton event $\mathbb{C}I$ can (and will) communicate with standard perturbation theory around the vacuum $\theta= 0$.

\section{Perturbation theory in the reduced quantum mechanics}
\label{sec:largeorder}
Now that we have a better knowledge of the non-perturbative sector of our $\eta$-deformed PCM and of the reduced quantum mechanics, we want to understand if these instanton (fracton) and complex-instanton (complex-fracton) events play any role in the semi-classical expansion of physical quantities.
To achieve this we consider a very simple quantity: the ground state energy of the quantum mechanics (\ref{eq:QM}) obtained from the maximally twisted reduction of our QFT.

\subsection{WKB Perturbation theory and Borel singularities} \label{sec:WKBstd}
We will compute the ground state energy in perturbation theory and we will see how this will be insufficient to obtain an unambiguous result because we will be missing some crucial non-perturbative effect.
To compute the large orders perturbative coefficients, we found that using a standard WKB approach was numerically faster than using uniform WKB.

We consider the time independent Schr\"odinger equation associated with the Hamiltonian (\ref{eq:QMWKB}),
\be
-g^4 \frac{\mathrm d^2}{\mathrm dx^2} \psi(x) +Q(x) \psi(x)=0\,,\label{eq:Sch}
\ee
where $Q(x)=V(x) - g^2 E$, with $E$ the rescaled energy and $V(x) = \sin x+\eta^2 \sin^2 x$.

We can use a WKB ansatz for the wave-function,
\be
\psi(x) = e^{-\frac{S(x)}{g^2}}\,,
\ee
reducing the time-independent Schr\"odinger equation to a Riccati equation for $S(x)$:
\be
g^2 S''(x) + S'(x)^2-Q(x)=0\,.\label{eq:Riccati}
\ee

We are interested in studying the resurgence properties of the energy levels of the quantum mechanical system above in the limit $g\to 0$ with $\eta \in \mathbb{C}$ fixed.
In particular for the ground state energy we can make the ansatz:
\begin{align}
S(x) &= \sum_{n=0}^\infty g^{2n} S_n(x)\,,\\
E^{(0)} &= \sum_{n=0}^\infty g^{2n} E_n\,.\label{eq:AnsatzWKB}
\end{align}
 with $E_0=1$, while the coefficients $E_n$ will be some factorially growing, $\eta$ dependent, complex numbers.
 The auxiliary functions $S_n(x)$ can be expanded as
 \be
 S_{n}(x) = \sum_{m=0}^\infty S_{n,m} x^{2m}\,,\label{eq:SnExpanded}
 \ee
 for some complex, $\eta$ dependent, coefficients $S_{n,m}$, since we have assumed that the functions $S_n(x)$ are regular 
  \footnote{For excited states this ansatz has to be changed since the wave-function $\psi(x)$ must have some nodes. In particular if we look for the $N$-th energy level, with $N=2\nu+1$ and $\nu$ a half integer, the energy will be $E^{(\nu)}=1+2\nu + O(g^2)$ while the WKB wave-function must take the form $S(x) = S^{(\nu)}_{0}(x)+ g^2 ( \nu \log x+ S_2^{(\nu)}(x))+O(g^4)$. The function $S^{(\nu)}_{0}(x)$ can be expanded has in (\ref{eq:SnExpanded}) while $S^{(\nu)}_{n\geq 2}$ will have both positive as well as negative integer powers of $x$.} at $x=0$. 
  
 We plugged the ansatz (\ref{eq:AnsatzWKB}) in (\ref{eq:Riccati}), expanded as a power series in $g^2$, and imposed that the Riccati equation is satisfied order by order in $g^2$.
 These are the first few terms in the small $g^2$ expansion of the Riccati equation
 \begin{align*}
 O(g^0):\qquad\qquad& S_0'(x)^2 -(  \sin^2 x +\eta^2 \sin^4 x) =0\,,\\ 
 O(g^2):\qquad\qquad& S_0''(x)+2 S_0'(x)\,S_1'(x) +E_0=0\,,\\
 O(g^4):\qquad\qquad& S_1''(x)+2 S_0'(x) S_2'(x)+ S_1'(x)^2+E_1=0\,.
  \end{align*} 

At this point we can use the small $x$ expansion (\ref{eq:SnExpanded}) and solve for the unknown coefficients $\{E_n,S_{n,m}\}$ order by order in $g$ and in $x$.
   
 Note that this procedure is completely independent from the argument of $\eta$, in particular we can find the perturbative coefficients $\{E_n\}$ for the energy levels of the quantum mechanical system (\ref{eq:Sch}) for all values of $\eta \in \mathbb{C}$. Obviously, the Hamiltonian associated with (\ref{eq:Sch}) is an Hermitian operator only for $\mbox{arg}(\eta) = 0 $ or $\mbox{arg}(\eta) = \pi/2 $. Nonetheless, the study of the large order behaviour of the coefficients $\{E_n\}$ will be extremely interesting if we allow $\mbox{arg}(\eta)$ to vary in $[0,\pi/2]$, despite the quantum mechanical model (\ref{eq:Sch}) not describing a physical system for generic $\eta \in \mathbb{C}$.

 We can compare the first few perturbative coefficients $E_n$ for different values of $\eta$
 \be\label{eq:WKBCoeff}
  \begin{aligned}
  E^{(0)}&= 1 -\frac{1}{4} g^2-\frac{1}{16}g^4-\frac{3}{64}g^6+O(g^8)\,,\qquad\qquad \eta=0\,, \\
  E^{(0)} &= 1 -\frac{1}{16} g^2-\frac{61}{256}g^4+\frac{777}{4096}g^6+O(g^8)\,,\qquad \eta=\frac{1}{2}\,,\\
    E^{(0)} &= 1 -\frac{7}{16} g^2-\frac{13}{256}g^4+\frac{15}{4096}g^6+O(g^8)\,,\qquad \eta=\frac{\ii }{2}\,,
  \end{aligned}
  \ee 
  obtained with standard WKB analysis as described above, against the coefficients (\ref{eq:UniWKBcoeff}), for $B=1/2$ (i.e. the ground state), that we will shortly obtain using uniform WKB approach and the two match perfectly for all values of $\eta\in \mathbb{C}$.
  
 As anticipated, we expect these coefficients $\{E_n\}$ to grow factorially with $n$, hence all the series in (\ref{eq:AnsatzWKB}) are only asymptotic. The first thing we want to understand is the structure of non-perturbative contributions that one must add to our ansatz (\ref{eq:AnsatzWKB}) to construct a proper transseries of the form (\ref{eq:TS}).
 
\begin{figure}[h!]
   \begin{center}
    \includegraphics[width=8cm]{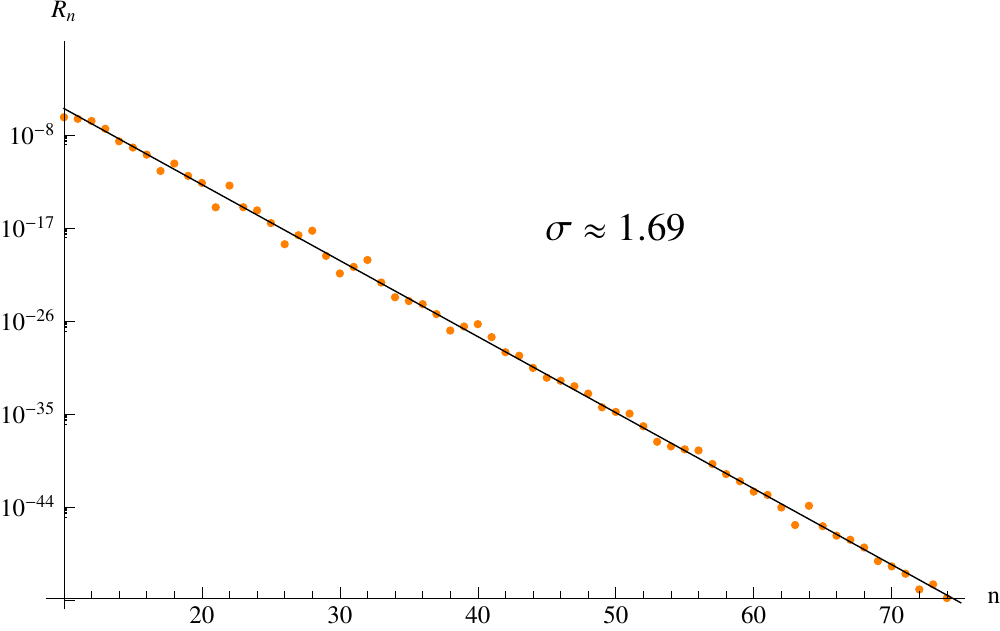}   
     \end{center}
  \caption{\footnotesize{The convergence of the Pad\'e approximant for $\eta =0.2$.  }}  \label{fig:conv}
\end{figure}

We first fix $\eta \in \mathbb{C}$ and then generate, using the procedure outlined above, the first $N_p=150$ perturbative coefficients $\{E_n\}$, while keeping $200$ significant precision digits. 
 We Borel transform the ground state energy, truncated to order $2N_p:\,E_T^{(0)}=E_0+...+g^{2N_p} E_{N_p}$ using,
 \be
 B_T(t) = \sum_{n=1}^{N_p+1} \frac{E_{n-1}}{n!} t^n\,.
 \ee
 Since we only kept a finite number $N_p$ of terms, this Borel transform is an entire function in the complex $t$-plane.
 For this reason we study the pole structure of the Borel-Pad\'e approximants (see e.g. \cite{BenderOrszag}),
 \be
 B_{n,m}(t) = \frac{\mathcal{P}_n(t)}{1+\mathcal{Q}_m(t)}\,,
 \ee
 where the polynomials $\mathcal{P}_n(t)$ and $\mathcal{Q}_m(t)$ can be constructed from $B_T(t)$ and have degrees $n$ and $m$ respectively, with $N_p=n+m$.  
 The idea of the   Borel-Pad\'e  technique is certainly not new and its applications to strongly coupled particle physics were summer school material in 1970  \cite{Bessis:1972bza}.  The use of Borel-Pad\'e approximants are supported by their strong convergence properties. For instance taking a diagonal approximant $m=n$, we can compare the first predicted value for the series with the known result as a function of $n$ and find an exponential convergence, 
\be
\exp R_n = \left| \frac{E_{2n+1}^{predicted} }{E_{2n+1}^{actual}}-1 \right| \sim C e^{- \sigma n}  \ . 
\ee 
This is illustrated by the log plot of  Figure \ref{fig:conv} in which we find $\sigma \sim 1.69$ for $\eta =0.2$.   Similar exponential efficacy of the Borel-Pad\'e method was conjectured  in \cite{Karliner:1998ge} and the method was used to predict quite successfully the then unknown $4$-loop contribution to the $\beta$-function  of QCD.  More recently  the exponential convergence of the  Borel-Pad\'e method was verified against explicit calculations using localisation techniques of correlation functions of Coulomb branch operators in four-dimensional ${\cal N} = 2$ superconformal field theories  in \cite{Gerchkovitz:2016gxx}.

  The Borel-Pad\'{e} approximants, being rational functions, will have poles in the complex $t$-plane. As we increase the number of coefficients $N_p$ we can obtain better and better Borel-Pad\'{e} approximants and their pole structure is supposed to converge to the branch-cut structure of the complete Borel transform,
 \be
 B(t) = \sum_{n=1}^\infty \frac{E_{n-1}}{n!} t^n\,.
 \ee

\begin{figure}[tb]
\begin{center}
\begin{tabular}{cc}
\resizebox{65mm}{!}{\includegraphics{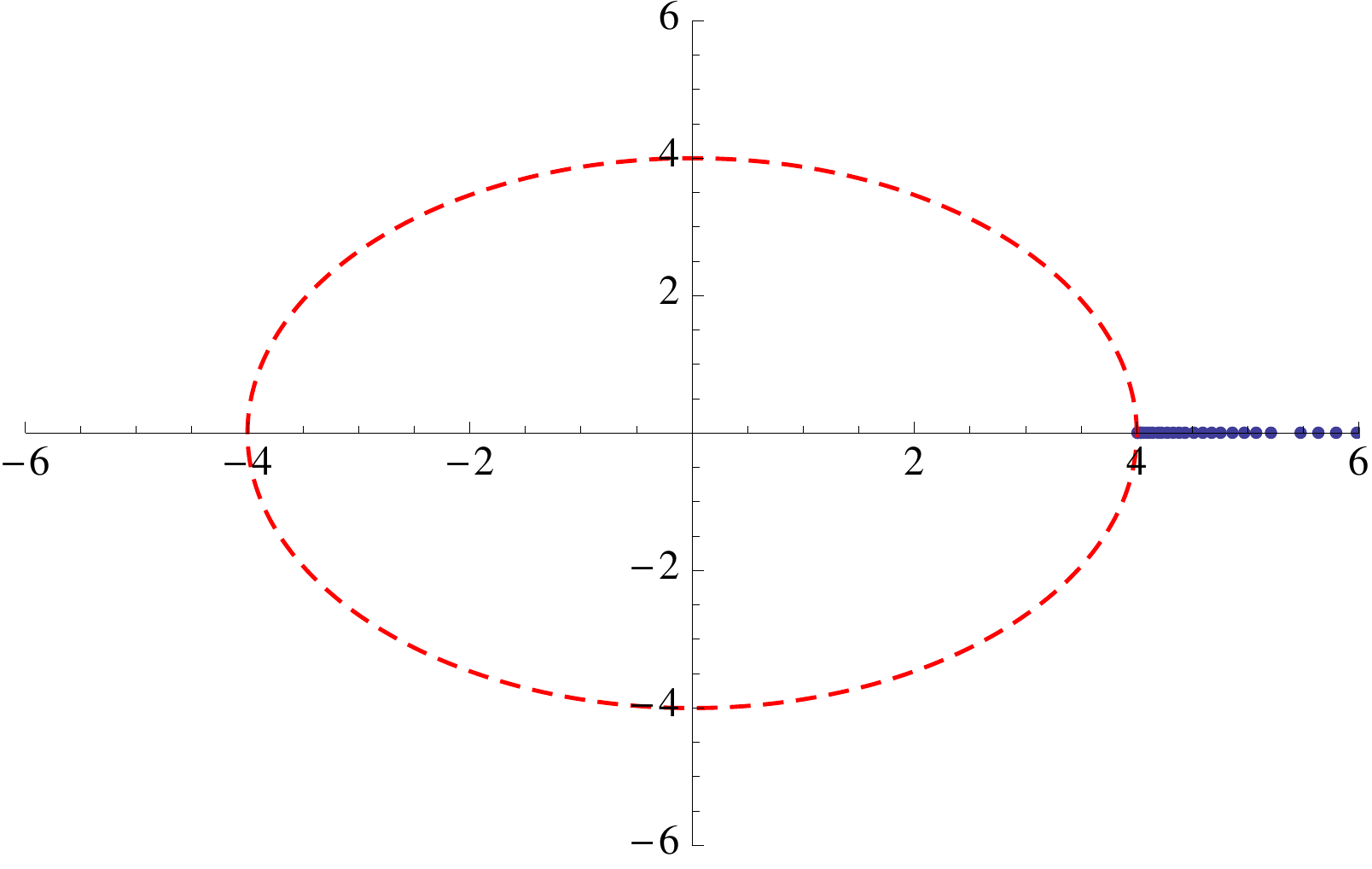}}
\hspace{2mm}
&
\resizebox{65mm}{!}{\includegraphics{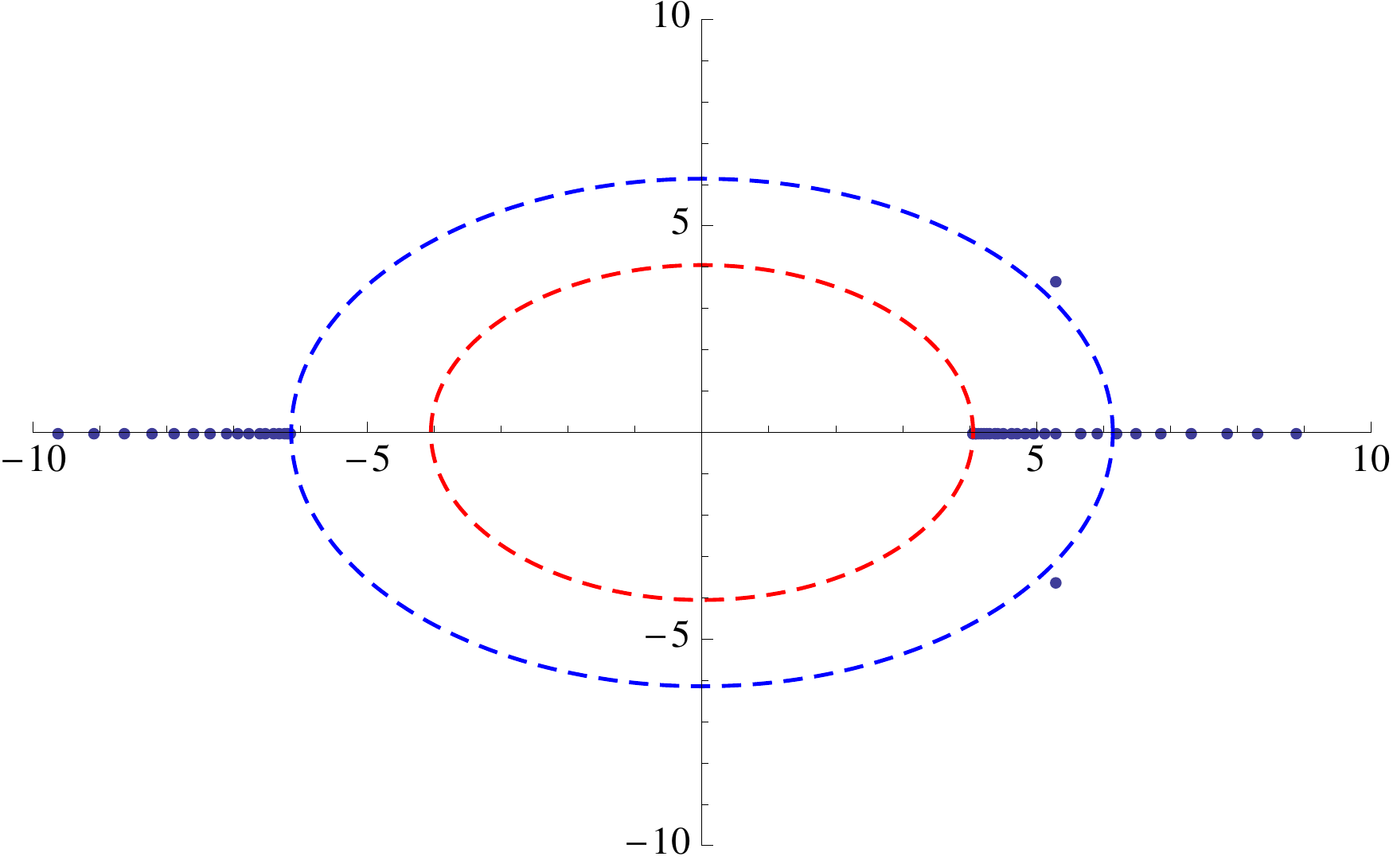}}\\
{\small $\eta=0$} & {\small $\eta=1/5$}
\vspace{5mm}
\end{tabular}
\begin{tabular}{c}
\resizebox{65mm}{!}{\includegraphics{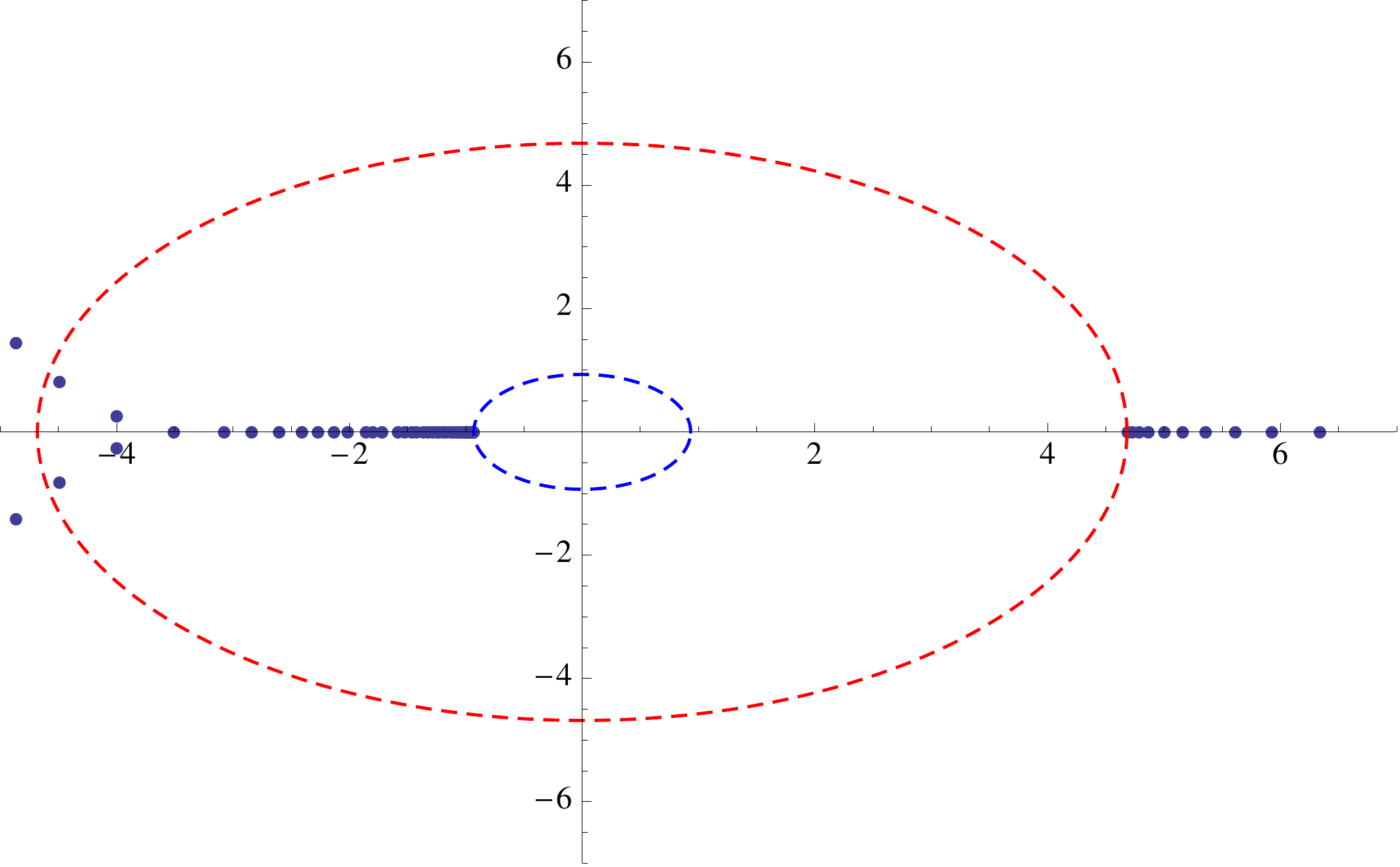}} \\
{\small $\eta=3/4$}
\vspace{-3mm}
\end{tabular}
\end{center}
  \caption{Singularities of the diagonal Borel-Pad\'e approximant $B_{75,75}(t)$ in the complex Borel plane for different values of $\eta=0,\,\eta =\frac{1}{5},\,\eta=\frac{3}{4}$. 
The dashed red circle denotes $\vert t\vert = \vert S_{I\bar{I}}\vert$, while the dashed blue circle denotes $\vert t\vert = \vert S_{\mathbb{C}I}\vert$.
}
\label{fig:polesReal}
\end{figure}

\begin{figure}[tb]
\begin{center}
\begin{tabular}{cc}
\resizebox{65mm}{!}{\includegraphics{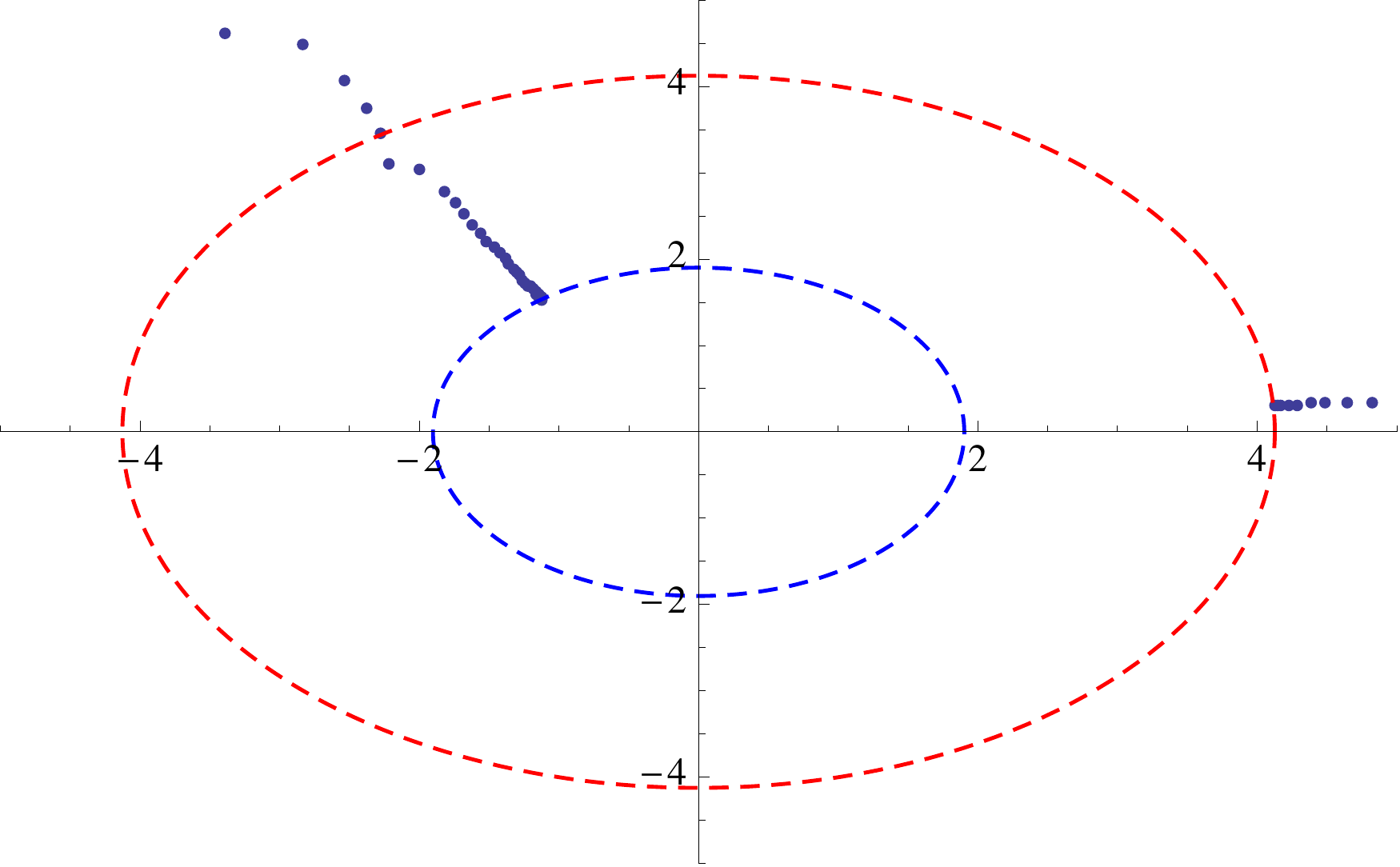}}
\hspace{2mm}
&
\resizebox{65mm}{!}{\includegraphics{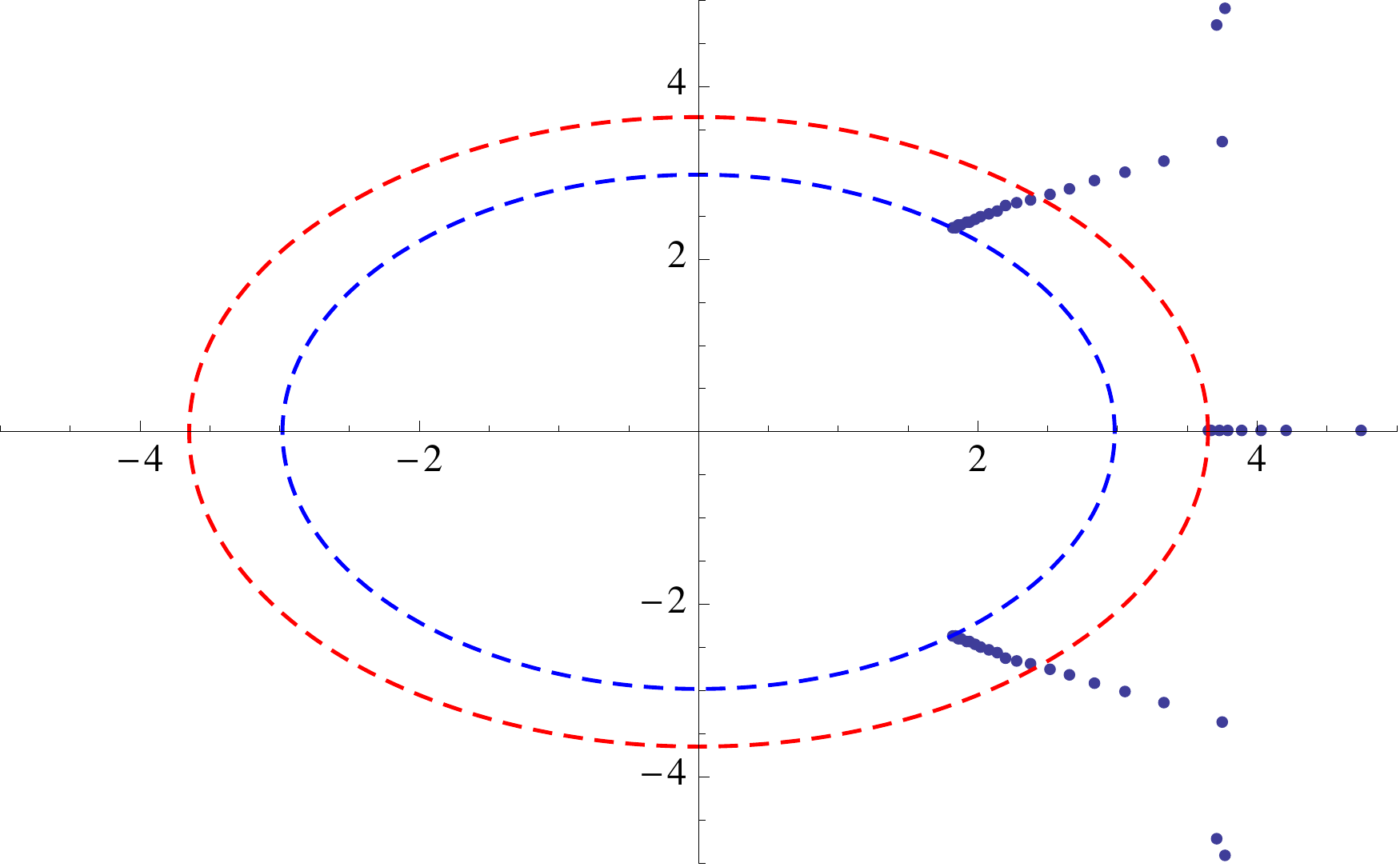}}\\
{\small $\eta=\frac{e^{\ii\,\pi/5}}{2}$ } & {\small $\eta=\frac{e^{\ii \,\pi/2}}{2}$}
\vspace{5mm}
\end{tabular}
\end{center}
  \caption{Singularities of the diagonal Borel-Pad\'e approximant $B_{75,75}(t)$ in the complex Borel plane for $\eta=\frac{e^{\ii\,\pi/5}}{2}$ (left) and $\eta=\frac{e^{\ii\,\pi/2}}{2}$ (right). 
The dashed red circle denotes $\vert t\vert = \vert S_{I\bar{I}}\vert$ while the dashed blue circle denotes $\vert t\vert = \vert S_{\mathbb{C}I}\vert$. On the left plot the branch cuts are in the directions $\mbox{arg}\,t=\mbox{arg}\,S_{I\bar{I}}$ and $\mbox{arg}\,t=\mbox{arg}\,S_{\mathbb{C}I}$ while on the right plot an additional branch is present in the direction $\mbox{arg}\,t=\mbox{arg}\,S_{\widetilde{ \mathbb{C} I } }$.}
\label{fig:polesComplex}
\end{figure}

 We can see from Figure \ref{fig:polesReal}-\ref{fig:polesComplex} that the singularity structures of the Borel-Pad\'{e} approximants change dramatically as we dial $\eta \in \mathbb{C}$. 
 
 For $\mbox{arg}(\eta) =0$, the only two Stokes directions, i.e. singular directions, for $B(t)$ are $\theta =\mbox{arg}(t)= 0$ and $\theta= \pi$. In this regime we know that there are two non-perturbative field configurations which can mix with perturbation theory, an instanton-anti-instanton configuration with action (\ref{eq:InstAct}), 
 \be
 S_{I\bar{I}} =2 S_I= 2 \left(1+ (\eta+\eta^{-1})\,\mbox{arctan} ( \eta ) \right)\,,\label{eq:SIIb}
 \ee 
 and a complex-instanton configuration with action (\ref{eq:SCI}),
 \be
 S_{\mathbb{C}I} =  \left(1- (\eta+\eta^{-1})\,\mbox{arccot}( \eta )\right)\,.
 \ee 
 Note that for $\eta \in\mathbb{R}$ both actions are real, $S_{I\bar{I}}>0$ while $S_{\mathbb{C}I}<0$.
 We see in Figure \ref{fig:polesReal} that the Borel transform has two branch-cuts starting precisely at $t= S_{I\bar{I}}$ and $t=S_{\mathbb{C}I} $.
 
 If we try to go back to the undeformed case, $\eta = 0$, we can see that the branch cut along the Stokes lines $\theta=\pi$ tries to disappear, moving towards $-\infty$, since $S_{\mathbb{C}I} \sim -\frac{\pi}{2\eta} +O(\eta^0) $ as $\eta \to 0$, while the branch cut along $\theta= 0$ starts precisely at $ t =  S_{I\bar{I}}  \sim 4+O(\eta)$, as expected from the undeformed Principal Chiral Model analysis carried out in \cite{Cherman:2014ofa} and shown in Figure \ref{fig:polesReal}.
 
 As we increase the argument of $\eta$, we see in Figure \ref{fig:polesComplex} that the two Stokes lines move from $\theta=0,\pi$ to the directions $\theta =\mbox{arg} (S_{I\bar{I}}) $ and $\theta = \mbox{arg}( S_{\mathbb{C}I}) $ and the two branch cuts start precisely at $t=S_{I\bar{I}} $ and $t=S_{\mathbb{C}I}$.

  When we reach $\mbox{arg}(\eta) =\pi/2$, i.e. $\eta = \ii \eta_R$ with $\eta_R \in\mathbb{R}$ and $\vert\eta_R\vert<1/\sqrt{2}$, the Borel transform of the ground state energy has three branch cut.
 The reason is that the complex-instanton action (\ref{eq:SCI}) has a branch cut discontinuity in the complex $\eta$ plane running precisely between $-\ii$ and $\ii$. This gives rise to a third type of non-perturbative saddle (see the discussion in Section \ref{sec:NPsaddles}) which can mix with perturbation theory. The actions (\ref{eq:InstActIm}-\ref{eq:SCIimag}) of the three relevant non-perturbative field configurations for $\eta = \ii\eta_R$ with $\eta_R \in\mathbb{R}$ and $\vert\eta_R\vert<1/\sqrt{2}$ are:
  \begin{align}
 S_{I\bar{I}} & = 2 \left(1- (\eta_R-\eta_R^{-1})\,\mbox{arctanh} ( \eta_R ) \right)\,,\label{eq:SFFbIm}\\
  S_{\mathbb{C}I} &=  \left(1- (\eta_R-\eta_R^{-1})\,\mbox{arctanh}( \eta_R )\right)-\ii \,\frac{\pi}{2}(\eta_R-\eta_R^{-1})\,,\label{eq:SCIIm}\\
S_{\widetilde{ \mathbb{C} I } }  &=   \left(1- (\eta_R-\eta_R^{-1})\,\mbox{arctanh}( \eta_R )\right)+\ii \,\frac{\pi}{2}(\eta_R-\eta_R^{-1})\,.\label{eq:SCItildeIm}
 \end{align}
 
 Note that $ S_{I\bar{I}} $ is once again real, while $S_{\mathbb{C}I}$ and $S_{\widetilde{ \mathbb{C} I } }  $  are complex and furthermore they are complex conjugates to one another: $S_{\widetilde{ \mathbb{C} I } }=\overline{S_{\mathbb{C}I}}  $. We have thus three Stokes lines at $\theta= 0$ and $\theta = \pm \mbox{arg} (S_{\mathbb{C}I})$, and the branch cuts of the Borel transform start precisely at $t=S_{I\bar{I}},$ $t= S_{\mathbb{C}I}$ and $t=S_{\widetilde{ \mathbb{C} I } }$ respectively, as shown in Figure \ref{fig:polesComplex}.
  
  Finally, if we crank up the modulus of $\eta$ such that $\vert \eta\vert \geq 1/\sqrt{2}$, while keeping $\mbox{arg}(\eta) =\pi/2$, we know from the analysis of Section \ref{sec:NPsaddles} that the complex-instantons cease to exist and we only have two types of non-perturbative saddles, both with real actions and the only Stokes line is for $\theta = 0$.

\subsection{Uniform WKB}\label{sec:UWKB}
To analyse the perturbative expansion in detail we follow the uniform WKB method that is well-suited to potentials with degenerate minima \cite{Dunne:2014bca}. This consists of two steps, first one is to solve the Schr\"odinger equation about a given minimum perturbatively and then one invokes a global boundary condition that will fully determine the energy levels.

Since the potential in (\ref{eq:QMWKB}) is locally harmonic, a sensible ansatz for the wave-function is,
\be \label{eq:UWKBwf}
\psi(\theta) = \frac{ D_\nu \left( \frac{  u(\theta)}{g}  \right)}{ \sqrt{u'(\theta)}}  \ ,
\ee
in which $D_\nu(z)$ is a parabolic cylinder function and $\nu =  B - \frac{1}{2}$ is parameter that will be determined via the global boundary condition but is exponentially close to an integer.  Substitution of this ansatz into the Schr\"odinger equation yields a non-linear equation,  
\be\label{eq:RiccatiUWKB}
V - \frac{1}{4}u^2 (u^\prime)^2 - g^2 E + g^2 B (u^\prime)^2 + \frac{g^4}{2} \sqrt{u^\prime} \left( \frac{u^{\prime\prime} }{ (u^\prime)^\frac{3}{2} } \right)^\prime = 0 \ ,
\ee
which is studied perturbatively with, 
 \be\label{eq:UWKBen}
E = \sum_{n=0}^\infty {\cal E}_{n}(B, \eta) g^{2 n} \ , \quad u(y) = \sum_{n=0}^\infty u_{n}(y,B,\eta ) g^{2 n} \ . 
\ee
 The first few terms of the energy series are for the potential of equation (\ref{eq:QMWKB}) are given by,  
 \be
 \label{eq:UniWKBcoeff}
\begin{aligned}
{\cal E}_{0} &= 2B \ , \\
 {\cal E}_{1} &= \frac{1}{8} \left(4 B^2+1\right) \left(3 \eta ^2-1\right)  \ , \\
{\cal E}_{2} &= \frac{B }{32} \left(-67 \eta ^4-22 \eta ^2-3\right)+\frac{B^3}{8} \left(-17 \eta ^4-2 \eta ^2-1\right) \ ,  \\
{\cal E}_{3} &= \frac{9 \left(171 \eta ^6+109 \eta ^4+\eta ^2-1\right)}{1024} + \frac{B^2}{128} \left(1707 \eta ^6+653 \eta ^4+17 \eta ^2-17\right)\\
&\phantom{-------------------}+\frac{5 B^4 }{64} \left(75 \eta ^6+13 \eta ^4+\eta ^2-1\right) \ . 
\end{aligned}
\ee
We can set $\eta=0$ and read the coefficients of the undeformed case already computed in \cite{Dunne:2014bca} and, for $B=1/2$ we can directly compare against the numerical coefficients (\ref{eq:WKBCoeff}) found using standard WKB for different values of $\eta$ with perfect agreement.

We compute the wave-function in perturbation theory and at zero'th order the wave-function is given by, 
 \be
 \begin{aligned}
 u_{0}(\theta )^2 & = 4 \int^\theta_0 \textrm{d}\tilde{\theta}  \sqrt{V(\tilde{\theta})} \\   
& =  \frac{2}{\eta} \left[\eta  \left(1-\cos (\theta ) \sqrt{\eta ^2 \sin ^2(\theta)+1}\right)-\left(\eta ^2+1\right) \tan ^{-1}\left(\frac{\eta  \cos (\theta )}{\sqrt{\eta ^2 \sin ^2(\theta )+1}}\right) \right.\\
&\qquad \left.+\left(\eta ^2+1\right) \tan ^{-1}(\eta )\right] \, . 
   \end{aligned}
 \ee
The first and second order wave-functions are given in terms of $u_0$ by,
\be
\begin{aligned}
u_1(\theta) &= \frac{B }{ u_0 }\left(\log \left(\frac{1}{8} \left(\eta ^2+1\right) u_0(\theta ){}^2\right)+2
   \tanh ^{-1}\left(\frac{\sin (\theta ) \cos (\theta )}{\sqrt{V(\theta )}}\right)\right) \ , \\
 u_2(\theta) &=-\frac{4 B^2+3}{4 u_0(\theta ){}^3}+\frac{2 B u_1(\theta )}{u_0(\theta
   ){}^2}+\frac{\chi(\theta )}{u_0(\theta )}-\frac{u_1(\theta ){}^2}{2 u_0(\theta )}  \ , 
\end{aligned}
\ee 
in which the function $\chi$ is quite unpleasant to write in full but is obtained by solving,
\be
16 V^\frac{5}{2} \chi' =4 V V^{\prime\prime}  -16 {\cal E}_1 V^2 -5 (V^\prime)^2 - 16 B^2 V \ ,
\ee
with the constant of integration such that $ u_2(0)=0$.   The values of these wave-functions at the mid-point $\theta= \frac{\pi}{2}$ will presently become useful and they are given by,
\be\label{eq:UWKBwftp}
\begin{aligned}
u_0\left(\frac{\pi}{2} \right) &= \sqrt{2 S_I}  \ , \quad  u_1(\pi /2 ) = (2 S_I)^{-\frac{1}{2}} B \log\left[ \frac{1}{4} S_I (1+\eta^2) \right] \ , \\
u_2\left(\frac{\pi}{2} \right) &= \frac{1}{96 \sqrt{2} S_I^\frac{3}{2} (1+\eta^2)} \left[ -24 B^2 \left(\eta ^2+1\right)
   \left(\log \left(\frac{\eta ^2+1}{4}
 S_I  \right)-4\right) \log \left(\frac{\eta ^2+1}{4}
S_I \right) \right. \\
& \left. -12 \left(4 B^2+3\right) \left(\eta ^2+1\right) + S_I \left(-12 B^2 \left(17 \eta ^4+6 \eta
   ^2-3\right)-67 \eta ^4-66 \eta ^2+9\right)   \right] \ ,
\end{aligned}
\ee
in which we recall that $S_I$ is precisely the instanton action defined in equation (\ref{eq:InstAct}). 
 
 \section{Resurgence Analysis} \label{sec:Resurgence}
 We are now in position to obtain a transseries ansatz for the ground state energy of the quantum mechanics  (\ref{eq:QMWKB}) and obtain some novel predictions.
 
 As explained in the Introduction and as we have just seen in practice, the Borel transform of the perturbative ground state energy (\ref{eq:AnsatzWKB}) will generically have branch cuts along some Stokes directions $\mbox{arg}(t)=\theta$.
 The two lateral resummations $\mathcal{S}_{\theta+\epsilon}(E^{pert})$ and $\mathcal{S}_{\theta-\epsilon}(E^{pert})$ will differ from one another and furthermore their difference will be exponentially suppressed,
 \be
 \mathcal{S}_{\theta+\epsilon}(E^{pert})-\mathcal{S}_{\theta-\epsilon}(E^{pert})\sim 2\pi\,\ii\,e^{-S_\theta/g^2} \Phi_\theta(g^2)\,,
 \ee
 with $S_\theta\in\mathbb{C}$ and $\mbox{arg}(S_\theta)=\theta$ and $\Phi_\theta(g^2)$ a new function with an asymptotic expansion for $g^2\sim 0$.
 
Thanks to a standard dispersion-like argument \cite{Bender:1973,Collins:1977dw} we can relate the large order behaviour of the perturbative coefficients $E_n$ in (\ref{eq:AnsatzWKB}) to all the discontinuities of the Borel transform $\mathcal{B}[E^{pert}]$:
\be\label{eq:DispRel}
E_n\sim \sum_{\theta_\ast}\frac{1}{2\pi \ii} \int_0^{e^{\ii\theta_\ast}\infty} \mathrm dw\,w^n\,\mbox{Disc}_{\theta_\ast}(E^{pert}(w))\,,
\ee 
where $\theta_\ast$ runs over all the Stokes directions while the discontinuity is precisely related to the jump in the lateral resummations $\mbox{Disc}_{\theta}(E^{pert}(w)) = \mathcal{S}_{\theta+\epsilon}(E^{pert})-\mathcal{S}_{\theta-\epsilon}(E^{pert})$.

From the discussion of Section \ref{sec:largeorder}, we know that $\mathcal{B}[E^{pert}]$ has two Stokes directions $\theta=0$ and $\theta= \pi$ for $\eta\in\mathbb{R}$ while for $\eta=\ii\eta_R$ with $0<\eta_R<1/\sqrt{2}$ we have three Stokes directions for $\theta=0$ and $\theta = \pm \mbox{arg}( S_{\mathbb{C}I})$ following (\ref{eq:SCIIm}-\ref{eq:SCItildeIm}).

Let us start with the $\eta\in\mathbb{R}$ case. 
A careful analysis of equation (\ref{eq:DispRel}) together with our numerical studies of the perturbative coefficients $E_n$ for numerous values of $\eta$, reveals that the large order perturbative coefficients behave for $n\gg1$ as,
\be\label{eq:LargeOrderRe}
\begin{aligned}
E_n&\sim A(\eta) \left(\frac{1}{2S_I}\right)^{n + 1}
  \Gamma(n+1) \left( 
  1 +a_{I\bar{I}}^{(1)}(\eta) \frac{2S_I}{n} + a^{(2)}_{I\bar{I}}(\eta) \frac{(2S_I)^2}{n(n-1)}+O(n^{-3}) \right)+\\
  &+ B(\eta) \left(\frac{1}{S_{\mathbb{C}I}}\right)^{n + 1/2}
  \Gamma\left(n+\frac{1}{2}\right)\left( 
  1 +a_{\mathbb{C}I}^{(1)}(\eta) \frac{S_{\mathbb{C}I}}{n-\frac{1}{2}} +  O(n^{-2}) \right)+...
  \end{aligned}\,,
\ee
where we omitted higher instanton contributions.
In the above equation $A(\eta)$ and $B(\eta)$ correspond to the Stokes constants in the singular directions $\theta=0$ and $\pi$ respectively, while $S_I$ and $S_{\mathbb{C}I}$ are precisely the instanton and complex-instanton actions (\ref{eq:aInst}-\ref{eq:SCI}). The coefficients $a_{I\bar{I}}^{(i)}(\eta) $ are the perturbative coefficients on top of the instanton--anti-instanton solution while $a_{\mathbb{C}I}^{(i)}$ are the perturbative coefficients on top of the complex-instanton solution.

For $\eta=\ii\eta_R$ with $0<\eta_R<1/\sqrt{2}$ the situation is very similar and the large order perturbative coefficients behave for $n\gg1$ as,
\be\label{eq:LargeOrderIm}
\begin{aligned}
E_n&\sim A(\ii \eta_R) \left(\frac{1}{2S_I}\right)^{n + 1}
  \Gamma(n+1) \left( 
  1 +a_{I\bar{I}}^{(1)}(\ii \eta_R) \frac{2S_I}{n} + a^{(2)}_{I\bar{I}}(i \eta_R) \frac{(2S_I)^2}{n(n-1)}+O(n^{-3}) \right)+\\
  &+ B(\ii  \eta_R) \left(\frac{1}{S_{\mathbb{C}I}}\right)^{n + 1/2}
  \Gamma\left(n+\frac{1}{2}\right)\left( 
  1 +a_{\mathbb{C}I}^{(1)}(\ii  \eta_R) \frac{S_{\mathbb{C}I}}{n-\frac{1}{2}} +  O(n^{-2}) \right)+\\
  &- B(\ii \eta_R) \left(\frac{1}{S_{\widetilde{ \mathbb{C} I } }}\right)^{n + 1/2}
  \Gamma\left(n+\frac{1}{2}\right)\left( 
  1 +a_{\mathbb{C}I}^{(1)}(\ii \eta_R) \frac{S_{\widetilde{ \mathbb{C} I } }}{n-\frac{1}{2}} +  O(n^{-2}) \right)+...
  \end{aligned}\,,
\ee
where the actions $S_{\mathbb{C}I} $ and $S_{\widetilde{ \mathbb{C} I } }$ are precisely given by (\ref{eq:SCIIm}-\ref{eq:SCItildeIm}) and we omitted again higher instanton contributions.
Note that in the above equation the Stokes constants $A(\eta)$ and $B(\eta)$, the instanton--anti-instanton perturbative coefficients $a_{I\bar{I}}^{(i)}(\eta) $, and the complex-instanton coefficients $a_{\mathbb{C}I}^{(i)}$ are exactly the same as in (\ref{eq:LargeOrderRe}), simply evaluated at purely imaginary $\eta =\ii\eta_R$. We have checked this also for generic complex values of $\eta$ and it remains correct.
 
 Making use of the uniform WKB approach\footnote{Roughly we need to use the resurgent
asymptotic behaviour of the parabolic cylinder functions in the complex plane, we refer to \cite{Dunne:2014bca} for all the technical details.} \cite{Dunne:2014bca} we can compute the Stokes constant $A(\eta)$ and the perturbative coefficients on top of the instanton-anti-instanton solution $a_{I \bar{I}}^{(i)}(\eta) $.

We first write the energy transseries for the $N^{th}$ level,
\be\label{eq:TSEn}
E^{(N)}(g^2) = E(B=N+\frac{1}{2},g^2) + \delta \nu\,\left[\frac{\partial E(B,g^2)}{\partial B}\right]_{B=N+\frac{1}{2}}+O(\delta \nu^2)\,,
\ee
where the asymptotic series $E(B=N+\frac{1}{2},g^2)$ is precisely the object we computed in (\ref{eq:UWKBen})  in terms of the coefficients (\ref{eq:UniWKBcoeff}).
At leading non-perturbative order, the parabolic cylinder function index $\nu$ in (\ref{eq:UWKBwf}) is exponentially close to the level number $N$ and $\delta \nu = \nu-N$ as explained in \cite{Dunne:2014bca}.

 As mentioned in Section \ref{sec:UWKB} to determine fully the energy levels we have to impose a global boundary condition.
 This global boundary condition is an extra constraint that the uniform WKB function (\ref{eq:UWKBwf}) must satisfy and it arises from the Bloch phenomenon. The eigenfunctions of a periodic potential must satisfy: $\psi(x +\pi)=e^{i\theta_{B} } \psi(x)$, where $\theta_B$ is the Bloch angle $\theta_B\in[0,\pi]$ that labels states in a given band of the spectrum.
 We can compute $\delta \nu$ by making use of this global boundary condition expressed by equation (63) of \cite{Dunne:2014bca} and if we consider the lowest part of the band and set the Bloch angle to $\theta_B=0$ we obtain,
 \be
 1\sim g\frac{\delta \nu} {u(\pi/2)} \exp\left(\frac{u(\pi/2)^2}{2g^2}\right)\,,
 \ee
 where $u(x)$ is the uniform WKB wave function, solution to (\ref{eq:RiccatiUWKB}), evaluated at mid turning point.
We can expand the wave-function at mid-turning point as an asymptotic power series in $g$ and the first few coefficients are given in (\ref{eq:UWKBwftp}). At leading order we have,
\be
\label{eq:deltanu}
\delta \nu \sim \sqrt{\frac{8}{g^2(1+\eta^2)}} \,\exp\left(-\frac{S_I}{g^2}\right)\,,
\ee
where $S_I$ is once again given by (\ref{eq:aInst}). By substituting (\ref{eq:deltanu}) in our transseries (\ref{eq:TSEn}) we see that the splitting of the lowest band is, as expected, a one instanton effect.

By expanding the global boundary condition to order $e^{-2S_I/g^2}$, we obtain an imaginary part of $\delta \nu$, 
\be
\label{eq:imnu}\begin{aligned}
&\mbox{Im}\,\delta \nu = \pm\pi\, \frac{16}{1+\eta^2}\left[1+ \left( 2B \sqrt{2S_I} u_1(\pi/2) - B^2\right. \right.\\
& \phantom{----------}\left.\left.-\frac{3}{4} -\sqrt{(2S_I)^3} u_2(\pi/2)- S_I u_1(\pi/2)^2\right) \frac{g^2}{S_I}\right] \exp\left(-\frac{2S_I}{g^2}\right)\,,
\end{aligned}
\ee
where the sign $\pm$ is correlated with $\mbox{arg}(g^2)\gtrless0$.
The leading imaginary part of the energy transseries (\ref{eq:TSEn}) coming from the two-instanton sector, including
the perturbative fluctuations around it, can be found from, 
 \be\label{eq:2deltanu}
 \begin{aligned}
 &\mbox{Im}\left[\ \delta \nu\,\frac{\partial E(B,g^2)}{\partial B}\right]_{B=\frac{1}{2}}=\\
 &\pm 2\pi\,\frac{16}{1+\eta^2}
 \left[1+\frac{1}{24} \left(-23 + 77 \eta^2 + \frac{8}{(1 + \eta^2)}\right)\frac{g^2}{2S_I}+O(g^{-4})\right]\exp\left(-\frac{2S_I}{g^2}\right)\,.
 \end{aligned}
 \ee
 
 By imposing that this purely imaginary term in the transseries (\ref{eq:TSEn}) is cancelled exactly by the discontinuity of the resummation of the perturbative expansion $\mbox{Disc}_{0}(E^{pert}(g))$ translates immediately into the large-order behaviour of the perturbative coefficients and we can read the Stokes constant $A(\eta)$ and the first perturbative correction on top of the instanton solution $a_{I\bar{I}}^{(1)}$:
\begin{align}
A(\eta) &\label{eq:A}= -\frac{16}{1+\eta^2}\,,\\
a_{I\bar{I}}^{(1)}(\eta) &\label{eq:aI}= \frac{1}{24} \left(-23 + 77 \eta^2 + \frac{8}{1 + \eta^2}\right)\,.
\end{align}
 As a check of our result we see that $a_{I\bar{I}}^{(1)}(0)=-\frac{5}{8}$ in precise agreement with equations (103)-(104) of \cite{Dunne:2014bca}.
 A path integral derivation, as opposed to our approach based on the Schr\"odinger equation, of (\ref{eq:aI}) would be an amazing check of the resurgent program. For the sine-Gordon model this agreement has been shown up to three loops \cite{Escobar-Ruiz:2015rfa}.
 
 Note that the uniform WKB is completely oblivious of the argument of $\eta\in\mathbb{C}$ so we use (\ref{eq:A}) and (\ref{eq:aI}) for generic $\eta$. It is also interesting to notice that from the uniform WKB is not at all obvious how to extract the Stokes constant $B(\eta)$ and the perturbative coefficients $a^{(i)}_{\mathbb{C}I}(\eta)$ associated with the complex-instanton. A superficial expansion of $\delta \nu$ seems to contain only powers of $\exp(-S_I/g^2)$ while we know from the singularities of $\mathcal{B}(E^{(pert)})$ that we should also find terms of the form $\exp(-S_{\mathbb{C}I}/g^2)$. Clearly these terms must be hiding in the asymptotic nature of the uniform WKB wave-function expansion (\ref{eq:UWKBen}-\ref{eq:UWKBwftp}) but for the moment we do not know how to extract them\footnote{We thank Gerald Dunne and Mithat Unsal for discussion on this issue.}.
 Thanks to our study of the large order behaviour of the perturbative coefficients we predict
 \be\label{eq:BCI}
 \begin{aligned}
 B(\eta) &= \frac{4\, \ii}{\sqrt{\pi^3\, \left( 1 +\eta^2\right)}}\,,\\
 a^{(1)}_{\mathbb{C}I}(\eta) &=\frac{1}{48}\left(-29 + 95 \eta^2 + \frac{8}{1 + \eta^2}\right)\,.
 \end{aligned}
 \ee
 A proper complexified path integral derivation of (\ref{eq:BCI}) would be an amazing check of the resurgent program.
  
\begin{figure}[tb]
\begin{center}
\begin{tabular}{cc}
\resizebox{65mm}{!}{\includegraphics{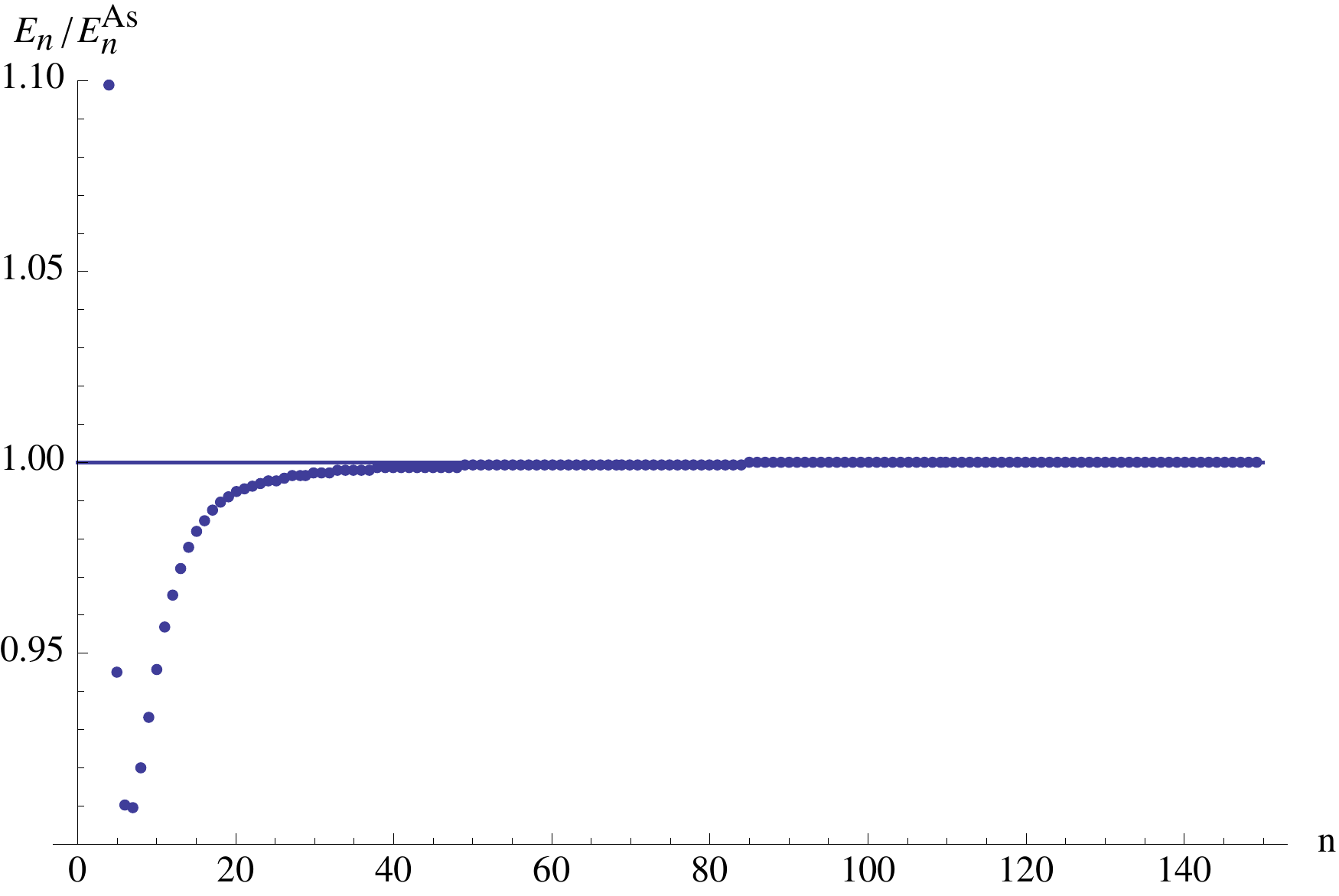}}
\hspace{2mm}
&
\resizebox{65mm}{!}{\includegraphics{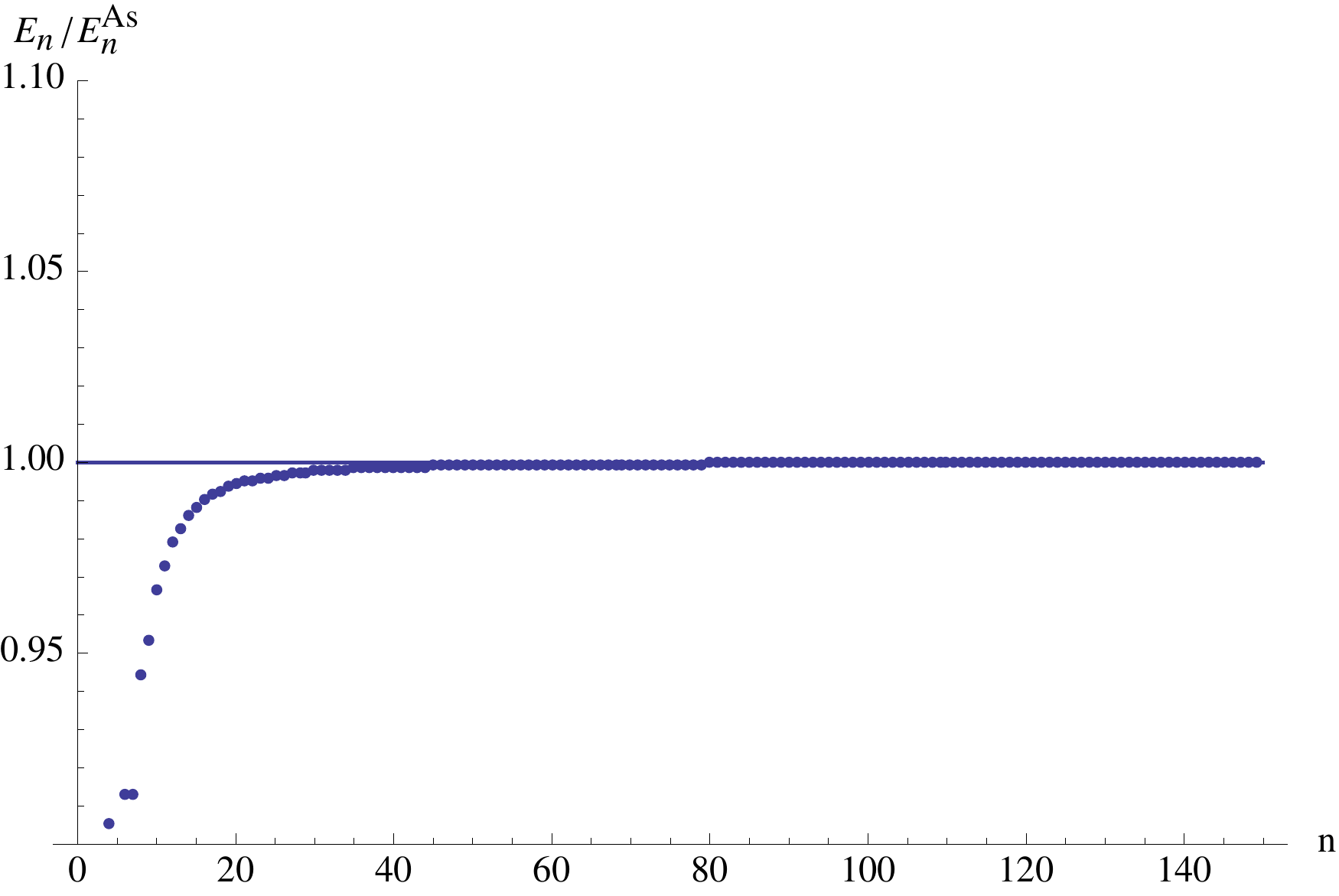}}\\
{\small $\eta=\frac{1}{10}$} & {\small $\eta=\frac{1}{2}$}
\vspace{5mm}
\end{tabular}
\end{center}
  \caption{Ratio between the exact perturbative coefficients $E_n$ and their asymptotic large order behaviours $E_n^{As}$ for different $\eta$. On the left plot $\eta<\eta_c$ and the instanton-anti-instanton action is the dominant effect, while on the right plot $\eta>\eta_c$ and the complex-instanton is dominant. 
}
\label{fig:LargeOrdRe}
\end{figure} 

\subsection{Large order behaviour in quantum mechanics}
 Let us analyse more in detail the large order behaviour of the perturbative coefficients (\ref{eq:LargeOrderRe})-(\ref{eq:LargeOrderIm}).
 Firstly, for $\eta\in\mathbb{R}$, we have that $\vert 2 S_I\vert \leq \vert S_{\mathbb{C}I}\vert$ for $0\leq \eta\leq\eta_{c}\approx 0.274$ so that the leading asymptotic behaviour of the perturbative energy coefficients (\ref{eq:LargeOrderRe}) is given by,
 \be
 E_n\sim E_n^{As}  =  A(\eta) \left(\frac{1}{2S_I}\right)^{n + 1}
  \Gamma(n+1) \left( 
  1 +a_{I\bar I}^{(1)}(\eta) \frac{2S_I}{n}\right).
 \ee
 We plugged in the above equation the Stokes constant (\ref{eq:A}) and the first instanton coefficient  (\ref{eq:aI}) obtained via resurgence theory and we can see from Figure \ref{fig:LargeOrdRe} the perfect agreement with the numerical coefficient $E_n$ evaluated in Section \ref{sec:WKBstd}.
 
 We can push it even further and consider the difference, 
 \be
 E_n-E_n^{As}\sim  A(\eta) \left(\frac{1}{2S_I}\right)^{n + 1}
  \Gamma(n+1) \left( a^{(2)}_I(\eta) \frac{(2S_I)^2}{n(n-1)}+O(n^{-3})\right)\,.
 \ee
 We applied Richardson extrapolation (see e.g. \cite{BenderOrszag}) and obtained predictions for $a^{(2)}_{I\bar I}(\eta)$ for different values of $\eta$ presented in Table \ref{tab:LOB-a2I}.
 The coefficient $a^{(2)}_{I\bar I}(\eta)$ can be in principle computed by expanding (\ref{eq:2deltanu}) to higher order in $g^2$ making use of the global boundary condition, although  it is necessary to know the uniform WKB wave-function $u_3(\pi/2)$ at mid turning point as in (\ref{eq:UWKBwftp}).
 
For $\eta>\eta_{c}$ we have that $\vert 2 S_I\vert > \vert S_{\mathbb{C}I}\vert$  so that the leading asymptotic behaviour of the perturbative energy coefficients (\ref{eq:LargeOrderRe}) is,
 \be
 E_n\sim E_n^{As}  = B(\eta) \left(\frac{1}{S_{\mathbb{C}I}}\right)^{n + 1/2}
  \Gamma\left(n+\frac{1}{2}\right)\left(1+a_{\mathbb{C}I}^{(1)}(\eta) \frac{S_{\mathbb{C}I}}{n-\frac{1}{2}}\right)\,.
 \ee
 We substituted in the above equation the predicted (\ref{eq:BCI}) Stokes constant\footnote{Note that, despite $B(\eta)$ being purely imaginary, since the complex-instanton action is negative the factor $(S_{\mathbb{C}I})^{-n - 1/2}$ produces an additional $i$ making everything real. } and the first perturbative correction to the complex-instanton and we can see from Figure \ref{fig:LargeOrdRe} the perfect agreement with the numerical coefficient $E_n$ evaluated in Section \ref{sec:WKBstd}.
 
  Turning to the complex sector, we can also push it further and consider the difference, 
 \be
 E_n-E_n^{As}\sim B(\eta) \left(\frac{1}{S_{\mathbb{C}I}}\right)^{n + 1/2}
  \Gamma\left(n+\frac{1}{2}\right)\left(a^{(2)}_{\mathbb{C}I}(\eta) \frac{(S_{\mathbb{C}I})^2}{(n-1/2)(n-3/2)}+O(n^{-3})\right)\,.
 \ee
 We applied Richardson extrapolation and obtained predictions for $a^{(2)}_{\mathbb{C}I}(\eta)$ for different values of $\eta$ presented in Table \ref{tab:LOB-a2CI}.
 Note that the predicted Stokes constant and the perturbative coefficients $a^{(1)}_{\mathbb{C}I}(\eta)\,,\,a^{(2)}_{\mathbb{C}I}(\eta)$ cannot be obtained from uniform WKB in any straightforward manner so they are genuine new predictions, coming from resurgent theory, for the perturbative expansion on top of the complex-instanton solution!
 
 \begin{table}[tb]
\begin{center}
  \begin{tabular}{clc}\hline
$\eta$ & Estimate of $a^{(2)}_{I\bar{I}}(\eta)$&\smallskip \\ \hline
$4\,\ii/20$ &$ -0.021810957$\\
$\ii/20$ & $-0.096233596$\\
$0$ &  $ -0.10156250000$\\
$1/20$ & $-0.1069366843$ \\
$1/10$ & $-0.1233298339$ \\
$3/20$ & $ -0.1515473901$ \\
$4/20$ & $-0.192911406$ \\
\hline
\end{tabular}
\end{center}
\caption{\label{tab:LOB-a2I}The numerical estimate of $a^{(2)}_{I\bar{I}}(\eta)$ from the large order behaviour of the perturbative expansion.
We use $150$ perturbative coefficients and the $12$th Richardson extrapolation. For $\eta=0$ we agree with the expected value $a^{(2)}_{I\bar{I}}(0)=-13/128$.}
\end{table}

 \begin{table}[tb]
\begin{center}
  \begin{tabular}{clc}\hline
$\eta$ & Estimate of $a^{(2)}_{\mathbb{C}I}(\eta)$&\smallskip \\ \hline
$2/5$ & $-0.371895139$ \\
$1/2$ & $-0.5871153429$ \\
$3/4$ & $-1.718457263$ \\
$1$ & $-4.280381944$ \\
$2$ & $-51.79280382$ \\
\hline
\end{tabular}
\end{center}
\caption{\label{tab:LOB-a2CI}The numerical estimate of $a^{(2)}_{\mathbb{C}I}(\eta)$ from the large order behaviour of the perturbative expansion.
We use $150$ perturbative coefficients and the $12$th Richardson extrapolation.}
\end{table}

\begin{figure}[tb]
\begin{center}
\begin{tabular}{cc}
\resizebox{65mm}{!}{\includegraphics{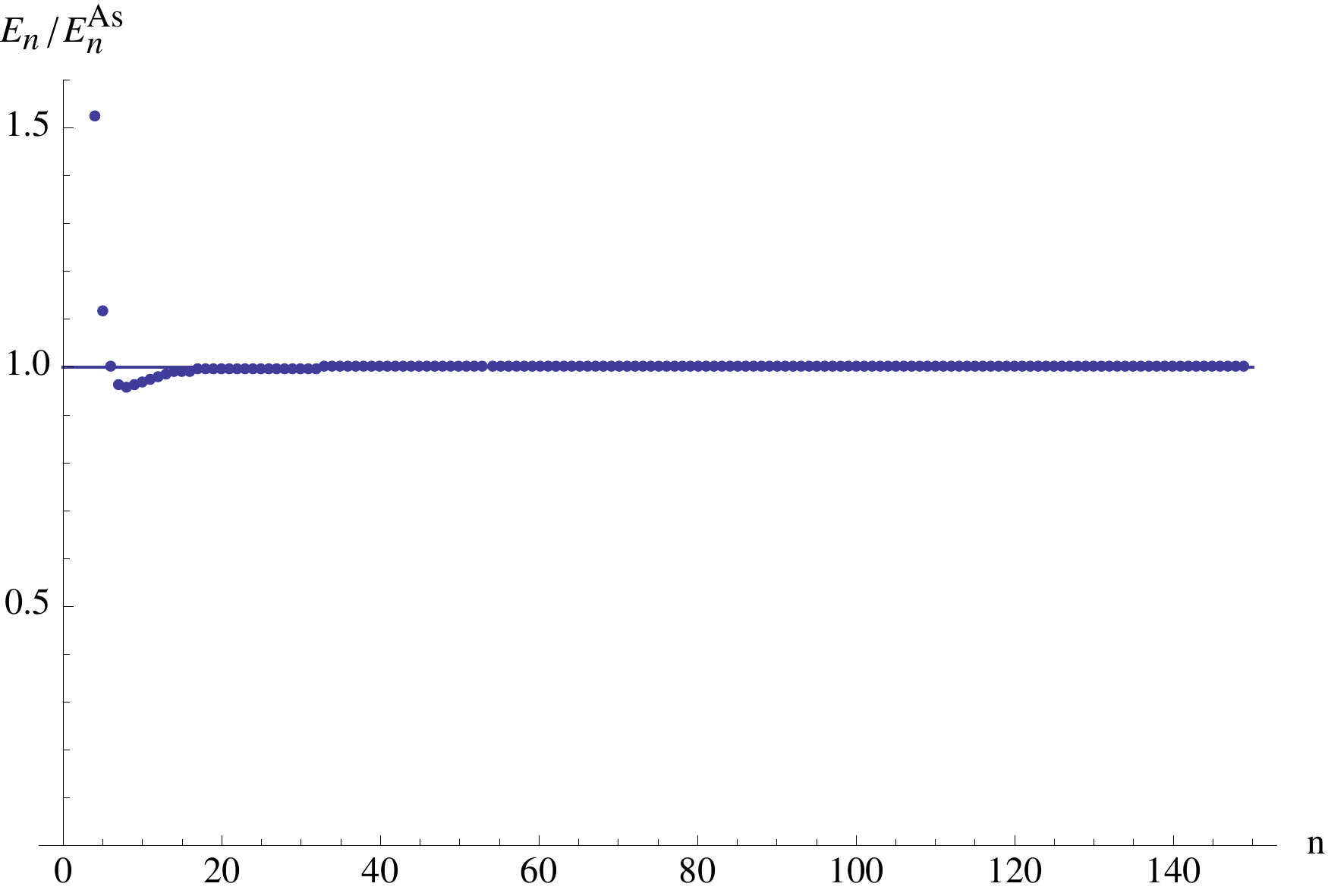}}
\hspace{2mm}
&
\resizebox{65mm}{!}{\includegraphics{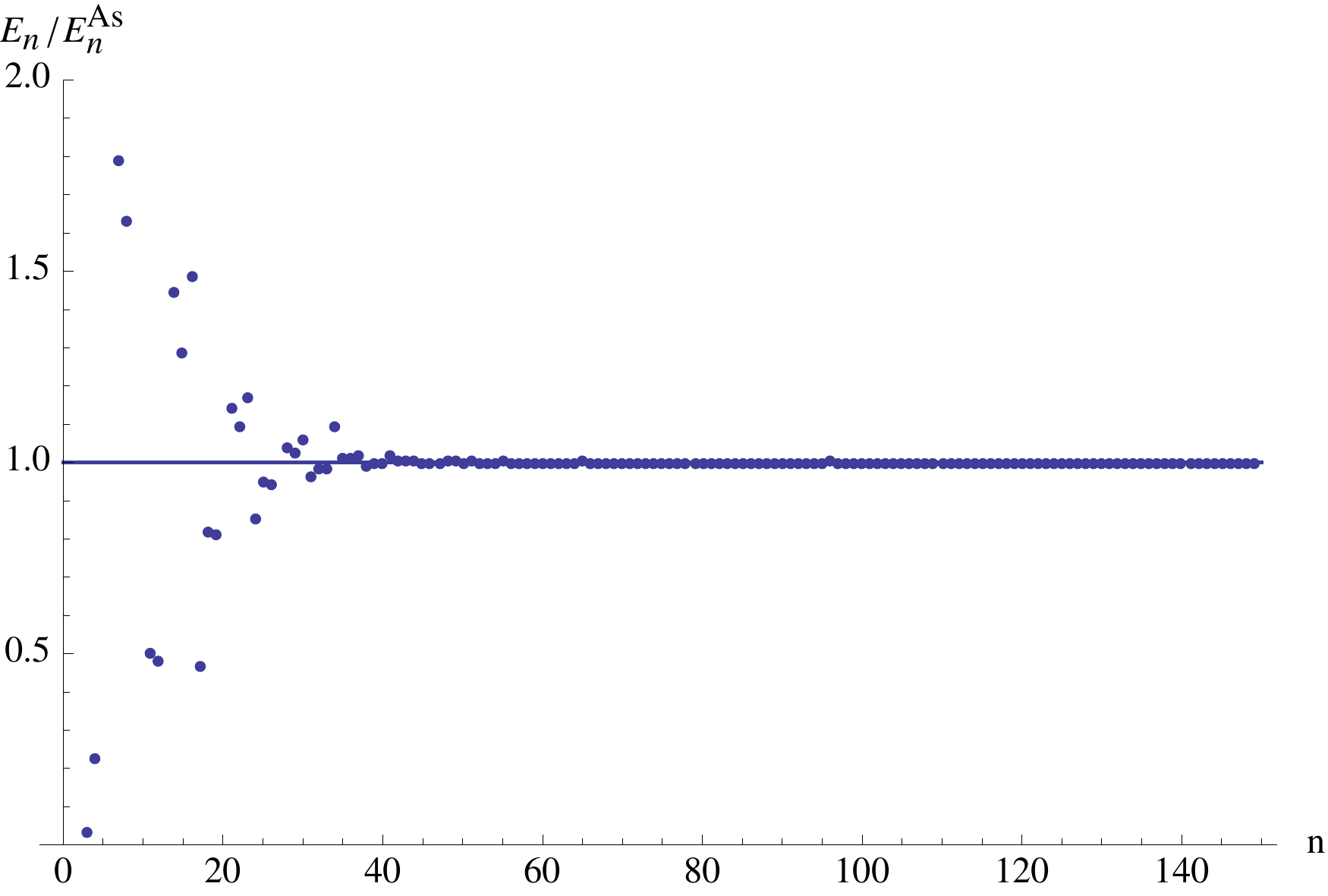}}\\
{\small $\eta_R=\frac{1}{5}$} & {\small $\eta_R=\frac{1}{2} $}
\vspace{5mm}
\end{tabular}
\end{center}
  \caption{Ratio between the exact perturbative coefficients $E_n$ and their asymptotic large order behaviours $E_n^{As}$ for different $\eta = \ii \eta_R$. On the left plot the instanton-anti-instanton action is the dominant effect, while on the right plot the two complex-instantons are the competing saddles producing an oscillating behaviour.
}
\label{fig:LargeOrdIm}
\end{figure}

We pass now to the complex realm with $\eta= \ii\eta_R$ and $\eta_R\in\mathbb{R}$.
For $0\leq\eta_R\leq\tilde{\eta}_c\approx 0.402$ we have that $\vert 2S_I \vert \leq \vert S_{\mathbb{C}I}\vert = \vert S_{\widetilde{ \mathbb{C} I } }\vert $.
In this regime the leading asymptotic behaviour of the perturbative energy coefficients (\ref{eq:LargeOrderIm}) is given by,
 \be
 E_n\sim E_n^{As}  =  A(\ii\eta_R) \left(\frac{1}{2S_I}\right)^{n + 1}
  \Gamma(n+1) \left( 
  1 +a_{I\bar I}^{(1)}(\ii\eta_R) \frac{2S_I}{n}\right).
 \ee
 We plugged in the above equation the Stokes constant (\ref{eq:A}) and the first instanton coefficient  (\ref{eq:aI}) obtained via resurgence theory and in Figure \ref{fig:LargeOrdIm} we see the perfect agreement with the numerical coefficient $E_n$ evaluated in Section \ref{sec:WKBstd}.
 
Once again we can consider the difference 
\be E_n-E_n^{As}\sim  A(\ii\eta_R) \left(\frac{1}{2S_I}\right)^{n + 1}
  \Gamma(n+1) \left(a^{(2)}_{I\bar I}(i \eta_R) \frac{(2S_I)^2}{n(n-1)}+O(n^{-3})\right)\,,
  \ee
and after Richardson extrapolation we can obtain predictions for $a^{(2)}_{I\bar I}(\eta)$ for different values of $\eta$, see Table \ref{tab:LOB-a2I}. 

Finally for $\eta_R>\tilde{\eta}_c\approx 0.402$ we have that $\vert 2S_I \vert > \vert S_{\mathbb{C}I}\vert = \vert S_{\widetilde{ \mathbb{C} I } }\vert $.
In this regime the leading asymptotic behaviour of the perturbative energy coefficients (\ref{eq:LargeOrderIm}) is given by,
 \be
 \begin{aligned}
 E_n\sim E_n^{As}  &=  B(\ii \eta_R) \left(\frac{1}{S_{\mathbb{C}I}}\right)^{n + 1/2}
  \Gamma\left(n+\frac{1}{2}\right)\left( 
  1 +a_{\mathbb{C}I}^{(1)}(\ii \eta_R) \frac{S_{\mathbb{C}I}}{n-\frac{1}{2}} ) \right)\\
  &- B(\ii \eta_R) \left(\frac{1}{S_{\widetilde{ \mathbb{C} I } }}\right)^{n + 1/2}
  \Gamma\left(n+\frac{1}{2}\right)\left( 
  1 +a_{\mathbb{C}I}^{(1)}(\ii \eta_R) \frac{S_{\widetilde{ \mathbb{C} I } }}{n-\frac{1}{2}} \right)\,.
  \end{aligned}
  \ee
From (\ref{eq:SCIIm}-\ref{eq:SCItildeIm}) we know that $\overline{S_{\mathbb{C}I}}  =S_{\widetilde{ \mathbb{C} I } }$ so that for $n\gg1$ the two saddles will compete with one another producing an oscillating asymptotic behaviour on top of the factorial growth, 
\be
 E_n^{As}  \sim \Gamma\left(n+\frac{1}{2}\right) \frac{\sin \left[(n + 1/2) \,\mbox{arg}(S_{\mathbb{C}I})\right] }{\vert S_{\mathbb{C}I}\vert^{g + 1/2} }\left(1+O(n^{-1})\right)\,,
 \ee
 matching beautifully the numerical coefficients $E_n$ as depicted in Figure \ref{fig:LargeOrdIm}.

 \subsection{The role of complex saddles}\label{sec:cSaddles}
 As already realised long time ago by Bogomolny and Zinn-Justin \cite{Bogomolny:1980ur,ZinnJustin:1981dx} (see also \cite{Balitsky:1985in}) and recently revisited in the context of Morse-Picard-Lefschetz theory \cite{ChermanUP,Behtash:2015loa}, when we consider instanton-anti-instanton events we have to be careful on the integration over quasi-zero modes.
 
In a path-integral approach we would be tempted to integrate over $\mathbb{R}$ for the separation $\tau$ between the instanton and the anti-instanton event but this would lead to a diverging contribution to the partition function. A more careful analysis suggests that the correct contour of integration, also called Lefschetz thimble, associated to the instanton-anti-instanton event lives in a complexification of the field space.
Depending on $\mbox{arg}(g)\gtrless0$ the correct contour of integration in the relevant direction of the complexified field space is given by $\tau\in \mathbb{R}\pm \ii \,\pi$.

The integral over the correct thimble for the instanton-anti-instanton contribution produces precisely the correct imaginary and exponentially suppressed term to compensate for the jump in lateral resummation of the perturbative vacuum contribution \cite{ChermanUP,Behtash:2015loa}.
Since we complexify the coupling constant to avoid Stokes lines it is only natural that we complexify also the field space to allow for more general field configurations.

This is also suggesting that in principle all finite action solutions to the complexified equation of motions could contribute to the transseries expansion of any physical observables.
This has been seen in the context of the analytic continuation of the path integral of quantum mechanics in phase space \cite{Witten:2010zr} and the analytic continuation of Chern-Simons \cite{Witten:2010cx}.
These two examples are rather peculiar since they have first order in time equation of motion.
For field theories with second order in time eoms the only examples we are aware of is given by the analytic continuation of Liouville theory \cite{Harlow:2011ny}.

In this work we have presented a new example of quantum field theory with second order in time equation of motions, for which there exist finite action solution living in a complexification of the field space: the complex-unitons (\ref{eq:defcuniton}).
We believe that their fractionalize constituents, the complex-instanton, will manifest themselves in the transseries expansion of some physical observables of the $\eta$-deformed $SU(2)$ PCM. We are currently computing such a candidate observable. 

\section{Stokes phenomena and Seiberg-Witten theory} \label{sec:Stokesgraphs}

 On a slightly different note from the previous Sections, we will analyse now the connection between our quantum mechanics  (\ref{eq:QMWKB}) and ${\cal N}=2$ gauge theories in $4$-d.
 
 As we have seen, the presence of poles in the Borel plane means that  lateral Borel resummations undergo jumping phenomena as one crosses Stokes lines. A helpful way to visualise these jumps is as topology changes of Stokes graphs (an elegant presentation with many more details than provided here can be found in \cite{Iwaki}).  As we shall detail shortly these Stokes graphs can be understood as spectral networks  \cite{Gaiotto:2009hg}  in ${\cal N}=2$ gauge theories.
 To explain this we first return to the general Schr\"odinger type equation (\ref{eq:Sch}) with a coupling $\epsilon = g^2>0$, 
 \be
 \left(  \frac{\mathrm d^2}{\mathrm d q^2} - \frac{1 }{\epsilon^2} Q(q;  \epsilon) \right) \Psi = 0  \ , 
 \ee
 and a WKB ansatz,  
 \be
 \Psi = \frac{1}{\sqrt{\chi}} \exp \int^q \chi(q') \mathrm dq' \ , 
 \ee 
 in which $\chi(q) = \sum_{i=-1}^{\infty} \epsilon^i \chi_i(q)$ and in general $Q= \sum_{i=0}^\infty \epsilon^i Q_i$. For quantum mechanics $Q=Q_0 = V-E$ and   $\chi_{-1} (q) = \sqrt{V(q) - E}$. 
 
 Viewing $q$ as a local coordinate on a Riemann surface one has that $\lambda = \epsilon \chi(q) \mathrm d q$ is a globally defined holomorphic differential one-form whose leading term is just the classical differential $\lambda_0 = \chi_{-1}  \mathrm dq = \sqrt{ V(q) - E }\mathrm dq$. The exact WKB quantisation condition reads,
\be\label{eq:exactWKB}
\int_\gamma \lambda = 2 \pi \epsilon \left( k + \frac{1}{2} \right) \ , \quad k \geq 0 \ , 
\ee  
where the regular perturbative series has an integration cycle running between two real turning points (zeros of $Q_0$).

 More generally we can analytically continue $\epsilon \rightarrow e^{-i \theta}\epsilon$ such that the Schr\"odinger equation becomes,
\be
\left(  \frac{\mathrm d^2}{\mathrm d z^2} - \frac{e^{2 \ii \theta} }{\epsilon^2} Q(z; e^{-\ii \theta}\epsilon) \right) \Psi = 0 \ ,
\ee
where now $z$ is recognised as a complex variable.  In effect, rotating $\theta$ corresponds picking out a ray in which we would like to perform a Borel resummation. 

 In what follows a crucial role is played by Stokes curves which are trajectories of the quadratic differential, 
 \be
 \phi  = Q_0(z) \mathrm dz\otimes \mathrm dz \ ,  
 \ee 
 and its square root  $\lambda_0 = \chi_{-1} (z) dz $ such that $\phi = \lambda_0 \otimes\lambda_0 $.    A directional Stokes curves emanating from a point $a$ is a trajectory $z(t)$ 
   of constant phase i.e.
\be
\lambda_0  \cdot \partial_t = e^{\ii \theta}\Leftrightarrow  {\rm Im}\, e^{\ii \theta} \int_{a}^z  \sqrt{Q_0(z)} \mathrm dz = 0  \ . 
\ee
A generic trajectory is asymptotic in both directions to a singular point (perhaps the same one and often at infinity).   A separating trajectory connects a turning point to a singular point and a finite trajectory is asymptotic in both directions to a turning point (perhaps the same one) or is closed. 

\begin{figure}[h!]
   \begin{center}
    \includegraphics[width=5cm]{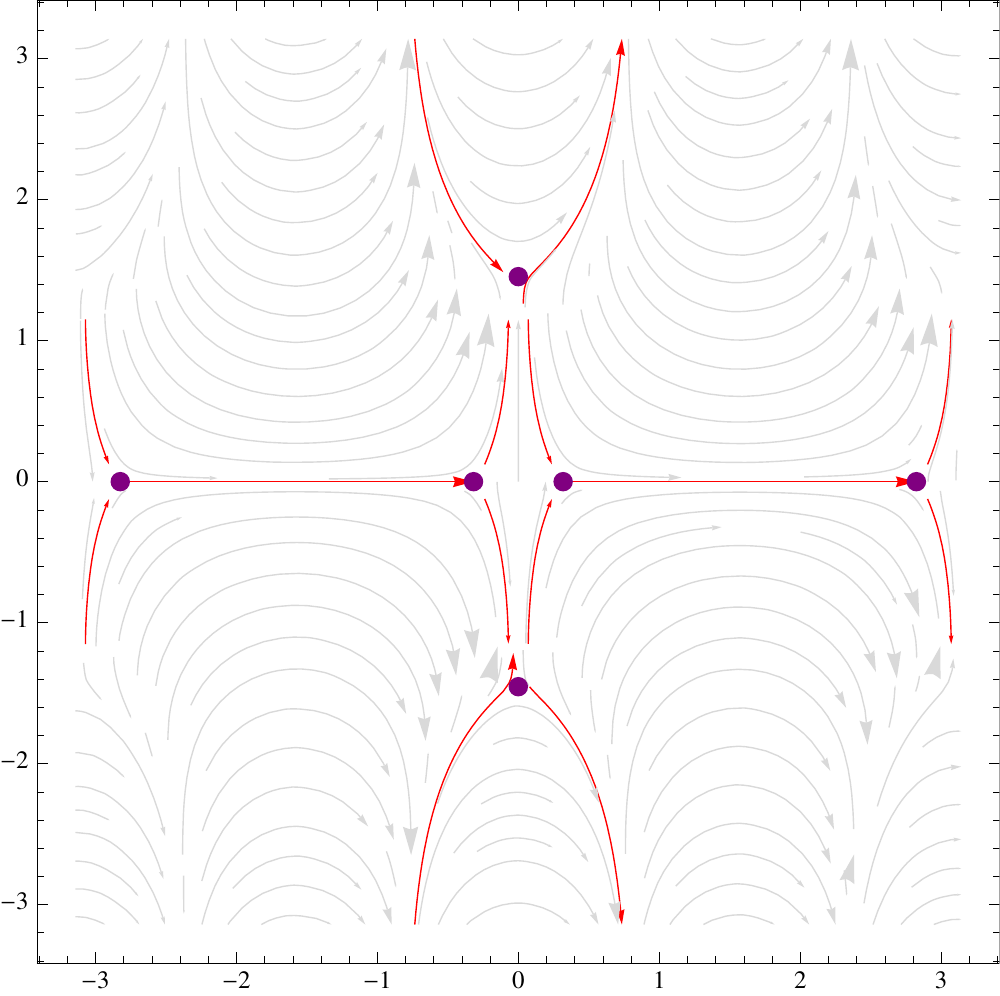} 
    \includegraphics[width=5cm]{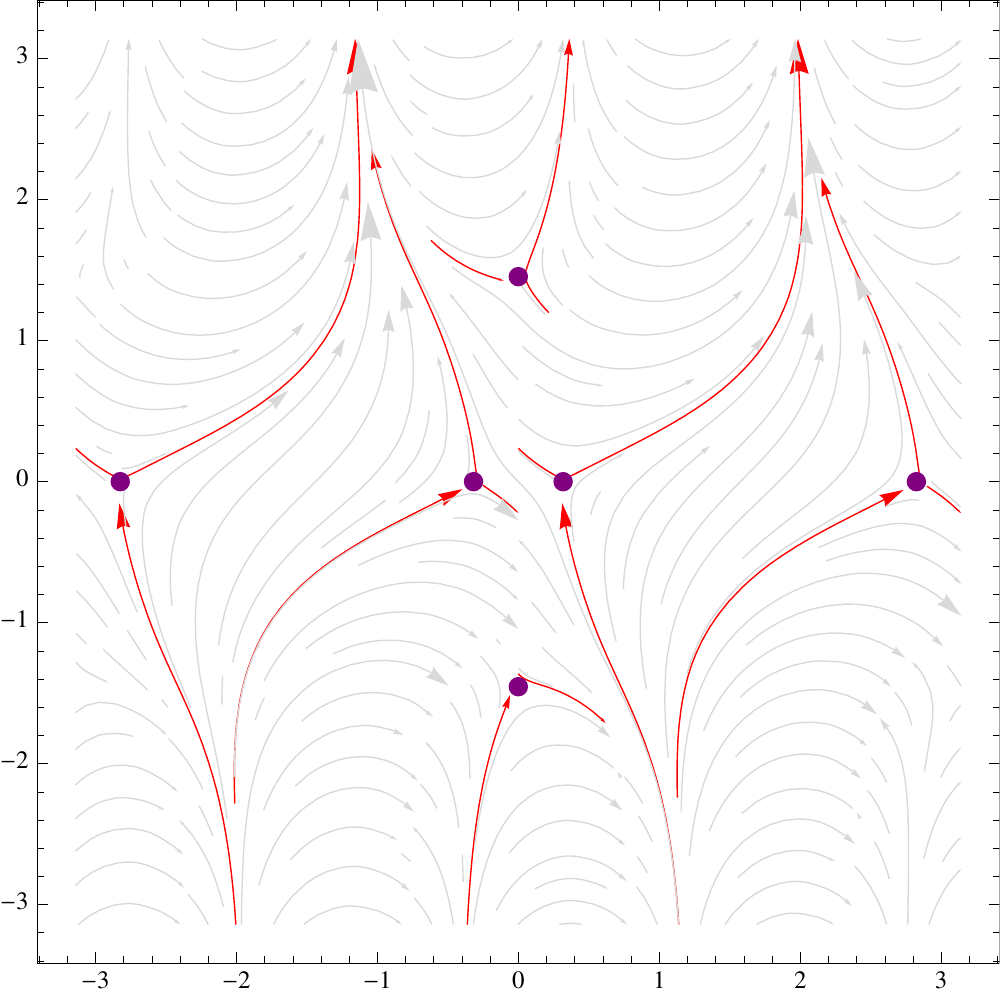} \\
    \includegraphics[width=5cm]{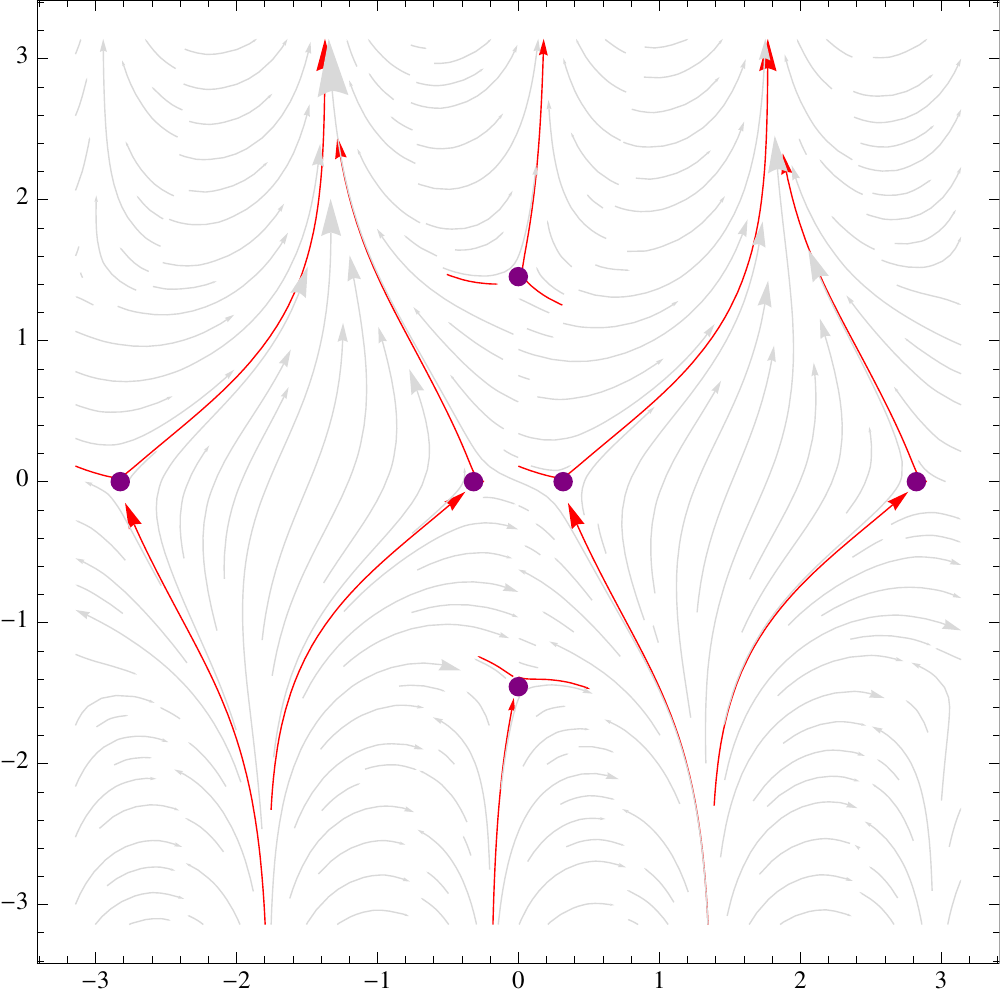} 
    \includegraphics[width=5cm]{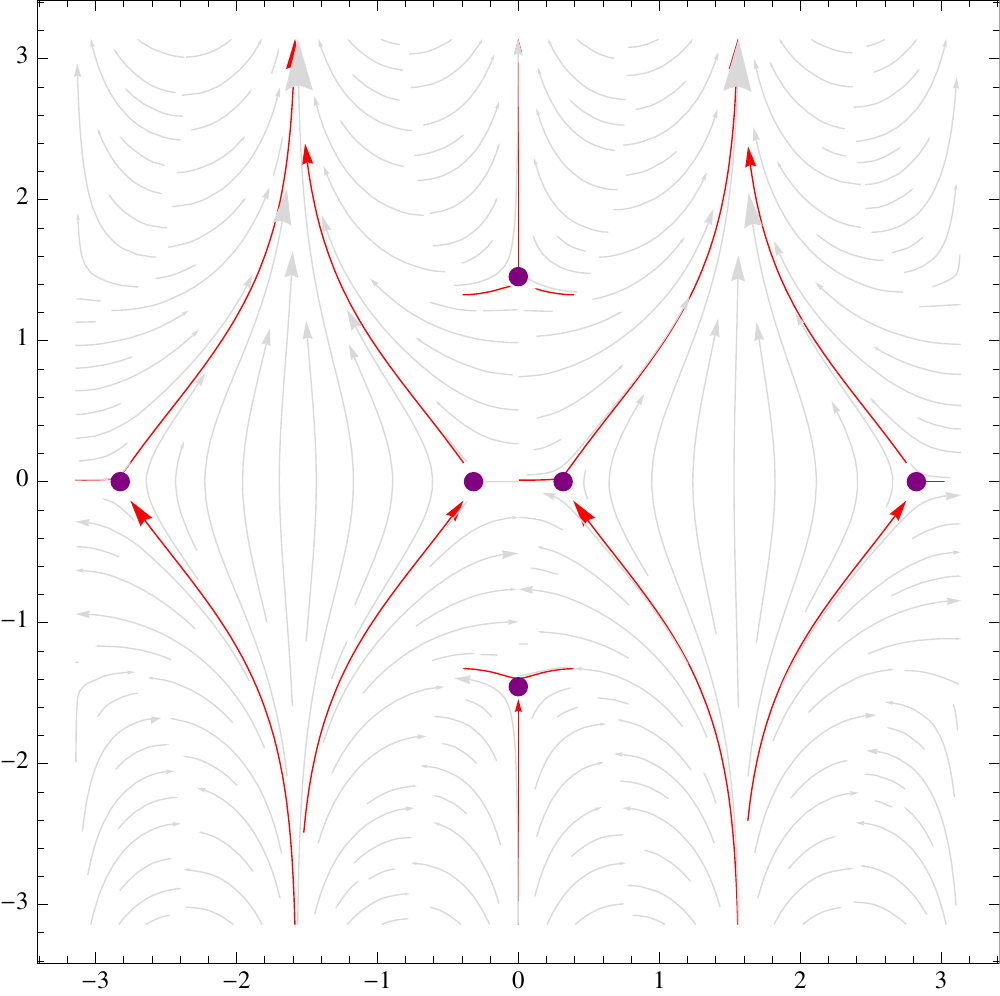}    
 \end{center}
  \caption{\footnotesize{Stokes graphs for $\eta = 1/2$ (with $E=0.1$) as we rotate the angle $\theta=\{0, \frac{\pi}{8}, \frac{3 \pi}{8}, \frac{\pi}{2} \} $ (top left to bottom right). Turning points are indicated in purple, generic trajectories in grey and separating/finite trajectories in red. At $\theta = 0 $ there are finite Stokes trajectories connecting the real turning points (along the horizontal axis) and simultaneous Stokes trajectories connecting the real turning points to the complex turning points arising from the $\eta$-deformation.}}  \label{fig:StokesgraphsEtaReal}
\end{figure}

\begin{figure}[h!]
   \begin{center}
    \includegraphics[width=5cm]{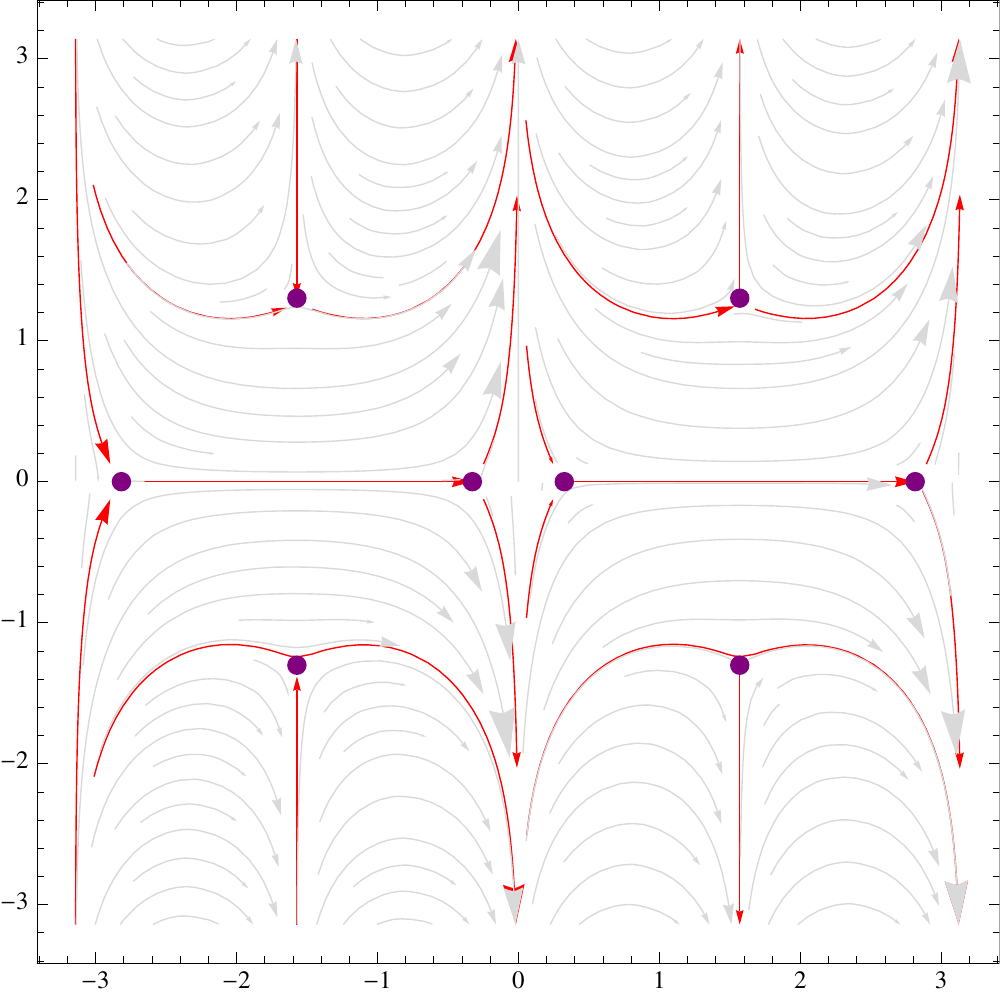} 
    \includegraphics[width=5cm]{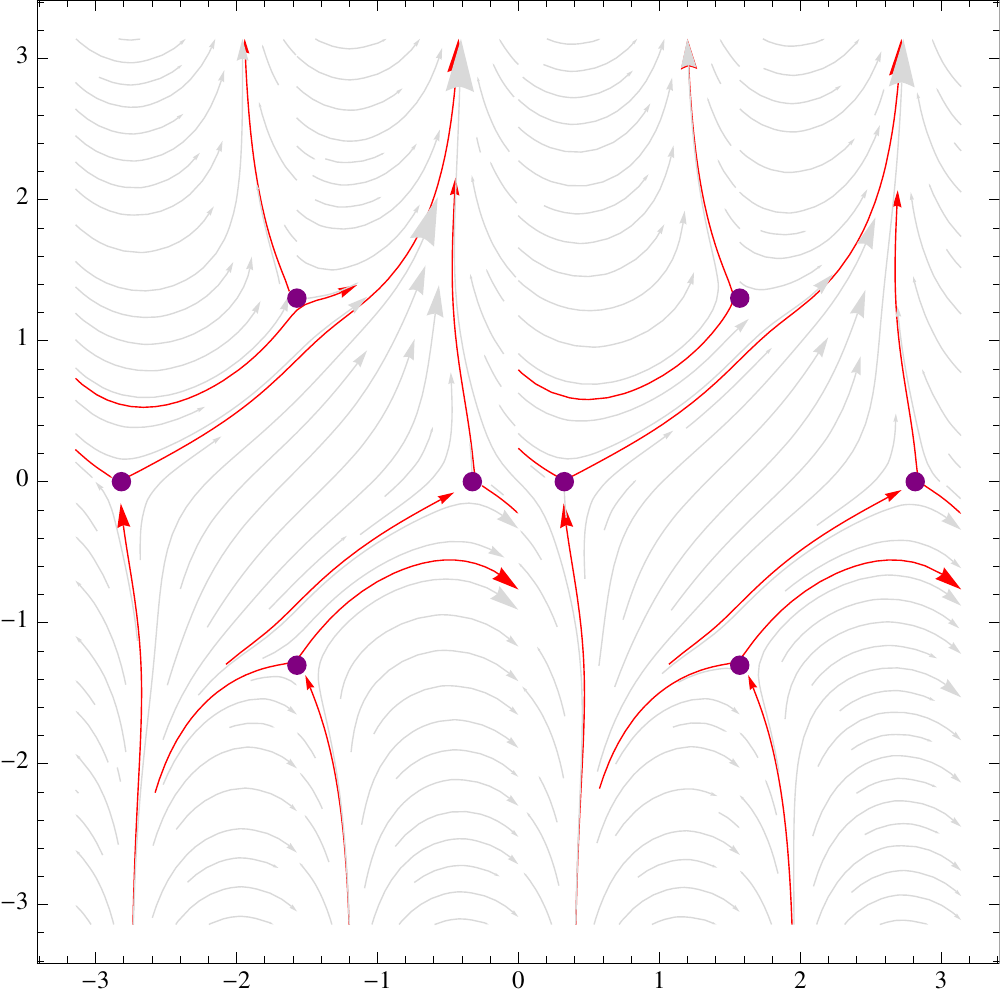} \\
    \includegraphics[width=5cm]{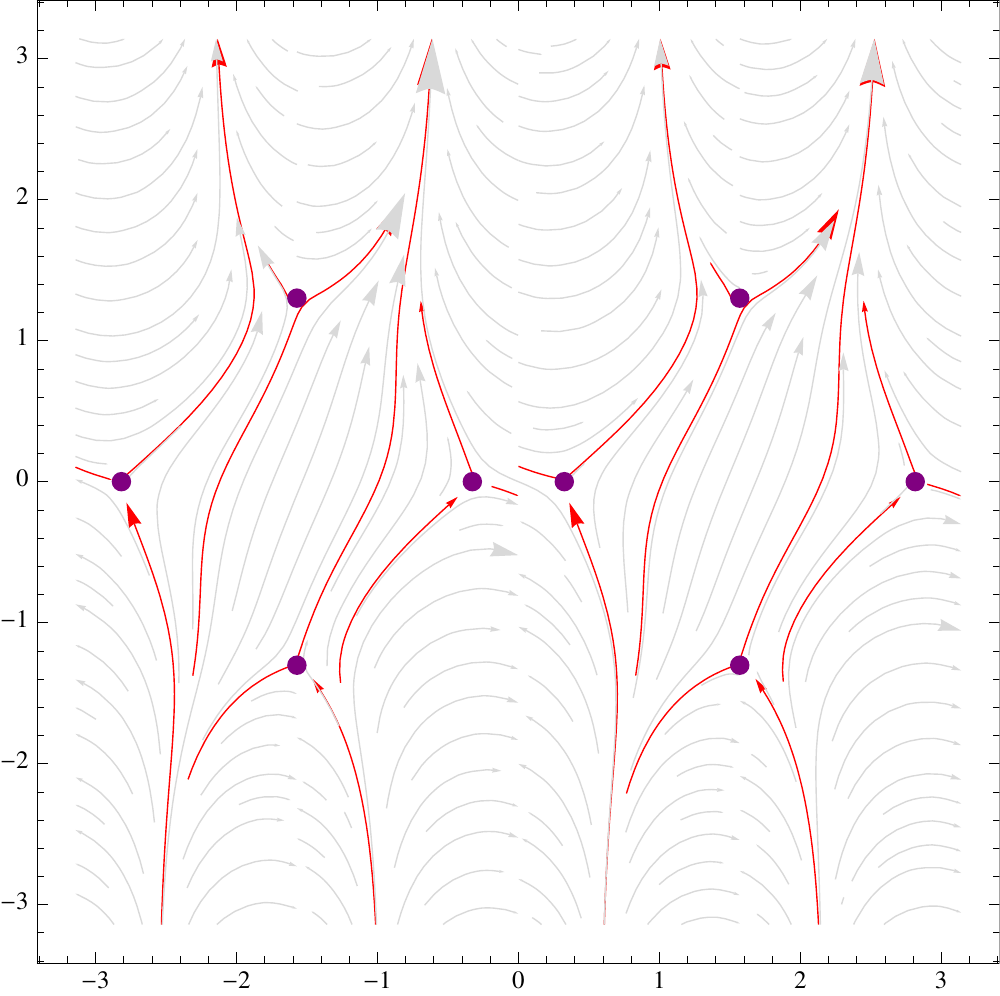} 
    \includegraphics[width=5cm]{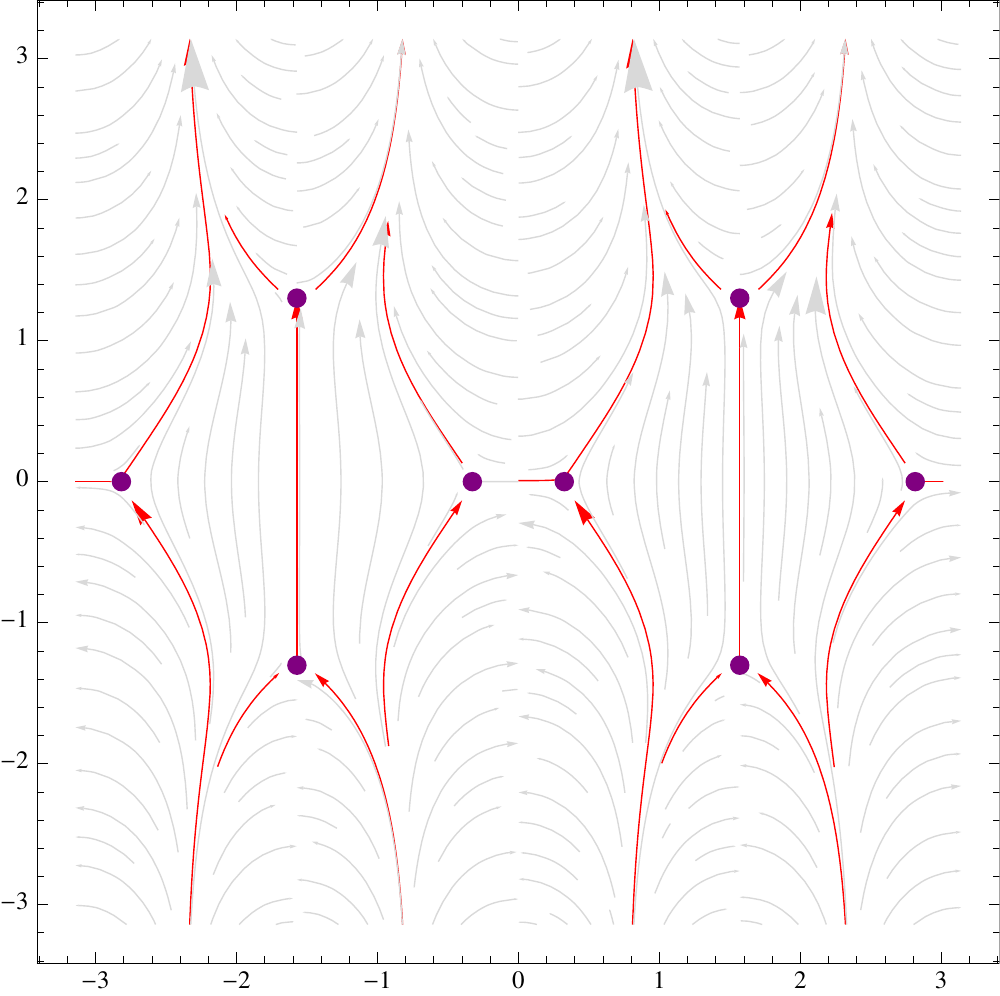}    
 \end{center}
  \caption{\footnotesize{Stokes graphs for $\eta =  \ii/2 $ (with $E=0.1$) as we rotate the angle $\theta=\{0, \frac{\pi}{8}, \frac{3 \pi}{8}, \frac{\pi}{2} \} $ (top left to bottom right).  At $\theta=0$ a finite Stokes line connects the real turning points.    Between the second and third panels we see a topology jump, in contrast to the case of $\eta \in {\mathbb R}$ displayed in Figure \ref{fig:StokesgraphsEtaReal},  indicating a Stokes direction with finite trajectory connecting the real turning point to the complex turning point. }}  \label{fig:StokesgraphsEtaIm}
\end{figure}

\begin{figure}[h!]
   \begin{center}
    \includegraphics[width=6.5cm]{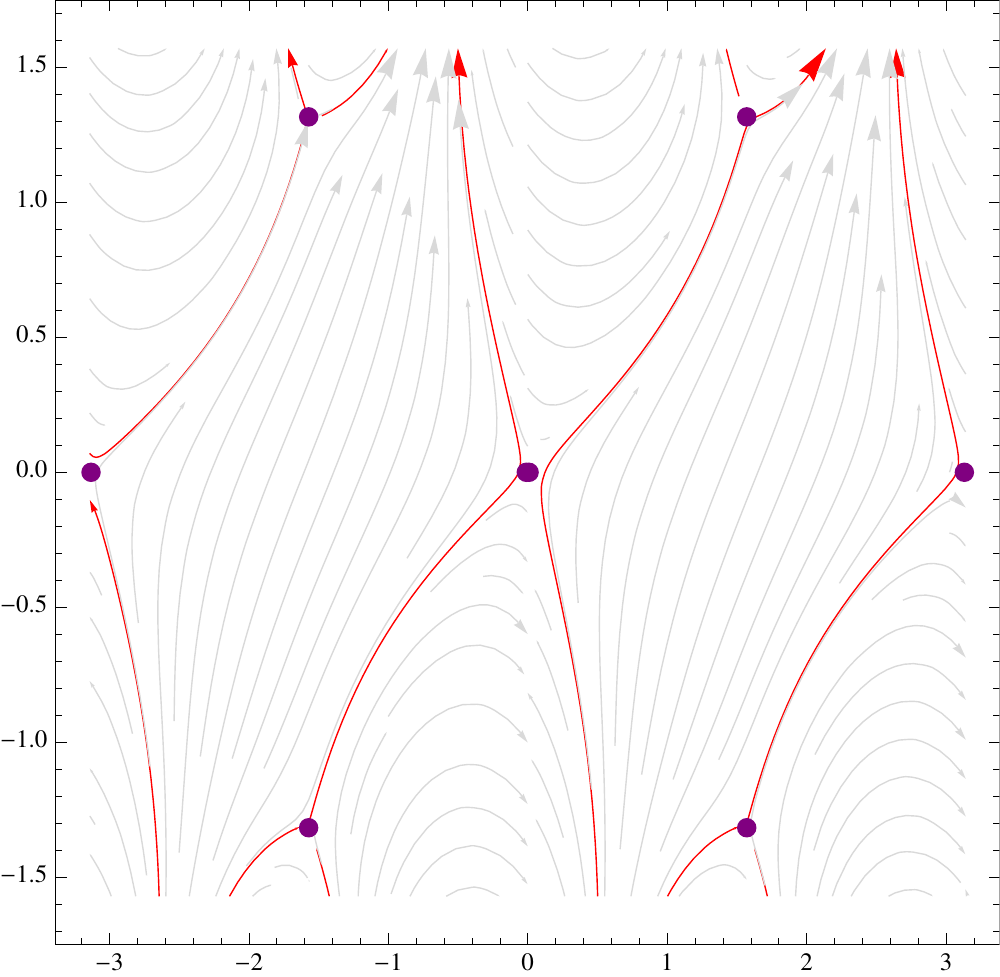}  
 \end{center}
  \caption{\footnotesize{Here we plot the critical Stokes graph  for $\eta = \ii/2 $ at the angle $\theta = \mbox{arg}(S_{\mathbb C I}) $ as we take $E=0$.   This reflects the location of singularities in the Borel plane of the perturbation theory illustrated in Figure \ref{fig:polesComplex}.}}  \label{fig:StokesgraphsCritical}
\end{figure}

For the case at hand with $V(z)=\sin^2(z)(1+\eta^2\sin^2(z))$, we plot in Figures \ref{fig:StokesgraphsEtaReal} and \ref{fig:StokesgraphsEtaIm} Stokes graphs for, respectively, $\eta$ real and $\eta$ pure imaginary. Both cases exhibit finite Stokes trajectories connecting real turning points indicating a Stokes direction at $\theta=0$. In the case of $\eta$ pure imaginary there are additional Stokes directions in which finite trajectories connect real and complex turning points. As we take the energy parameter of the quadratic differential to zero, this Stokes directions occurs precisely at the angle set by the complex uniton configuration $\theta= \mbox{arg}(S_{\mathbb C I})$ and $\theta= \mbox{arg}(S_{\widetilde{\mathbb C I}})$, as can be seen in Figure \ref{fig:StokesgraphsCritical}. This is in line with the locations of poles in the Borel plane, see Figure \ref{fig:polesComplex} and surrounding discussion.

There is a beautiful connection between Stokes graphs and  $N=2$ gauge theories \cite{Gaiotto:2009hg} where the differential $\lambda_0$ is identified with the Seiberg-Witten differential $\lambda_{SW}$.   As the angle $\theta$ progresses these Stokes graphs can undergo topology changing morphs and for particular values of $\theta$, where the topology jumps, a finite Stokes curve appears.  These finite Stokes curves correspond exactly to the appearance of BPS states, and can be directly identified with BPS strings configuration \cite{Klemm:1996bj} in the M-theory brane construction of the $N=2$ gauge theory.   What then of the full quantum differential $\lambda$?   The conjecture in \cite{Nekrasov:2009rc,Mironov:2009uv} is that periods give the exact prepotential of the $\Omega$-deformed theory in the NS limit i.e. 
\be
{\cal F}(a ; \epsilon_1) = \lim_{\epsilon_2 \rightarrow 0} \{ \e_1 \e_2 \log Z(a, \e_1 , \e_2)  \}
\ee
obeys
\be\label{eq:cycles}
a=  \oint_A  \lambda  \ , \qquad a_D = \frac{\partial {\cal F}} {\partial a} = \oint_B \lambda 
\ee
with $\lambda =\epsilon \chi(q) dq$.  The idea is that the $\hbar =\epsilon$ expansion of the quantum mechanics exactly matches the $\e_1$ expansion of the prepotential.   

This connection has been most clearly exemplified by considering the quantum mechanics associated to the Mathieu equation \cite{Mironov:2009uv,He:2010xa,He:2010if,Basar:2015xna,Kashani-Poor:2015pca}  for which the corresponding gauge theory is the $SU(2)$ $N_f=0$ theory.   Since the quantum mechanics associated to the undeformed Principal Chiral Model is that of the Mathieu equation and that of the  $\eta$-deformed PCM is  the Whittaker-Hill equation it is natural to wonder if there is some similar connection at play here.  Indeed this is the case; the $\eta$-deformed quantum mechanics can be related to a mass deformation of the $SU(2)$ gauge theory.\footnote{ Whilst this manuscript was in the final stages of preparation two preprints \cite{Piatek:2016xhq} and \cite{Ashok:2016yxz} studied the Whittaker-Hill equation, in the context of the 4-d/2-d correspondence, where it emerges as a limit of the null vector decoupling equation associated to irregular conformal blocks for the $N_f=2$ $SU(2)$ gauge theory (as previously noted in \cite{Rim:2015tsa}). The preprint \cite{Ashok:2016yxz} also provides insights into the topology changes exhibited in the Stokes graphs relevant to these gauge theories.  
     } 
 
For the Whittaker-Hill equation eq.~\eqref{eq:WH} we have a quadratic differential, 
\be
\phi_{WH} =  (-a +2 q \cos(2\theta) + 2 p \cos(4 \theta) ) \mathrm d \theta \otimes \mathrm d \theta \ . 
\ee
The quadratic differential of the first realisation of the $SU(2)$ $N_f =2$ gauge theory whose differential is \cite{Gaiotto:2009ma}, 
\be
\phi_{N_f=2} =\left(  -\frac{\Lambda}{z^4} - 2 \frac{\Lambda m_1} {z^3} + 2 \frac{u}{z^2} - 2  \frac{\Lambda m_2} {z }  + \Lambda^2 \right) \mathrm dz \otimes \mathrm dz \ ,
\ee
where $\Lambda$ is the scale of the theory, $u$ the parameter along the Coulomb branch and $m_i$ are masses of two flavour hypermultiplets.   These two differentials coincide after the transformation $z= e^{2 \ii \theta}$ and the identification,
\be
u = \frac{a}{8} \ , \quad  M= m_1 = m_2 = \frac{q}{ 4 \sqrt{p}} \ , \quad  \Lambda = \frac{\sqrt{p}}{2} \ . 
\ee
In terms of the parameters of the $\eta$ PCM, eq.~(\ref{eq:WH2}), we see the nice relation, 
\be
M= \frac{1}{4t\eta} \ , \quad  \Lambda = - \frac{ \eta}{8 g^2}  \ . 
\ee
It is striking that the mass parameter is the simple RG invariant combination of parameters in the $\eta$-deformed model.  One interesting feature of the  $SU(2)$ $N_f =2$ gauge theory is that for a certain value of the bare mass, Seiberg-Witten singularities coincide and the theory enhances to an Argyres-Douglas SCFT  \cite{Bilal:1997st}.  In the conventions above  this occurs at $M= 2\Lambda$  \cite{Gaiotto:2009hg} and in our variables this happens for $\eta^2= - \frac{1}{2}$, precisely the value for which the complex critical points coalesce into a single real value as discussed in Section \ref{sec:NPsaddles}.
 
 From the quantum mechanics point of view we expect a drastic change in the transseries expansion once $\eta^2$ passes this critical value, while in the gauge theory side we have that the Argyres-Douglas point corresponds to the critical point of a  phase transition \cite{Russo:2014nka}.
In \cite{Russo:2014nka} the free energy of the $Nf=2~ SU(2)$ theory has been computed exactly for $M<2\Lambda$ in a suitable decompactification limit and it corresponds to the imaginary part of the Seiberg-Witten prepotential evaluated at $a_D=0$, i.e. at a point where the $B$ cycle shrinks to zero (\ref{eq:cycles}). 
If we increase $M$ while keeping $a_D=0$, we can reach the point $M=2\Lambda$ where the cycle A also shrinks to zero size. Precisely at the Argyres-Douglas point the Riemann surface defined by the Seiberg-Witten differential develops a cusp and the free energy manifests a non-analytic behaviour signalling the presence of a phase transition.
It would be very interesting to understand if we can learn anything new about the two phases of the transition from the quantum mechanical model\footnote{We thank Jorge Russo for bringing these important aspects to our attention.}.   
 
As a final comment we note that the $\hbar$ expansion of the quantum mechanics corresponds to the small $\Omega$-deformation limit $\epsilon_1\to0$ while $\epsilon_2=0$. A natural question would be to understand the connections between this expansion and the weak gauge coupling constant expansion performed in \cite{Aniceto:2014hoa,Honda:2016mvg} (see also the earlier work \cite{Russo:2012kj}) for a similar class of $\mathcal{N}=2$ theories.

\newpage
 \section{Conclusions}
 
 In this paper we analyse the $\eta$-deformed $SU(2)$ principal chiral model. We find uniton-like solutions to the second order equation of motions that reduce to the standard uniton solutions in the PCM model when we send the deformation parameter $\eta\to0$. Furthermore we find new solitons that are solutions to the complexified equation of motions. These complex-unitons have quantized actions and live in a natural complexification of the field space $Sl(2,\mathbb{C})\supset SU(2)$.
 
 We put the theory on a cylinder $\mathbb{R}\times S_L^1$ with twisted boundary conditions related to adiabaticity and subsequently performed a KK-reduction to obtain an effective quantum mechanics for the low energy degrees of freedom. The quantum Hamiltonian thus obtained is related to the Whittaker-Hill equation and displays a large variety of non-perturbative finite action soliton solutions. These non-perturbative objects are instantons modified by the $\eta$-deformation, forming the fractionalized constituent of the previously discussed uniton solutions. Furthermore we find additional non-perturbative objects living in a complexification of the quantum mechanical model. These complex-instanton solutions are precisely the fractionalized constituent of the new complex-uniton solutions discussed.
 
 Resurgence theory combined with a Morse-Picard-Lefschetz approach to the path integral of quantum field theory suggests that all finite action solutions living in a complexification of the field space should contribute to the semi-classical expansion of physical observables. 
 We strengthen this claim by studying the asymptotic perturbative expansion of the ground state energy of our reduced quantum mechanics. We identify all the Stokes lines of the Borel transform of the perturbative ground state energy and relate them precisely to the instanton-anti-instanton and complex-instantons events.
 
 We expect these fractionalized constituents of the unitons and the complex-unitons to contribute to the transseries representation of the path integral for the $\eta$-deformed principal chiral model.
 Note however that generically, due to the running of the coupling constant, the location of the singularities in the Borel plane might change as we perform this adiabatic continuation from $\mathbb{R}^2$ to $\mathbb{R}\times S^1_L$ as explained in \cite{Cherman:2014ofa}. In the model under consideration we have an additional $RG$-invariant parameter that might allow us to understand quantitatively how this Borel flow behaves. We are currently investigating in perturbation theory some bona-fide quantum field theory observable to match against the quantum mechanics expectations of the Borel plane structure.
 
 We make use of uniform WKB and resurgence theory to extract non-perturbative information out of the large order behaviour of the perturbative energy coefficients and we predict the first few non-perturbative corrections on top of these solitonic objects. A proper complexified path integral derivation of these perturbative corrections is not present at the moment and it would a beautiful confirmation of the resurgence program.
 
 Finally we relate the Stokes phenomena associated with the jumps in lateral resummation for the perturbative expansion of the ground state energy of our quantum mechanics with the Stokes graph associated with the spectral network of a very particular $4$-dimensional $\mathcal{N}=2$ supersymmetric gauge theory.
 We relate the Schr\"odinger equation associated with the Whittaker-Hill equation to the Seiberg-Witten quadratic differential of a $4$-d $\mathcal{N}=2$ gauge theory with gauge group $SU(2)$ and $N_f=2$.
 As we dial the $\eta$-deformation parameter to the particular value $\eta=\ii/\sqrt{2}$ we have that multiple complexified classical vacua of the quantum mechanics coalesce and the corresponding $4$-d $\mathcal{N}=2$ theory is driven to an Argyres-Douglas SCFT.

 \section{Acknowledgment}
 We thank Gleb Arutyunov, Gerald Dunne, Oleg Evnin, Frank Ferrari, Ben Hoare, Ctirad Klimcik, Marcos Marino, Jorge Russo, Konstantinos Sfetsos, Konstantinos Siampos, Tin Sulejmanpasic, Paul Sutcliffe, Mithat Unsal, Stijn van Tongeren and Wojtek Zakrzewski for useful discussions. 
 
 D.D. is grateful to Hamburg University and DESY for their hospitality and to the Collaborative Research Center SFB 676 ``Particles, Strings, and the Early Universe'' for financial support during the final stages of this project. The work of D.C.T. was supported in part by FWO Vlaanderen through project G020714N and postdoctoral mandate 12D1215N, by the Belgian Federal Science Policy Office through the Interuniversity Attraction Pole P7/37, and by the Vrije Universiteit Brussel through the Strategic Research Program ``High-Energy Physics''. S.D. is supported by a PhD Fellowship of the Research Foundation Flanders (FWO).
 
 \begin{appendix}
 \section{Algebraic Details}
 \subsection{Conventions} 
   For a compact semi-simple Lie group $G$ corresponding to an algebra $\frak{g}$, we parametrise a group element $g\in G$ by local coordinates  $X^\m$, $\m=1,2,\dots , \dim(G)$. The right and left invariant Maurer--Cartan forms, as well as the orthogonal
matrix (or adjoint action) relating them, are defined as,
\be
\begin{aligned}
  & L^{a}_{\pm} = L^a_\m \del_\pm X^\m  =  {\rm Tr}(T_a g^{-1}\del_\pm g )\ ,
 \quad R^a_\pm =  R^a_\m \del_\pm X^\m = {\rm Tr}(T_a  \del_\pm g g^{-1} ) \ ,
\\
 & R^a_\m = D_{ab}\,L^b_\m\ ,  \quad D_{ab}={\rm Tr}(T_a g T_b g^{-1})\ ,
 \label{jjd}
 \end{aligned}
\ee
where,
\be
\mathrm{d}L=-L\wedge L\,,\quad \mathrm{d}R=R\wedge R\,,\quad \mathrm{d}D\,D^T=-F\,,\quad F_{ab}:=f_{abc}\,R_c\,.
\ee
The generators $T_a$ obey $[T_a,T_b]= f_{ab}{}^{c} T_c$, are normalised as ${\rm Tr}(T_a T_b)=\d_{ab}$,  and with respect to the Killing metric, defined by $f_{ac}{}^{d}f_{bd}{}^{c}=-c_G\,\d_{ab}$,  the structure constants with lowered indices  $f_{abc}$ are totally antisymmetric. Group theoretic indices are frequently raised  by using $\delta_{ab}$.   World-sheet light cone coordinates are defined as $\s^{\pm} = \tau \pm \sigma$.

 \subsection{Drinfeld Double and the ${\cal R}$-matrix}
It may be helpful to the reader to summarise some salient facts about the ${\cal R}$-matrix that is used to define the $\eta$-deformations.  
 Consider a semisimple Lie group $G$, a Lie algebra $\frak{g}$, and a
matrix ${\cal R}$ (an endo-morphism of $\frak{g}$),  assumed to be anti-symmetric  with respect to the Killing form on $\frak{g}$, which defines a bracket,
\be
\label{Rbracket}
[A,B]_{\R}=[\R A,B]+[A,\R B] \ ,  \quad \forall  A,B \in \frak{g} \, .
\ee
A sufficient condition for \eqref{Rbracket} to satisfy  the Jacobi identity is
the modified classical YB equation (mYB),
\be
[\R A, \R B] - \R[A,B]_\R = -c^2[A, B]  \ ,  \quad \forall  A,B \in \frak{g} \,,\quad c\in\mathbb{C}\,.
\ee
Up to trivial rescaling there are three distinct choices for the parameter $c$; $c^2=1$, $c^2 =-1 $ and $c^2 = 0$.   Here we shall restrict to a compact bosonic group for which the only solutions are $c^2=-1$.  

The two Lie-brackets (the usual one $[A,B]$ and $[A,B]_{\cal R}$) over the same vector space define a bi-algebra.   Thinking of the $[A,B]_{\cal R}$ as defining an algebra $\frak{g}_{\cal R}$  we have a Drinfeld double defined by $\frak{d}  = \frak{g} \oplus \frak{g}_R=\frak{g}^\mathbb{C}$  viewed as a Lie algebra. Indeed,    $\frak{g}^\mathbb{C}$ may be equipped with an inner product,
 \[
 \langle A+ \ii B , A'   + \ii B' \rangle = \textrm{Im} (A+\ii B, A'+ \ii B')  \ ,
 \]
 with respect to which $\frak{g}$ is a maximal isotropic {\em and} when $\R$ is anti-symmetric w.r.t. $( \cdot, \cdot)$ so is $\frak{g}_R$.

We can specify a Cartan basis of $\frak{g}^\mathbb{C}$: $H^{\mu}$ are a Hermitian basis for the Cartan sub algebra and $E^{\a}$ and $E^{-\a}$ are the raising and lowering operators for a root $\vec{\a}$ (see e.g \cite{Klimcik:2008eq} section 2.2 for details).  Then a basis for the real algebra $\frak{g}$ is given by $\{ T^{\mu}, X_{+}^{\a}, X_{-}^{\a} \}$ with $\a >0 $) where,
\be
T^{\mu}= \ii H^{\mu}  \ , \quad X_{\pm}= \frac{\ii}{\sqrt{2}} (E^{\a} \pm E^{-\a}) \ . 
\ee
In this basis a canonical choice for the ${\cal R}$ matrix is given by a simple action, 
\be
{\cal R}: \{ T, X_{+},X_{-} \} \mapsto \{ 0, X_{-}, -X_{+} \} \ . 
\ee 
Although we shall not consider this in the present context, it is possible to augment this ${\cal R}$  with ``homogenous'' solutions to the  classical $c^2=0$ Yang-Baxter Equation, i.e.   rotations amongst Cartan directions associated to so-called Drinfeld-Reshetikhin twist.  

 \end{appendix}
\setcounter{equation}{0}
\renewcommand{\theequation}{\thesection.\arabic{equation}}

 \bibliography{biblio}

 \end{document}